\documentclass[journal]{IEEEtran}

\usepackage{epsfig}
\usepackage{multirow}
\usepackage{makecell}
\usepackage{cite}
\usepackage{amsmath}
\usepackage{hyperref}
\usepackage{xcolor}
\usepackage{subfigure}
\usepackage{amsfonts}
\usepackage{balance}
\usepackage{xparse}

\newcommand{\etal}{\emph{et al.}}

\begin{document}

\title{\Huge Survey on Aerial Radio Access Networks: Toward a Comprehensive 6G Access Infrastructure}

\author{Nhu-Ngoc~Dao,
        Quoc-Viet Pham,
        Ngo Hoang Tu,
        Tran Thien Thanh,\\
        Vo~Nguyen~Quoc~Bao,
        Demeke~Shumeye~Lakew,
        and Sungrae~Cho
    \thanks{N.-N. Dao is with the Department of Computer Science and Engineering, Sejong University, Seoul 05006, Republic of Korea (e-mail: nndao@sejong.ac.kr).}
    \thanks{Q.-V. Pham is with the Korean Southeast Center for the 4th Industrial Revolution Leader Education, Pusan National University, Busan 46241, Republic of Korea (e-mail: vietpq@pusan.ac.kr).}
    \thanks{N. H. Tu and T. T. Thanh are with the Department of Computer Engineering, Ho Chi Minh City University of Transport, Ho Chi Minh City 710372, Vietnam (e-mail: tu.ngo@ut.edu.vn, thanh.tran@ut.edu.vn).}
    \thanks{V. N. Q. Bao is with the Posts and Telecommunications Institute of Technology, Ho Chi Minh City 710372, Vietnam (e-mail: baovnq@ptithcm.edu.vn).}
    \thanks{D. S. Lakew and S. Cho are with School of Computer Science and Engineering, Chung-Ang University, Seoul 06974, Republic of Korea (e-mail: demeke@uclab.re.kr, srcho@cau.ac.kr).}
}

\maketitle

\begin{abstract}
Current network access infrastructures are characterized by heterogeneity, low latency, high throughput, and high computational capability, enabling massive concurrent connections and various services. Unfortunately, this design does not pay significant attention to mobile services in underserved areas. In this context, the use of aerial radio access networks (ARANs) is a promising strategy to complement existing terrestrial communication systems. Involving airborne components such as unmanned aerial vehicles, drones, and satellites, ARANs can quickly establish a flexible access infrastructure on demand. ARANs are expected to support the development of seamless mobile communication systems toward a comprehensive sixth-generation (6G) global access infrastructure. This paper provides an overview of recent studies regarding ARANs in the literature. First, we investigate related work to identify areas for further exploration in terms of recent knowledge advancements and analyses. Second, we define the scope and methodology of this study. Then, we describe ARAN architecture and its fundamental features for the development of 6G networks. In particular, we analyze the system model from several perspectives, including transmission propagation, energy consumption, communication latency, and network mobility. Furthermore, we introduce technologies that enable the success of ARAN implementations in terms of energy replenishment, operational management, and data delivery. Subsequently, we discuss application scenarios envisioned for these technologies. Finally, we highlight ongoing research efforts and trends toward 6G ARANs.
\end{abstract}


\IEEEpeerreviewmaketitle

\section{Introduction}
The Internet of things (IoT) requires most electronic devices used in our daily lives to have Internet capability. In recent decades, macro shifts in the technological requirements of mobile services have occurred with each network generation~\cite{david20186g}. The first wave realizes service heterogeneity transformation from voice-only services in the first generation (1G) to multimedia services in the third generation (3G). The second wave focused on improvements in quality of service (QoS) in terms of peak throughput maximization, end-to-end latency minimization, network connectivity maximization, and user serviceability, which have been the key challenges for third to fifth-generation (5G) networks~\cite{dao2020ic,dao2017adaptive}. In recent years, a third wave has begun to address concerns regarding access to ubiquitous mobile technology services which have become integral to our lives, developing from a foundation being established by the IoT paradigm in (sub)urban areas with 5G toward Internet globalization in the sixth generation (6G) and beyond. For this purpose, the International Telecommunication Union (ITU) has launched the Network-2030 initiative to discuss potential technologies and innovation~\cite{network2030}. 

Internet globalization aims to equip all users with Internet capability~\cite{bi2019ten}. Primary examples include event-based communications where vast numbers of users assemble in one location (e.g., concerts, festivals, stadiums, and squares), which require a tremendous increase in network service connections for a short time, thereby potentially overloading existing communication infrastructure. Another important case is disaster communications in which network infrastructure may be damaged or destroyed and thus rendered unable to provide service, nonetheless communication remains critical to support search and rescue (SAR) operations. From another perspective, scheduled communications may iteratively establish services as planned periodically, such as environmental sensing reports in intelligent agriculture and aerial forest surveillance. A fixed infrastructure is not necessarily suitable for such scenarios, which call for a mobile portable infrastructure instead. Finally yet importantly, sparsely populated communities (e.g., in isolated areas and at sea) especially would benefit from improved Internet access capabilities for emergency and informational communications~\cite{chen2020vision}.

\subsection{Motivation} \label{sec1a}
Unfortunately, current access infrastructure designs have not comprehensively considered network serviceability in the aforementioned scenarios. Such serviceability has not been included in the requirements for the most recent telecommunication technology, 5G~\cite{david20186g,samdanis2020road}. In particular, the ITU IMT-2020 specification defines that standard 5G infrastructure is to target either network performance improvements or QoS such as connectivity, throughput, latency, reliability, energy efficiency, and mobility~\cite{imt2020}. However, the responsibility of broadening access to Internet service is commonly seen as a business concern instead of a crucial technical requirement. As a result, mobile services in underserved areas have not been given significant attention to accommodate these communication needs.
\begin{figure*}[!t]
\centering
\includegraphics[width=1.0\textwidth]{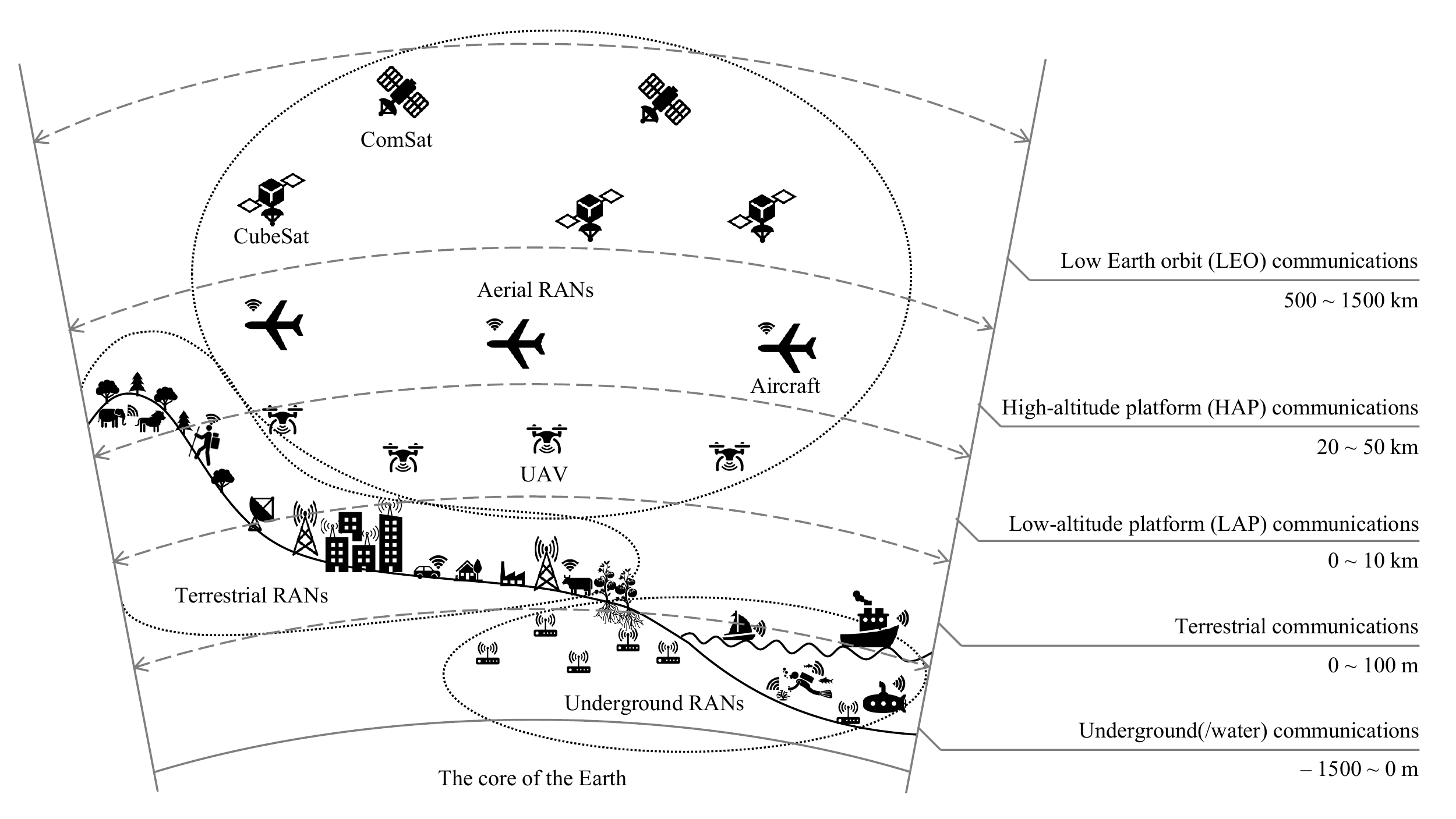}
\caption{ARANs in a comprehensive 6G access infrastructure.} 
\label{fig:6gaccess}
\end{figure*}

In reality, these scenarios have been only partially and incompletely addressed, in an unsystematic fashion. The fundamental advantages of aerial networking infrastructures such as mobility, better channel access, improved coverage, and a higher probability of line-of-sight (LoS) signal propagation compared to terrestrial networks serve as a promising foundation to overcome the current challenges. For instance, owing to their quick and flexible three-dimensional (3-D) deployment capabilities, unmanned aerial vehicles (UAVs) have been involved in wireless systems for multiple purposes such as communication relaying, coverage expansion, and traffic alleviation~\cite{zhang2020uav}. To support various application scenarios, multiple UAVs may collaborate and establish different networking topologies including mesh, star, bus, and hierarchical architectures with respect to the communication requirements and the capabilities of the UAVs. Conversely, satellite communications have been exploited to provide users with internetworking services in remote areas for the last decade~\cite{abo2019survey}. In this system, ground stations typically act as transfer nodes that connect to the satellite constellation and relay communications to user devices. Recently, an ambitious project (namely Starlink\footnote{https://www.starlink.com}) being constructed by SpaceX Corp. began establishing a constellation of thousands of small satellites in low Earth orbit (LEO) to deliver high-speed broadband Internet directly to end users across the globe. Individually, these solutions have supported several specific applications with a basic degree of success. However, there has not been a systematic integration of such infrastructures into mobile networks as native components that can seamlessly provide global communications.

The above observation leads to the necessity of studying current proposals and reorganizing such aerial access architectures toward a comprehensive access infrastructure for 6G networks. Because airborne objects are involved in these systems as primary transceiver components, we introduce an appropriate definition for this type of infrastructure, i.e., aerial radio access network (ARAN) (see Fig.~\ref{fig:6gaccess}). ARANs provide a radio access medium from the sky to end users for Internet services using aerial base stations (ABSs). Typical ABSs include aircraft and airships such as UAVs, drones, balloons, and airplanes equipped with wireless transceiver antennas, while the backhaul links may be provided by (miniaturized) satellites and terrestrial macro base stations. Conceptually, unification of existing systems to form a multitier and hierarchical ARAN is expected to help provide a comprehensive and standard reference model for future research.

\subsection{6G Visions Toward a Comprehensive Access Infrastructure} \label{sec1b}

Specifically, visions of 6G wireless systems are discussed in some recent surveys~\cite{Saad2020AVision, giordani2020toward, Yaacoub2020AKey6G}. Saad~\etal~\cite{Saad2020AVision} provided a vision of 6G applications and technologies and predicted that completely integrating terrestrial, airborne, and satellite communications will play a vital role in 6G wireless systems. In this context, low-altitude platforms (LAPs) complement underserved areas on the ground with additional connectivity, while both LAPs and terrestrial networks may benefit from high-altitude platforms (HAPs) and satellite communications for backhaul links. Giordani~\etal~\cite{giordani2020toward} speculated on a number of important 6G use cases (e.g., augmented/virtual reality, teleoperation, unmanned mobility, and Industry 4.0) and associated technologies (e.g., full-duplex communications, out-of-band channel estimation, and sensing-based localization). To provide Internet connectivity to rural areas, Yaacoub~\etal~\cite{Yaacoub2020AKey6G} reviewed technologies for front/back-end interface connectivity (e.g., 5G, drones/UAVs, satellite, and free-space optics) and the challenges of efficiently deploying electrical grids. 

Motivated by the successes of artificial intelligence (AI) in many engineering disciplines, many believed that AI would revitalize 6G wireless systems by providing intelligent solutions~\cite{Shafin2020Artificial, Song2020Artificial}. Holographic beamforming, orbital angular momentum multiplexing, and the Internet of (Bio-)Nano-Things were also identified as promising 6G technologies~\cite{zhang20196g}. To accelerate 6G research, the Finnish 6G Flagship program\footnote{https://www.6gchannel.com} recently published 12 specific white papers on various research themes, for instance, applications of machine learning for 6G in~\cite{ali20206g}, edge intelligence in~\cite{peltonen20206g}, 6G broadband connectivity in~\cite{rajatheva2020white}, and massive IoT connectivity in~\cite{mahmood2020white}. To efficiently provide operational environments for these foundational technologies, existing mobile systems will need to be supplemented toward a comprehensive infrastructure in 6G~\cite{you2021towards}.

With an expected extraordinary popularization of Internet capability on daily-life devices in future 6G networks, vertical and horizontal communication coverage has never stopped expanding to improve serviceability to accommodate massive numbers of users and services. To this end, a ubiquitous 3-D coverage model was introduced in~\cite{zhang20196g}~and~\cite{huang2019survey} integrating different aspects of the architecture such as aerial, terrestrial, and underground communications into a unified access platform. A combination of existing infrastructures and supplemented components forming such a unified platform defines comprehensive 6G access infrastructures. Unfortunately, the aforementioned technical recommendation documents have not yet focused on standardizing requirements toward these 6G infrastructures.

Directly interacting with end users, radio access networks (RANs) play important roles in infrastructure evolution. Some surveys over the last decade have been dedicated to reviewing RANs, and recently there have been some speculative studies on 6G. However, they fail to provide an up-to-date survey on ARANs. Several aspects of RAN in 5G networks have been reviewed in the existing literature, for example, radio resource management in~\cite{olwal2016survey} and control-data separation architecture in~\cite{mohamed2015control}. Considered a key enabling technology to enhance 5G performance and cost efficiency, cloud RANs (CRANs) were well-reviewed in some survey articles such as~\cite{Peng2016Recent, Alimi2018Toward,habibi2019comprehensive}. In particular, Peng~\etal~\cite{Peng2016Recent} presented various CRAN architectures, key techniques (e.g., front-end interface compression and collaborative processing), and cooperative resource allocation studies for CRAN systems. A comprehensive tutorial on CRAN optical fronthaul in 5G cellular networks was provided in~\cite{Alimi2018Toward}. As 5G-and-beyond (B5G) networks provide communication, computation, caching, and control~\cite{pham2020survey}, 
the pairing of edge computing and RAN is a promising solution and has been investigated in many research works~\cite{peng2016fog, reznik2018cloud}. In particular, a Fog-RAN architecture was proposed in~\cite{peng2016fog}, and the integration of multiaccess edge computing (MEC) and RAN was investigated in~\cite{reznik2018cloud}. Moreover, Because end devices are increasingly capable of accessing multiple RANs, heterogeneous RANs have been studied in many research works~\cite{vu2015energy}. Recently, various technologies and network architectures have been proposed for B5G and 6G RANs, for example, millimeter Wave (mmWave) and terahertz (THz) communications in~\cite{wang2018millimeter, han2019terahertz}, new radio and unlicensed spectrum in~\cite{Lagen2020NewRadio}, blockchain-enabled RAN (BRAN) in~\cite{ling2019blockchain}, and AI-defined RAN in~\cite{yao2019artificial}.

There have been surveys on aerial communications, but they are limited and/or specific to particular topics. For instance, satellite and CubeSat communications were discussed in~\cite{kodheli2020satellite} and~\cite{saeed2020cubesat}, respectively, opportunities and challenges from the integration of terrestrial and aerial communications were reviewed in~\cite{wang2019convergence}, UAV wireless communications and networking in~\cite{fotouhi2019survey}, hierarchical airborne (i.e., satellites, HAPs, and LAPs) wireless communications in~\cite{cao2018airborne}. 
Although the demand for a comprehensive 6G access infrastructure in the next 10 years is a matter of urgency, we are not aware of any surveys on ARANs, and this motivates us to conduct the present survey. To summarize, the state-of-the-art surveys on aerial communications and major contributions offered by this work are provided in Table~\ref{Table:Summary_ExistingSurveys}.

\begin{table*}[h]
    \renewcommand{\arraystretch}{1.2}
	\caption{Summary of existing surveys on aerial wireless communications and RANs.}
	\label{Table:Summary_ExistingSurveys}
	\centering
	\begin{tabular}{|c|c|c|p{13.7cm}|}
		\hline 
		\multirow{2}{*}{\textbf{Reference}}  & \multicolumn{2}{c|}{\textbf{Main theme}} & \multirow{2}{*}{\textbf{Contributions}}  \\ 
		
		\cline{2-3}
		{} & \textbf{5G/6G} & \textbf{RANs} & {} \\
		\hline
		\hline
		
		\multirow{1}{*}{\cite{Saad2020AVision}} & \multirow{1}{*}{\checkmark} &  &  A discussion of emerging applications and key technologies in 6G wireless systems \\ \hline
	
	    \multirow{1}{*}{\cite{giordani2020toward}} & \multirow{1}{*}{\checkmark} &  &  Speculation about a set of 6G use cases and technologies \\ \hline
		
		\multirow{1}{*}{\cite{Yaacoub2020AKey6G}} & \multirow{1}{*}{\checkmark} &  & Opportunities and challenges of providing Internet connectivity for rural areas in 6G   \\ \hline
		
		 \multirow{1}{*}{\cite{Shafin2020Artificial, Song2020Artificial}} & \multirow{1}{*}{\checkmark} &  &  Roles and applications of AI for 6G wireless systems \\ \hline
		 
		 \multirow{1}{*}{\cite{ali20206g, peltonen20206g, rajatheva2020white, mahmood2020white}} & \multirow{1}{*}{\checkmark} &  &  White papers from the Finnish 6G Flagship program on various research aspects of 6G \\ \hline
		 
		~\cite{olwal2016survey} &  & \checkmark &  Radio resource management in 5G RANs \\ \hline
		 
		 \multirow{1}{*}{\cite{mohamed2015control}} &  & \multirow{1}{*}{\checkmark} &  Separation of data and control planes in 5G RANs \\ \hline
		 
		 \multirow{1}{*}{\cite{Peng2016Recent,habibi2019comprehensive}} &  & \multirow{1}{*}{\checkmark} &  Various aspects of CRAN, including network architectures, key techniques, and radio resource allocation \\ \hline
		 
		 \multirow{1}{*}{\cite{peng2016fog, reznik2018cloud}} &  & \multirow{1}{*}{\checkmark} &  Integration of edge computing and RANs, Fog-RAN in~\cite{peng2016fog}, and MEC-RAN in~\cite{reznik2018cloud} \\ \hline
		 
		 \multirow{2}{*}{\cite{wang2018millimeter, han2019terahertz, Lagen2020NewRadio, ling2019blockchain, yao2019artificial}} &  & \multirow{2}{*}{\checkmark} &  Technologies and architectures for B5G and 6G RANs such as mmWave RANs~\cite{wang2018millimeter}, THz communications~\cite{han2019terahertz}, blockchain RANs in~\cite{ling2019blockchain}, and AI RANs in~\cite{yao2019artificial} \\ \hline
		 
		 \multirow{2}{*}{\cite{kodheli2020satellite, saeed2020cubesat, wang2019convergence, fotouhi2019survey, cao2018airborne}} & \multirow{2}{*}{\checkmark} &  &  Aerial communications in 5G/B5G/6G wireless systems such as satellite communications in~\cite{kodheli2020satellite}, UAV communications in~\cite{fotouhi2019survey}, and airborne communications in~\cite{cao2018airborne} \\ \hline
		 
		 \multirow{1}{*}{Ours} & \multirow{1}{*}{\checkmark} & \multirow{1}{*}{\checkmark} &  A survey on aerial RANs toward a comprehensive 6G access infrastructure \\ \hline
	\end{tabular}
\end{table*}

\subsection{Research Scope and Methodology} \label{sec1c}
To provide a comprehensive overview of ARAN technology in the context of 6G, this study focuses on clarifying a multitier and hierarchical ARAN architecture from multiple perspectives. The survey framework and major topic organization are illustrated in Fig.~\ref{fig:framework}. In brief, our survey discovered:
\begin{figure}[!t]
\centering
\includegraphics[width=1.0\linewidth]{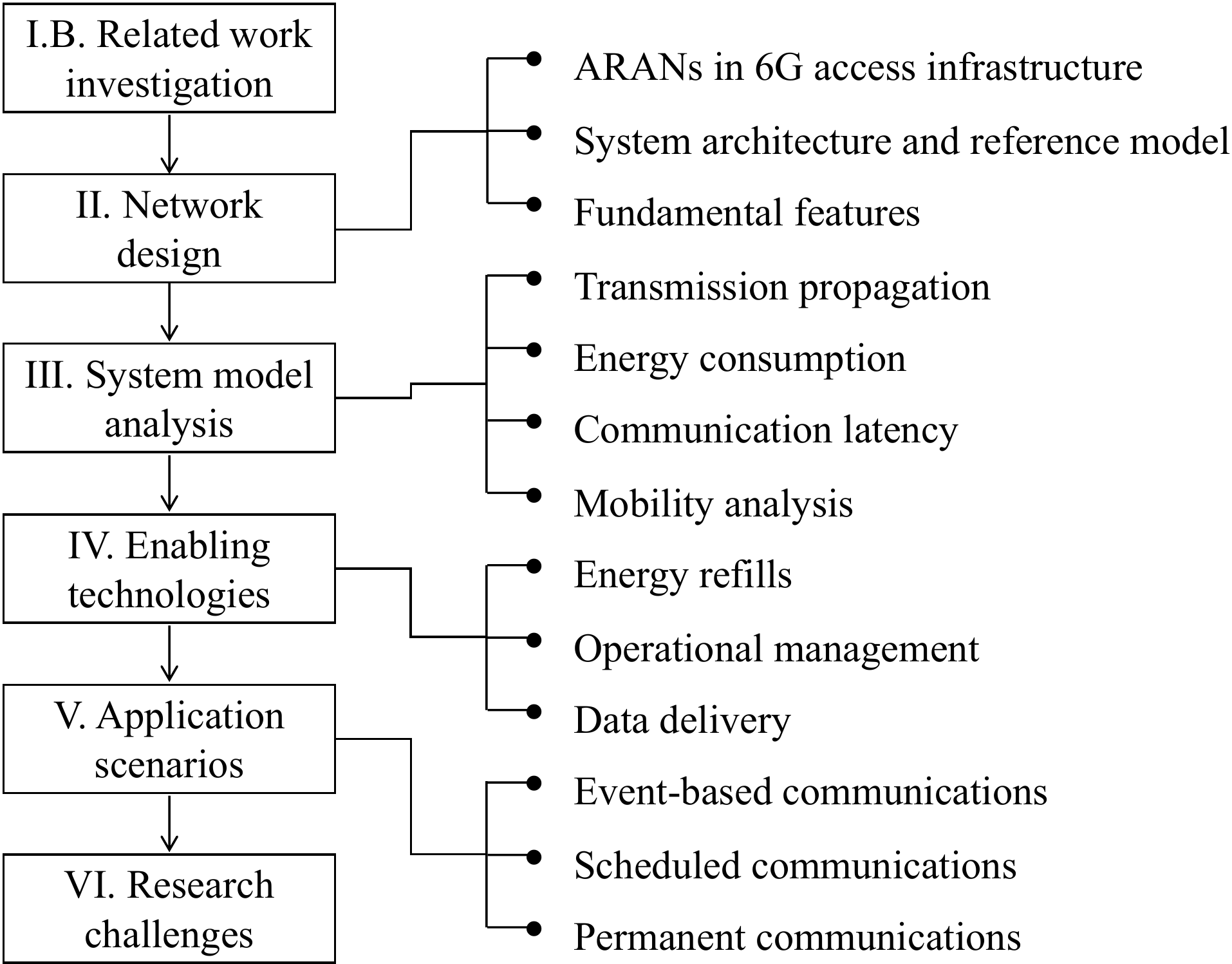}
\caption{Survey framework and major topic organization.} 
\label{fig:framework}
\end{figure}
\begin{itemize}
    \item \textit{Network design:} An extensive review of similar access infrastructure designs was conducted. To provide a convenient reference for readers with different backgrounds and interests, we adopt a logical narrative structure, initially providing a high-level view of ARAN contribution to 6G access infrastructure. Here, the positions, roles, and relations of ARANs are investigated. Afterward, we discuss technical ARAN characteristics regarding multitier and hierarchical system architecture as well as its standard reference model. From these analyses, the fundamental features that distinguish ARANs from other RANs are derived. Details are provided in Section~\ref{sec2}.
    \item \textit{System model:} We analyze four key technical aspects of ARANs including transmission propagation, energy consumption, communication latency, and system mobility. In particular, transmission propagation modeling deals with path loss formulation in a 3-D wireless environment. As energy efficiency is essential to maintain sustainable flight operations, we analyze energy consumption in behaviors such as data transmission, data computation, an airborne object's motion, and hovering. In addition, multihop communication latency and on-the-move data computation time are jointly investigated to identify potential end-to-end service delay in ARAN model. Finally, we study system mobility and trajectory scheduling for ARANs, which is expected to be helpful for future research regarding learning algorithms for system optimization. These details are covered in Section~\ref{sec3}.
    \item \textit{Enabling technologies:} Regarding ARAN realization, we review technologies that could feasibly support ARAN implementations on three planes: energy replenishment, operational management, and data delivery. In an ARAN, energy replenishment is a critical first step toward system sustainability. The flight components must have either a self-recharging or a wireless recharging capability besides the basic charging station method. Accordingly, these typical approaches to charging are discussed. On the operational management plane, we focus on recent advances in foundational pillars including network softwarization, mobile cloudization, and data mining technologies supporting ARANs. As the airborne components in ARANs interconnect via different wireless spectra on the front and back-end interfaces, enabling technologies for data delivery on these segments have distinguishable features. The frequency spectrum, communication protocol, and multiaccess schemes especially are within the scope of our survey. Details are provided in Section~\ref{sec4}. 
    \item \textit{Application scenarios:} We envision emerging application scenarios that effectively exploit ARAN infrastructures in the context of 6G. The applications are classified into three categories, including event-based, scheduled, and permanent communications. Here, we clarify the requirements of these applications and describe how the advances of ARANs can support them well. In addition, we note the common problems inherent in these application scenarios that may attract research and interest from specialist industrial communities. Details can be found in Section~\ref{sec5}.
    \item \textit{Research challenges:} To drive potential research on 6G ARANs, we highlight essential challenges for the successful development and implementation of ARANs. As a native component of the comprehensive 6G access infrastructure, ARANs face challenges in terms of intelligent radio design, extremely high spectrum exploitation, and network stability to enable emerging technologies and applications in 6G networks. Moreover, security and privacy issues are considered as critical issues to protect user data and system operations. Finally, reputable and efficient evaluation tools are needed to validate proposals for ARANs. Details are discussed in Section~\ref{sec7}.
\end{itemize}

In summary, the main contributions of this paper are as follows. The survey serves as a reference framework for interested readers, providing state-of-the-art knowledge and studies regarding the emerging ARAN model. Technical foundations of ARANs are taxonomized systematically from different perspectives such as network design, system models, enabling technologies, and application scenarios. Further, an appropriate lesson learned is derived from related work analyses to conclude each perspective survey. Finally, future challenges are highlighted to illustrate research trends in ARANs toward a comprehensive 6G access infrastructure. A flowchart of our research methodology is shown in Fig.~\ref{fig:framework}. Common acronyms used in this survey are summarized in Table~\ref{tab:abbreviation}.

\begin{table}[!t]
    \caption{List of abbreviations}
    \centering
    \begin{tabular}{|p{2cm}|p{6cm}|}
        \hline
        \textbf{Abbreviation} & \textbf{Description}\\
        \hline
        {3GPP} & {3rd Generation Partnership Project organization} \\ \hline
        {AI} & {Artificial intelligence} \\ \hline 
        {ABS} & {Aerial base station} \\ \hline 
        {ARAN} & {Aerial radio access network} \\ \hline 
        {BRAN} & {Blockchain enabled radio access network} \\ \hline 
        {CNN} & {Convolutional neural network} \\ \hline
        {CNPC} & {Control and non-payload communication} \\ \hline        
        {CRAN} & {Cloud enabled radio access network} \\ \hline 
        {DA-RAN} & {Drone assisted radio access network} \\ \hline 
        {DL} & {Deep learning} \\ \hline 
        {EH} & {Energy harvesting} \\ \hline 
        {ETSI} & {European Telecommunications Standards Institute} \\ \hline
        {FANET} & {Flying ad hoc network} \\ \hline  
        {FCC} & {Federal Communications Commission} \\ \hline 
        {GEO} & {Geostationary Earth orbit} \\ \hline 
        {GPS} & {Global positioning system} \\ \hline 
        {HAP} & {High altitude platform} \\ \hline  
        {ITS} & {Intelligent transportation system} \\ \hline 
        {ITU} & {International Telecommunication Union} \\ \hline                        
        {IoT} & {Internet of things} \\ \hline 
        {LAP} & {Low altitude platform} \\ \hline 
        {LEO} & {Low Earth orbit} \\ \hline 
        {MEC} & {multiaccess edge computing} \\ \hline         
        {MEO} & {Medium Earth orbit} \\ \hline 
        {MIMO} & {Multiple-input multiple-output} \\ \hline 
        {ML} & {Machine learning} \\ \hline
        {NFV} & {Network function virtualization} \\ \hline         
        {NOMA} & {Non-orthogonal multiple access} \\ \hline
        {NR} & {New radio} \\ \hline 
        {PV} & {Photovoltaic} \\ \hline    
        {QoS} & {Quality of service} \\ \hline
        {RAN} & {Radio access network} \\ \hline              
        {RF} & {Radio frequency} \\ \hline 
        {RIS} & {Reconfigurable intelligent surface} \\ \hline 
		{SAR} & {Search and rescue} \\ \hline                
        {SDN} & {Software defined networking} \\ \hline
        {SISO} & {Single-input single-output} \\ \hline
        {STIN} & {Satellite terrestrial integrated network} \\ \hline
        {T-UAV} & {Tethered unmanned aerial vehicle} \\ \hline 
        {U-UAV} & {Untethered unmanned aerial vehicle} \\ \hline
        {UAV} & {Unmanned aerial vehicle} \\ \hline 
        {UE} & {User equipment} \\ \hline 
 		{UPF} & {User plane function} \\ \hline        
        {URLLC} & {Ultra-reliable low-latency communications} \\ \hline 
        {VSAT} & {Very small aperture terminal} \\ \hline
        {eNB} & {Evolved node B} \\ \hline
        {gNB} & {Next-generation node B} \\ \hline  
        {mmWave} & {Millimeter wave} \\ \hline  
    \end{tabular}
    \label{tab:abbreviation}
\end{table}

\begin{table*}[h]
    \renewcommand{\arraystretch}{1.2}
	\caption{Related terms regarding aerial communication assisted mobile networks.}
	\label{tab:relatedwork}
	\centering
	\begin{tabular}{|c|c|c|c|p{10.2cm}|p{2.5cm}|}
		\hline 
		\multirow{2}{*}{\textbf{Name}}  & \multicolumn{3}{c|}{\textbf{Access infrastructure}} &
		\multirow{2}{*}{\textbf{Definition}} &
        \multirow{2}{*}{\textbf{Applications}}  \\ 
		\cline{2-4}
		{} & \textbf{LAP} & \textbf{HAP} & \textbf{LEO} & {} & {}\\
		\hline
		\hline
		
		\multirow{5}{*}{DA-RAN} & \multirow{5}{*}{\checkmark} & & & DA-RAN stands for a drone/UAV assisted radio access network, where drones help to extend the coverage area and capacity of terrestrial access infrastructure. In this context, drones can connect either directly to terrestrial base stations or via a head node in (multihop) mesh, star, tree, and chain topologies. Unlicensed/licensed spectrum can be exploited by drones in DA-RANs~\cite{he2017drone,shi2018drone,zhao2019uav}. 
		& Aerial surveillance, smart farm, smart city, environment studies, temporary capacity boost, etc. \\ \hline
	
	    \multirow{5}{*}{FlyRAN} & \multirow{5}{*}{\checkmark} & \multirow{5}{*}{\checkmark} & & FlyRAN stands for flying radio access network, where aircraft and airships equipped with radio transceivers are utilized as ABSs to provide mobile services in underserved areas. In FlyRAN, the LAP tier exploits the unlicensed/licensed spectrum, while the HAP tier typically exploits the licensed spectrum to avoid unmanaged conflicts with terrestrial systems~\cite{shah2017association,huang2019airplane,wang2019deployment,lee2020path}. The term FANET is covered by this definition~\cite{chriki2019fanet,lakew2020routing}. & Remote populations, rescue \& surveillance, environment studies, temporary capacity boost, etc. \\ \hline
	    
	    \multirow{5}{*}{STIN} & & & \multirow{5}{*}{\checkmark} & STIN represents a satellite-terrestrial integrated network by which satellite communications supplement the terrestrial systems to offer global seamless and ubiquitous Internet services to users in isolated areas. STIN utilizes the licensed spectrum for ground-air connections. The airborne infrastructure includes multiple tiers such as LEO, medium Earth orbit (MEO), and geostationary Earth orbit (GEO) satellite constellations~\cite{lin2019joint,guidotti2019integrated,zhu2019cooperative,li2020energy}. & Isolated populations, emergency~\&~alarm, global roaming, large-scale Earth studies, etc.\\ \hline

	    \multirow{8}{*}{ARAN} & \multirow{8}{*}{\checkmark} & \multirow{8}{*}{\checkmark} & \multirow{8}{*}{\checkmark} & Referred to the definition in Section~\ref{sec1a}, ARAN is a multitier and hierarchical aerial access infrastructure combining the FlyRAN and LEO communication systems of STIN to provide radio access medium from the sky to end users for Internet services through ABSs equipped with heterogeneous wireless transceivers. The ABSs can be UAVs, drones, balloons, and airplanes. The definition of ARAN was first introduced by Song~\etal~\cite{song2020icc} with the original name AAN (aerial access network) at the IEEE International Conference on Communications (ICC) in June 2020. & Smart farm/city, temporary capacity boost, underserved populations, rescue \& aerial surveillance, emergency \& alarm, scalable environment studies, etc.\\ \hline

	\end{tabular}
\end{table*}

\section{Network Design} \label{sec2}
In this section, we position ARANs in a comprehensive 6G access infrastructure and analyze them from the perspectives of system architecture and reference model.
\subsection{ARANs in 6G Access Infrastructure}
In the context of comprehensive 6G access infrastructure, ARANs are positioned in the aerial communication layer to serve high-altitude and terrestrial users. As illustrated in Fig.~\ref{fig:6gaccess}, ARANs encompass three systems including LAP communications at the altitude of 0--10~km, HAP communications at the altitude of 20--50~km, and LEO communications at the altitude of 500--1500~km above relative sea-level~\cite{song2020icc,dao2021aran}. Compared to terrestrial and underground(water) RANs, ARANs serve a large coverage space with highly dynamic and mobile users for in-flight infotainment, aerial surveillance, flying vehicular control, and isolated populations.

The literature review reveals several aerial communication classes that are closely related to ARANs. The terms related to aerial communications that support mobile networks are summarized in Table~\ref{tab:relatedwork}. Briefly, these are drone/UAV assisted radio access networks (DA-RANs) focusing on LAP communications~\cite{he2017drone,shi2018drone,zhao2019uav}, flying radio access networks (FlyRANs) aiming at LAP/HAP communications~\cite{shah2017association,huang2019airplane,wang2019deployment,lee2020path}, and satellite-terrestrial integrated networks (STINs) involving satellite communications~\cite{lin2019joint,guidotti2019integrated,zhu2019cooperative,li2020energy}. Because each of the similar terms represents a partial ARAN tier that is used for special applications, the interconnection among these systems is weak and asynchronous. Hence, the introduction of an ARAN definition is a significant contribution toward unifying the aforementioned systems into a common model.

\begin{figure*}[!t]
\centering
\includegraphics[width=1.0\textwidth]{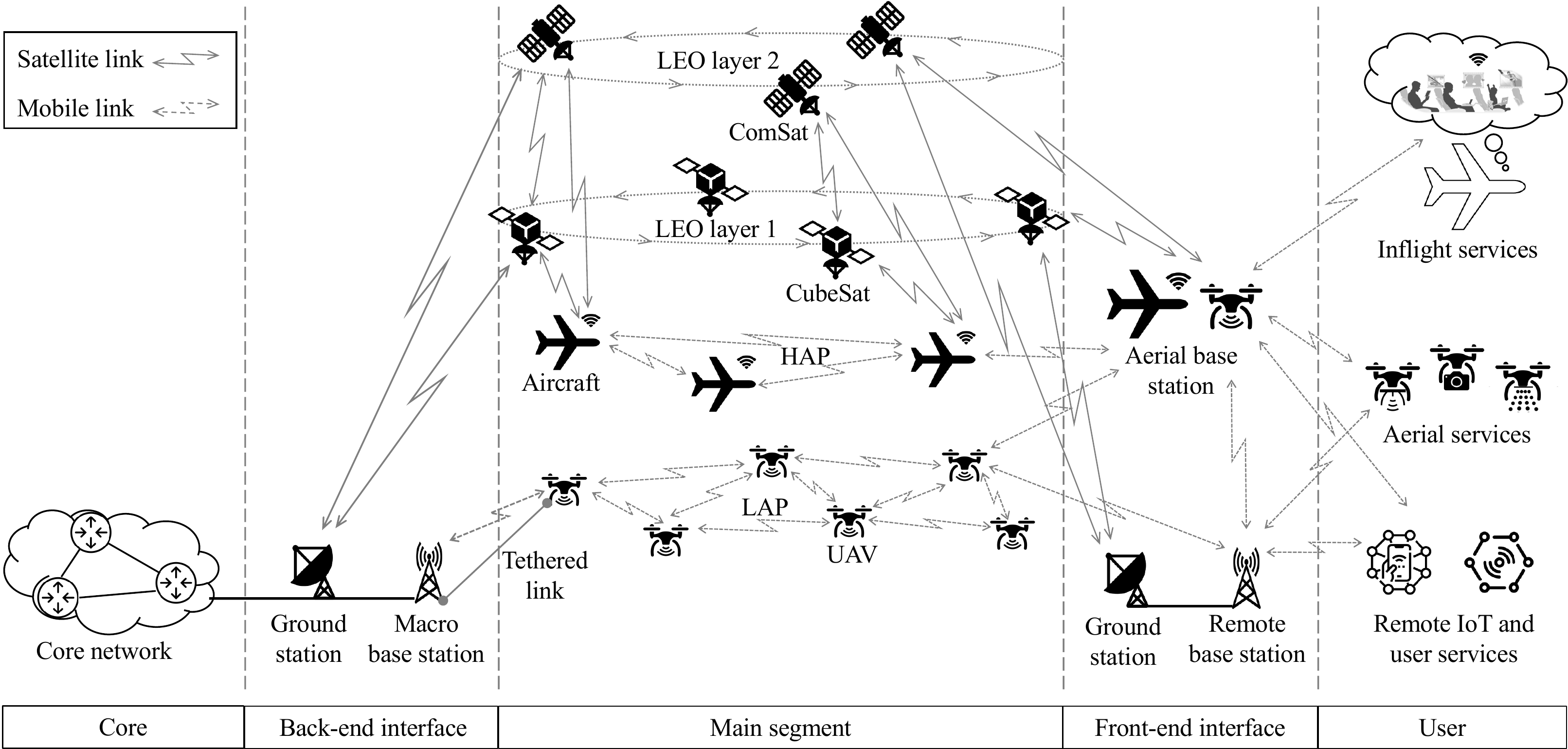}
\caption{System architecture of ARANs.} 
\label{fig:architecture}
\end{figure*}

\subsection{System Architecture}
Figure~\ref{fig:architecture} shows a wide perspective to demonstrate a detailed ARAN architecture in the context of a complete user-core path. Typically, an ARAN architecture comprises (\textit{i}) a main segment that is a cross-tier networking infrastructure shared among ABSs at the LAP, HAP, and LEO altitudes, (\textit{ii}) a front-end interface providing terrestrial and aerial access points that gather user connections, and (\textit{iii}) a back-end interface bridging ARAN infrastructure to the terrestrial core networks. Note that LEO communication systems use satellite links to contact the terrestrial network through ground stations, while LAP and HAP systems use mobile (wireless) links to contact ABSs and the terrestrial base stations (e.g., gNBs and eNBs) directly.

In the main segment of an ARAN, LAP systems are at the lowest tier. Drones and UAVs in LAP systems act as ABSs, providing connectivity directly to aerial and terrestrial end users or remote base stations using wireless technologies such as 5G new radio (NR) and Wi-Fi. Conversely, these ABSs may utilize either satellite technology to connect with LEO communication systems or wireless technologies to the terrestrial macro base stations for backhaul transmission to the core networks. LAPs can be classified by size, range, speed, and endurance. According to the US Department of the Army~\cite{uavclassification}, LAPs are classified into five categories, i.e., small~$<$~medium~$<$~large~$<$~larger~$<$~largest groups. Most LAPs are relatively lightweight and cost-effective devices that can be deployed quickly and flexibly. However, they have relatively low endurance with limited energy and networking resources. To mitigate these issues, tethered technology can be considered a feasible approach to either establish a reliable broadband backhaul through a fiber cable or energize LAPs by a powerline connection between the LAPs and terrestrial stations~\cite{kishk2020,kishk2020aerial,alzidaneen2019resource}. Nevertheless, LAPs effectively support time-sensitive and event-based scenarios such as emergency and rescue, aerial surveillance, traffic offloading, and mobile hotspots at public gatherings.

In the middle tier of the ARAN main segment, HAPs are defined by the ITU Radio Regulations (RR) as radio stations located at a specified, nominal, fixed point relative to the Earth\footnote{https://www.itu.int/en/mediacentre/backgrounders/Pages/High-altitude-platform-systems.aspx}. 2, 6, 27/31, and 47/48~GHz frequency bands were assigned for HAP communications in bidirectional HAP-terrestrial links~\cite{pelton2020high} at three world radio communication conferences (WRC-97, WRC-2000, and WRC-12). HAP systems serve aerial and remote terrestrial end users with wireless technologies in a wide coverage area from high altitude. For backhaul transmission to the terrestrial core networks, HAP systems mostly utilize satellite technology via the LEO communication systems. In practical situations, some industries have implemented trial projects using lightweight solar-powered aircraft and airships to provide stable broadband services in rural and remote areas~\cite{arum2020review}.

At the top of ARAN main segment, two-layer LEO communication systems consist of miniaturized satellites below (i.e., CubeSats) and communication satellites above (i.e., ComSats), orbiting at an altitude of 500--1500 km. CubeSats aim to provide low-latency and high-throughput Internet services, while ComSats are designed for high coverage and service availability. These two LEO layers interact with each other through interlinks, which provide redundancy, backup, and collaboration interfaces for dynamic network organization. Compared with other satellite classes, LEO satellites are characterized by lower latency, cost-efficiency, and quick production and deployment. Most LEO satellites operate on Ku, Ka, and V bands to provide Internet connection to aerial and ground stations within tens of ms latency~\cite{su2019broadband}. Unlike the LAP/HAP systems, LEO communication systems typically do not provide services to the end users directly. Both fronthaul to the end users and backhaul to the core networks are satellite transmissions through ground stations and very small aperture terminals (VSATs). To orchestrate intertier networking operations in an ARAN, LEO communication systems additionally support backhaul tunnels allowing LAP/HAP systems to connect to the core networks. Recent years have witnessed several emerging commercial satellite projects on LEO communication systems such as OneWeb, Telesat, and Starlink, with hundreds of satellites successfully launched into orbit and thousands of satellite launches planned for the near future~\cite{davoli2019small}.

In ARANs, the front-end interface includes ABSs at the LAP/HAP tiers, remote base stations, and ground stations enabling end users network access the networks. The ABSs and remote base stations are ARAN access points that provide wireless links directly to aerial and terrestrial users. Meanwhile, the ground stations are end-points of satellite links from LEO communication systems that deliver traffic to and from the remote base stations. Depending on prevailing circumstances, the ABSs and remote base stations may cross-serve both aerial and terrestrial services, as illustrated on the left side of Fig.~\ref{fig:architecture}. Conversely, the back-end interface includes macro base stations and ground stations to accommodate backhaul traffic toward the core networks. Typically, the macro base stations help transfer traffic flows from LAP systems, while the ground stations support traffic forwarding from (HAP systems to) LEO communication systems; see the right side of Fig.~\ref{fig:architecture}. 

\begin{figure*}[!t]
\centering
\includegraphics[width=.85\textwidth]{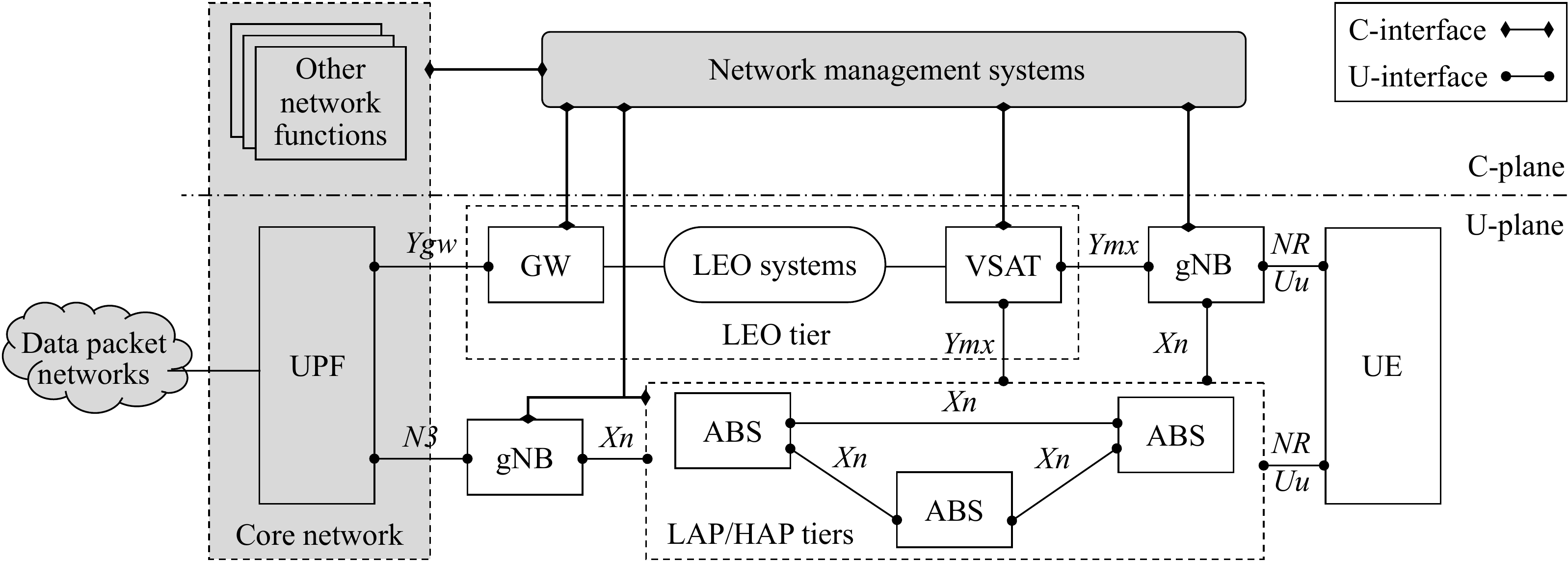}
\caption{Reference model of ARAN adopting the 3GPP 5G standards.} 
\label{fig:model}
\end{figure*}

\subsection{Reference Model}
The ARAN reference model is built by jointly adopting the 3GPP TS 23.501 (version 16.5.0 Release 16)~\cite{3gpp2020} and the ETSI TR 103 611 (V1.1.1) standards~\cite{etsi2020} that were released recently by the 3rd Generation Partnership Project (3GPP) organization and the European Telecommunications Standards Institute (ETSI), respectively. Fig.~\ref{fig:model} depicts the ARAN reference model in detail. 
This reference model is based on the 3GPP 5G standard, where ABSs perform the functions of the next-generation Node B (gNB), and the LEO communication tier is considered a trusted non-3GPP component~\cite{rinaldi2020non}. It is worth noting that this model does not cover non-trusted LEO system interaction because LEO networks are considered critical components in ARAN architecture. In particular, mobile service management and control through the LEO networks must be handled by the mobile networks. For this purpose, only trusted LEO networks are considered satisfactory for network integration. In Fig.~\ref{fig:model}, the description of the 5G core components is abstracted significantly, except the user plane function (UPF), which is the contact point to access the core networks and other systems (i.e., intranet and Internet). Mapped on the system architecture illustrated in Fig.~\ref{fig:architecture}, the components of the reference model are represented as follows:
\begin{itemize}
    \item User equipment (UE): inflight, aerial, remote IoT, and terrestrial user services.
    \item ABS: LAPs and HAPs
    \item VSAT: ground stations at the front-end interface
    \item Gateway (GW): the ground stations at the back-end interface
    \item LEO systems: CubeSats, ComSats, and their control systems
    \item gNB: remote and macro base stations.
\end{itemize}

In the control plane, a common 5G network management system is in charge of managing, controlling, and monitoring operations of almost all the ARAN components such as ABS, VSAT, GW, and gNB via C-interfaces of corresponding reference points. In the user plane, ABSs are considered as 5G gNB-functional components. Therefore, the \textit{NR Uu} reference points are utilized between ABSs and UEs, while the \textit{Xn} reference points are defined for ABSs' interactions and between ABSs and gNBs. As a standard 5G reference model, gNBs and UPFs are connected via \textit{N3} reference points. A detailed description of the \textit{NR Uu}, \textit{Xn}, and \textit{N3} reference points can be found in the 3GPP TS 23.501 (version 16.5.0 Release 16)~\cite{3gpp2020} and the 3GPP TS 38.300 (version 16.2.0 Release 16)~\cite{3gpp2020nr} standards. Conversely, to standardize the translation between 3GPP messages and non-3GPP messages (i.e., IP packets) on interfaces gNB/ABS--VSAT and GW--UPF, the \textit{Ymx} and \textit{Ygw} reference points are defined in the ETSI TR 103 611 (V1.1.1) standard~\cite{etsi2020}.

\subsection{Fundamental Features}
The fundamental features that distinguish ARANs from the other RAN architectures and help to reinforce emerging 6G services are described as follows:
\begin{itemize}
    \item \textit{Ubiquity:} The LEO communication tier guarantees service continuity across the globe with three advantages, including wide coverage, networking backup/resilience, and emergency broadcast, for which the terrestrial infrastructure has limited capacity.
    \item \textit{Mobility:} The dynamicity of aerial LAP/HAP topology implementation and the networking overlay among the LAP, HAP, and LEO communication systems help ARANs to adapt flexibly to the requirements of end users anywhere on the ground and in the air. 
    \item \textit{Availability:} Operating in the air at various altitudes, ARANs are not commonly affected by natural and man-made disasters capable of rendering terrestrial communication infrastructures vulnerable and interrupting service. In other words, better service availability is provided regardless of the recipient’s terrain (e.g., mountain, sea, desert, etc.).
    \item \textit{Simultaneity:} Multitier LAP/HAP/LEO communication systems can adaptively self-organize to forward information-centric services effectively across discrete (aerial and terrestrial) locations on simultaneous multicast and broadcast streams using various wireless access technologies.
    \item \textit{Scalability:} Because ABSs interlink to each other using aerial wireless ad hoc technologies and there are hierarchical networking overlays among LAP, HAP, and LEO communication tiers, ARANs can quickly establish scalable topologies for local sites without service interruptions.
\end{itemize}

\section{System Model} \label{sec3}
In this section, we discuss four key aspects of ARANs: transmission propagation, energy consumption, latency analysis, and system mobility. 

\subsection{Transmission Propagation} \label{sec3a}

A key challenge in aerial communications is accurately modeling radio propagation. Indeed, channel modeling is necessary for designing correct waveforms, resource allocation, modulation order, multiantenna techniques, and interference management~\cite{khawaja2019survey}. It is widely known that channel characterization for cellular networks has been modeled and empirically verified. Propagation channels for satellite communications at various frequency bands (i.e., Ku, Ka, and V) were also well-studied~\cite{panagopoulos2004satellite}. Moreover, many HAP projects for wireless communications were undertaken in the 1990s and 2000s. For instance, the theoretical model of small-scale fading for HAP propagation channels was investigated in~\cite{dovis2002small}, and the application of HAPs for broadband communications, e.g., channel modeling, interference, coding techniques, and resource allocation was presented in~\cite{karapantazis2005broadband}. Owing to recent technological advancements, investment in the industry, and the involvement of major tech corporations like Google and Facebook, a significant amount of research has been conducted to integrate HAPs into B5G wireless systems. As the characterization of HAP and satellite communications channels has been well established, in this subsection we focus on discussing channel modeling for LAP-enabled communications. 

Depending on the channel model, there are two main modeling approaches, including empirical channel models formulated on field measurements, and analytical models that analyze the transmission channel under certain conditions and/or assumptions. 
Measurement campaigns play an important role in developing empirical channel models. Various measurement campaigns with various environmental and measurement settings have been conducted, e.g., the configuration of antenna arrays as either single-input single-output (SISO) or multiple-input multiple-output (MIMO), channel sounding for narrow-band and wide-band channels, propagation scenarios (e.g., open space and mountains), and evaluation of elevation angles~\cite{khawaja2019survey}.
Another modeling classification based on propagation environments includes three approaches~\cite{cao2018airborne}: a deterministic model, stochastic model, and geometric-based stochastic model. In particular, the first approach relies on the assumption of certain network layouts and is thus suitable for large-scale fading effects, the second approach takes into account multipath components, and the third approach is to study spatio-temporal channel characteristics in 3-D stochastic environments.

\subsubsection{Deterministic Models} 
\begin{figure}[!t]
\centering
\includegraphics[width=1.0\linewidth]{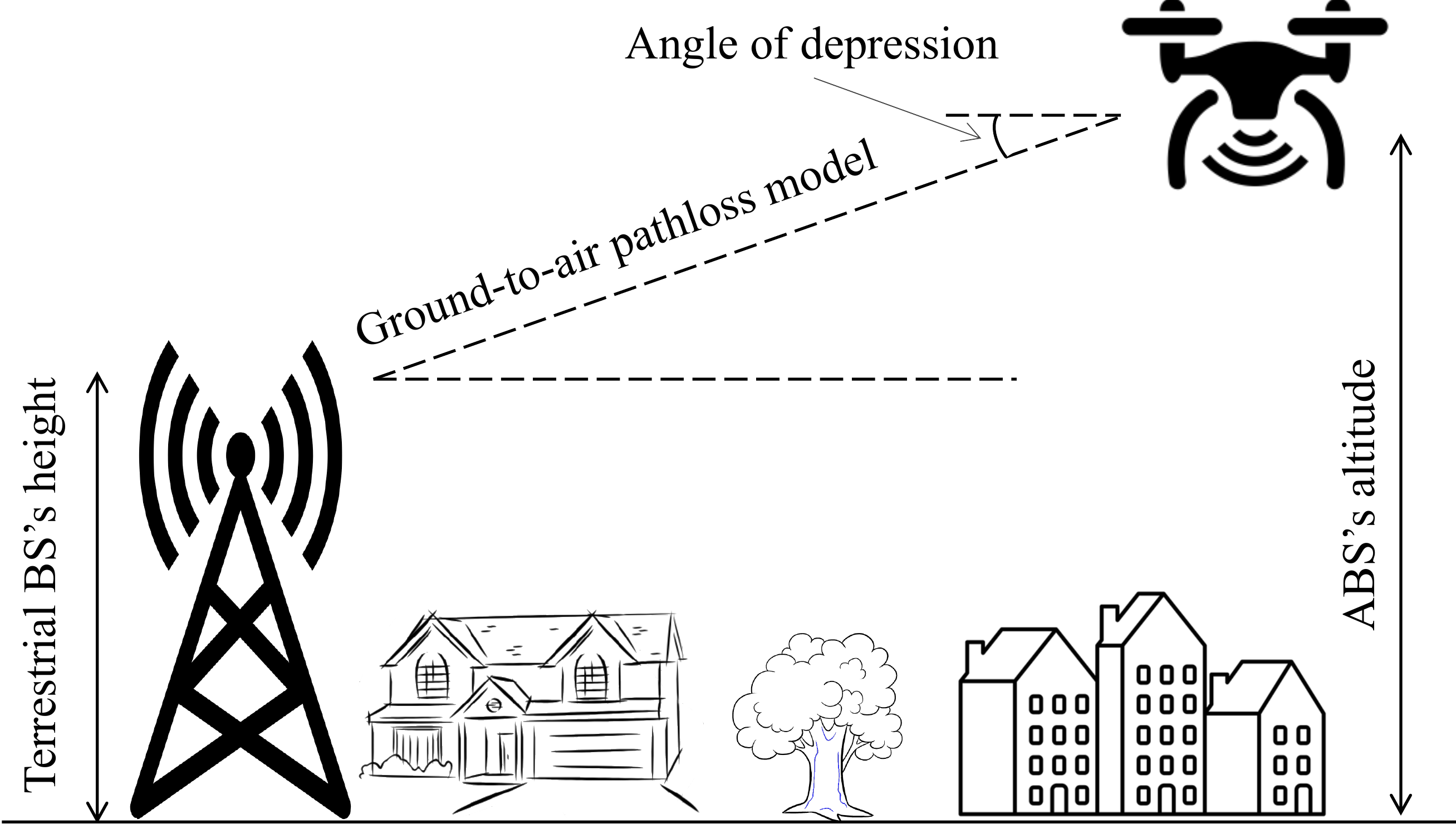}
\caption{Illustration of an experimental setting for ground-to-air path loss models.} 
\label{fig:channelmodel_suburban}
\end{figure}

Deterministic channel models are characterized by information on the propagation environment such as terrain topography and the composition of buildings or obstacles. Moreover, ray-tracing software is usually employed in the literature to analyze deterministic channel models. 
For example, Feng~\etal~\cite{feng2006path} investigated a statistical model for evaluating the path-loss and shadowing of air-to-ground channels in urban areas. The proposed model is analyzed by ray-tracing simulations over different frequency bands, including 200 MHz, 1 GHz, 2 GHz, 2.5 GHz, and 5 GHz. Unlike the log-distance channel models of terrestrial communications, the channel model proposed in~\cite{feng2006path} is dependent not only on the distance from the ABS (i.e., UAV in LAP systems) but also on the elevation angle. 
To generalize the channel model in~\cite{feng2006path} to multiple urban environments, Al-Hourani~\etal~\cite{al2014modeling} proposed a generic model to estimate the path-loss of air-to-ground channels and facilitate radio frequency (RF) planning. 
In particular, the path loss, $\eta$, for a given ground user is computed as
\begin{equation}
    \eta = \text{PL} - \text{FSPL},
\end{equation}
where PL and FSPL are the total power loss and free space path loss, respectively. Here FSPL can be calculated from the Friis transmission equation as
\begin{equation}
    \text{FSPL} = 20\log(d) + 20\log(f) - 27.55,
\end{equation}
where $d$ refers to the distance in 3D space between the ABS and ground user, and $f$ denotes the carrier frequency (in MHz).
Bor-Yaliniz~\etal~\cite{bor2017environment} proposed a path loss model for modern metropolitan scenarios, which may include many skyscrapers with different heights, and ray-tracing simulations were conducted to evaluate the performance of the proposed model. A ground-to-air channel model for suburban environments was proposed in~\cite{al2018modeling}. In particular, the total path loss was composed of two components, including terrestrial path loss and aerial excess path loss, which is heavily dependent on the depression angle, as illustrated in Fig.~\ref{fig:channelmodel_suburban}. Another channel model for suburban areas can be found in~\cite{cui2019low}. The measurement was carried out at a radio frequency of 3.9 GHz, and the path loss was calculated by a two-ray model. 

\subsubsection{Stochastic Models} In stochastic channel models, the fading characteristics of multipaths are taken into consideration. In the literature, the performance of these stochastic models have been evaluated by empirical measurements or by geometric analyses via computer simulations. Compared to the deterministic model, the accuracy of a stochastic model depends on environmental information and the nonstationary propagation channel. 
In the multipart papers~\cite{matolak2017airI, sun2017airII, matolak2017airIII, sun2017airIV}, different air-to-ground channels for LAP communications were proposed. In~\cite{matolak2017airI}, a wideband statistical channel model over two frequency bands (970 MHz and 5 GHz) was proposed for off-shore scenarios. For other terrain types, channel models for hilly and mountainous environments can be found in~\cite{sun2017airII}, and channel models for suburban and peri-urban environments can be found in~\cite{matolak2017airIII}. In the fourth part~\cite{sun2017airIV}, shadowing loss, shadowing duration, and small-scale fading statistics were analyzed. By using empirical measurement data, it was shown that shadowing loss can be modeled as a function of an aircraft’s roll angle, but it does not correlate with shadowing duration. For airport environments, Rieth~\etal~\cite{rieth2019aircraft} conducted two measurement campaigns to characterize small-scale results (e.g., delay and Doppler spread statistics) and large-scale effects (shadowing and antenna misalignment) of air-to-ground radio channels. Several interesting results are found by this work, e.g., max and min delay resolutions of 2184 and 52 ns, respectively, which correspond to the respective coherence bandwidth of 92 and 3846 kHz (i.e., coherence bandwidth = 1/(5 $\times$ delay resolution)). Such findings are necessary for selecting payload symbols, especially, in frequency-selective fading LAP systems.
A recent channel model proposed in~\cite{gutierrez2019comparison} was dedicated to residential and mountainous desert scenarios. By comparing statistical channel models for these two environments, it was found that the fading effects observed in residential settings were less than those observed in mountainous desert settings. This is reasonable because LoS links are typically available in residential, and the signals reflected from the ground can be obstructed well by residential buildings.

\subsubsection{Geometric-based Stochastic Models} Compared to the two above modeling approaches, a geometric-based stochastic model is suitable for considering spatio-temporal channel characteristics and for deriving analytical performance expressions such as coverage radius and channel capacity. For example, Cheng~\etal~\cite{cheng2019A3D} investigated a 3-D geometric-based stochastic model for MIMO nonstationary propagation channels. Particularly, the nonstationarity of LAP communications was overcome by integrating time-varying angles of arrival and departure into the reference model and by considering both line-of-sight and non-line-of-sight components. An interesting observation from this work is that ABS-specific parameters (e.g., moving direction and antenna setup) and altitude significantly affect channel stability, demanding more efficient control strategies for the ABS trajectory to render the proposed channel model more stable and less nonstationary. 
Recently, Jiang~\etal~\cite{jiang2020novel} considered the presence of interfering objects in designing air-to-ground channels and developed a computationally-efficient method to estimate the angular parameters such as azimuth/elevation angles of departure and arrival. 
Unlike~\cite{cheng2019A3D, jiang2020novel}, the work in~\cite{ma2020wideband} does not fix the moving speed of the ABS, thus complicating the channel model by the directions of transceivers’ motion.

\subsection{Energy Consumption} \label{sec3b}
Energy-efficient communications have been one of the most important design requirements in B5G wireless systems. It was recently reported in~\cite{stolaroff2018energy} that the use of drones for package delivery is relatively energy-efficient compared with ground-based delivery provided that the warehouse system is designed properly and the drones are used within their endurance and ferry ranges. Despite many potentials and applications, airborne objects, especially those working in the LAP tiers, are typically powered by onboard batteries that usually have limited capacity and lifetime. Several promising technologies have been developed to mitigate the impact of limited energy and battery lifetimes such as wireless power transfer and energy harvesting (EH). However, the implementation of these technologies to massive airborne objects is in practice still questionable and will not be available in the foreseeable future. Owing to the importance of energy efficiency in maintaining sustained flying operations for ARANs, the modeling of energy consumption during flight is a critical challenge. 

Various factors must be considered when developing energy consumption models for aerial communications. The obvious examples are UAV types, flying speed, acceleration, payload, and external factors such as weather conditions. The first factor is reasonable because UAVs usually travel in highly dynamic environments (e.g., weather and wind conditions) that significantly affect flight capabilities. These factors may enhance system performance in some cases. For instance, when a UAV is flying with the wind, it can fly faster but consumes less energy. Moreover, external temperatures can also directly affect performance by causing battery lifetime and drain issues. As reported in~\cite[Fig. 4]{tseng2017autonomous}, UAV consume less energy when they fly with a tailwind, but they consume more energy when flying into a headwind. Moreover, within a wind speed threshold, the higher the wind speed, the more energy-efficient the UAV operation. Second, the energy consumption of a UAV significantly depends on its flying state, e.g., hovering, horizontal, and vertical motion~\cite{thibbotuwawa2018energy}. Finally, as the weight of a UAV increases according to the payload, the energy consumption also increases as a function of the payload. As empirically tested in~\cite[Fig. 3]{tseng2017autonomous} when a UAV is hovering, the power consumption increases linearly with increments in the payload. 
Because various factors may affect the energy consumption of aerial systems, it is very challenging to model energy consumption for all environmental conditions. 

\subsubsection{Energy Consumption in Hover and Vertical Movement}
To model the power consumed during hovering for delivery services, Dorling~\etal~\cite{dorling2017vehicle} considered modeling the power consumption of a multirotor helicopter as a function of its total weight, which includes the frame weight, battery, and payload. The power, $P$, required for single-rotor copters to hover was mathematically established in~\cite{leishman2006principles} as
\begin{equation}
    P = \frac{T^{3/2}}{\sqrt{2 \sigma A}},
\end{equation}
where $\sigma$ is the fluid density of the air, $A$ is the area of the rotor disk, and $T = (M + m)g$ is the rotor thrust where $M$ is the frame weight, $m$ is the battery and payload weight, and $g$ is the gravitational acceleration. 
Based on this model, Dorling~\etal~\cite{dorling2017vehicle} proposed a new power-consumption calculation for multirotor copters on the assumption that each rotor bears an equal amount of frame, battery, and payload weight. Thus, the power consumed by a rotor is modeled as
\begin{equation}
    P_{0} = \frac{(\frac{M}{N} + \frac{m}{N})^{3/2}g^{3/2}}{\sqrt{2 \sigma A}},
\end{equation}
where $N$ is the total number of rotors.
As a result, the total power consumed by the multirotor copter can be given as 
\begin{equation} \label{Eq:Power_mr_copter}
    P = N\left( \frac{M}{N} + \frac{m}{N} \right)^{3/2}\frac{g^{3/2}}{\sqrt{2 \sigma A}} = (M + m)^{3/2}\sqrt{\frac{g^{3}}{2 \sigma A N}}.
\end{equation}
To simplify the model further \eqref{Eq:Power_mr_copter}, Dorling~\etal~\cite{dorling2017vehicle} considered a linear approximation composed of two parts: one represents the power consumed to bear the battery and payload weight and another accounts for the power needed for the vehicle to hover. 
These models are also used to calculate power consumption during takeoff and landing, that is, the power consumed during takeoff and landing is set to be equivalent to that of hovering~\cite{thibbotuwawa2018energy}.
Many models have been proposed to consider other factors, for example, motor and propeller efficiencies were integrated in~\cite{Marins2018AClosed}, and multiple dynamics of UAVs (e.g., electrical dynamics of the battery and aerodynamics of the rotor-propeller assembly) were studied in~\cite{Michel2019Multiphysical}. 

\begin{figure}[!t]
\centering
\includegraphics[width=0.5\linewidth]{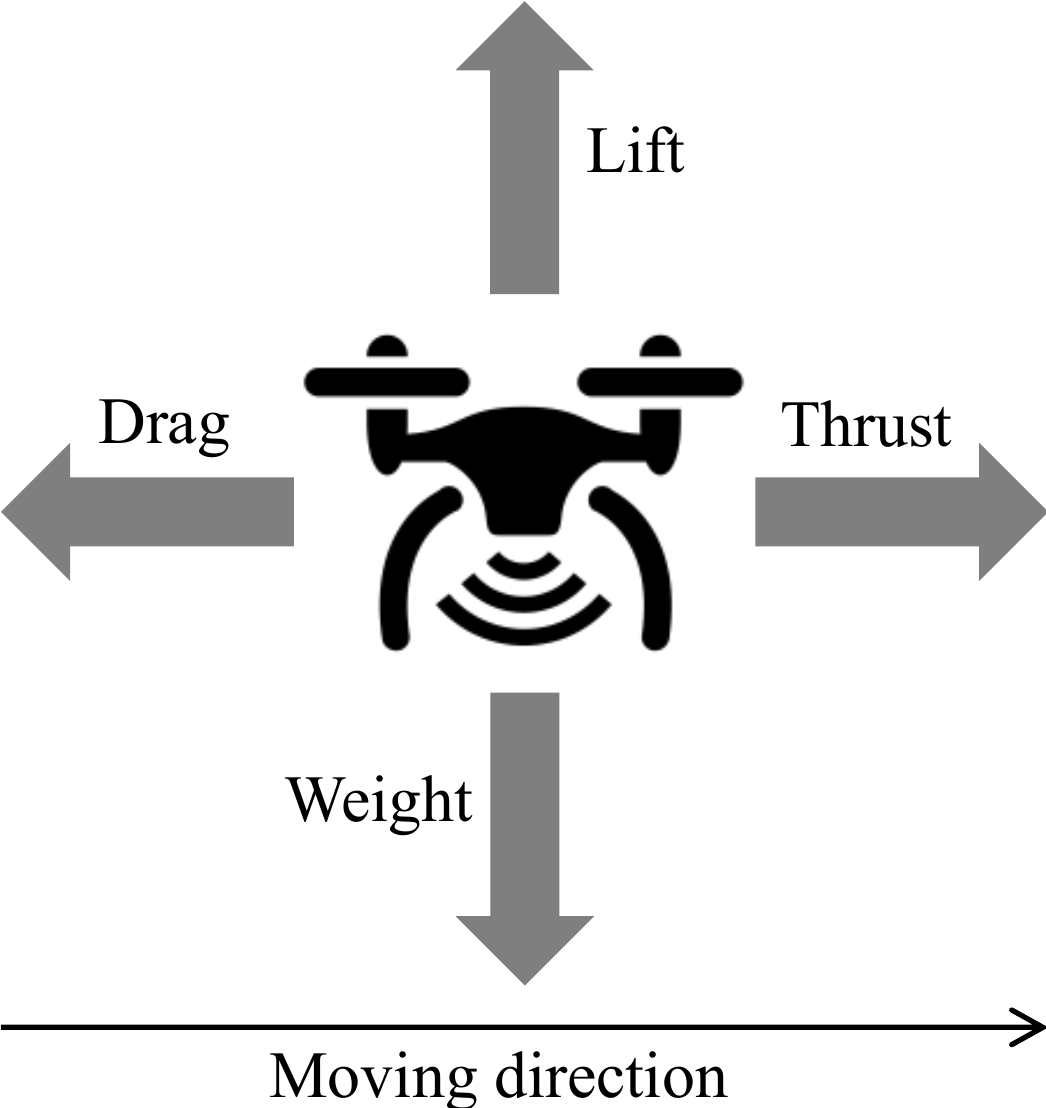}
\caption{Four forces occur in UAVs.} 
\label{fig:4forces}
\end{figure}

\subsubsection{Energy Consumption in Horizontal Movement} The four principal forces lift, weight, thrust, and drag should be considered for a drone's flight. Fundamentally, drag consists of lift-induced and parasitic drags and power is needed to overcome them. The former is directly prosegmental to the air density, $\sigma$, the square of the drone velocity, $V$, and the square of the span loading (i.e., the ratio between the lift and wingspan in N/m) as~\cite{austin2011unmanned}
\begin{equation} \label{Eq:lift_induced_drag}
    D_{i} = \kappa (L/b)^{2} / \left( \frac{1}{2} \sigma \pi V^{2}\right),
\end{equation}
where $\kappa$ is a non-dimensional coefficient, $L$ is the lift generated by the drone, $b$ is the wingspan, and $V$ is the forward velocity. Because air density decreases at higher altitudes, more lift-induced drag is generated, and accordingly, more power and a greater wingspan are required for the drone to operate.  
Another form of drag is parasitic drag, which can be estimated as~\cite{austin2011unmanned}
\begin{equation}
    D_{p} = C_{D} S \sigma V^{2}/2,
\end{equation}
where $C_{D} $ is the parasitic drag coefficient, and $S$ is the wing area. Thus, the respective power required to overcome lift-induced and parasitic drags can be given as 
\begin{align}
    & P_{i} = D_{i} V = \kappa L^{2} / \left( \frac{1}{2} \sigma \pi V b^{2} \right), \\
    & P_{p} = D_{p} V = C_{D} S \sigma V^{3}/2.
\end{align}
As a result, the total power needed for the drone to stay afloat is calculated as 
\begin{equation} \label{Eq:power_drag}
    P = P_{i} + P_{p} = \frac{\kappa L^{2}}{\frac{1}{2} \sigma \pi V b^{2}} + \frac{C_{D} S \sigma V^{3}}{2}.
\end{equation}
The power consumption model \eqref{Eq:power_drag} has been employed by Zeng and Zhang in~\cite{zeng2017energy} to optimize UAV flight trajectory subject to various design constraints imposed by the initial and final locations as well as maximum and minimum velocity and acceleration values. This work has played a seminal role in research on UAV wireless communications. We note that the power consumption models above are for fixed-wing UAVs, while those of rotary-wing UAVs were investigated in~\cite{zeng2019energy}. For more details, we invite interested readers to refer to this work and the references cited therein.

\subsection{Latency Analysis} \label{sec3c}
Along with propagation models and energy consumption, communication latency is another key feature of aerial communications requiring a thorough study. 

\subsubsection{Communication Latency}
In addition to on-demand communications, aerial communications have also been implemented to support low-latency services. 
An example of this is the use of ABSs as aerial caching servers. In this scenario, ground users can directly request content from an aerial caching server instead of sending the request to a remote server, thus reducing latency and avoiding a network bottleneck.
Conversely, achieving low latency is also important in aerial communications. Along with high reliability and high security, low latency is a critical requirement for control and non-payload communication (CNPC) links for supporting the management of LAP/HAP systems and avoiding crashes. 
Since LEO satellites typically have lower altitude orbits compared to GEO and MEO, LEO communication is highly attractive to the industry owing to its ability to provide latency on the order of tens of ms. The OneWeb system can provide Internet services of 400 Mbps with an average latency of 32 ms, while the Telesat system promises to have latency between 30 and 50 ms, and the Starlink system can offer broadband Internet with a latency from 15 to 35 ms, as reported in~\cite{su2019broadband}. These aerial platforms (i.e., LAP, HAP, and LEO) should be integrated into ARANs to provide Internet services with different latency ranges. While low-tier LAP systems aim to provide for event-based scenarios and low-latency services, high-tier LEO communication systems are designed to provide Internet access across the globe. In the middle tier of the ARAN model, HAP systems maintain a balance between service availability, latency, and deployment cost. 

In general, end-to-end latency can be approximated as twice the sum of radio, backhaul, core, and transport latencies. Radio latency is composed of queuing latency, frame alignment delay, transmission latency, and processing latency. In long-distance transmission, LEO communications take advantage of the speed of light in space, whereas in optical cables, the speed of light is relatively lower, typically 180,000 to 200,000 km/s.
Using queuing theory, Horani~\etal~\cite{horani2018latency} showed that the use of ABSs can reduce the queuing delay compared to that of terrestrial communications and that an ABS's altitude significantly affects latency by contributing to the likelihood of line-of-sight communication links. An analysis of latency in uplink mmWave-caching networks was provided in~\cite{liao2020end}, on the assumption that queuing latency accounts for the majority of total latency. Based on theoretical analyses, it was shown that end-to-end latency is reduced when more ABSs are deployed in the network, while it increases when user density increases. Wu~\etal~\cite{wu2020latency} leveraged the concept of ``physical layer security" to guarantee low latency for secure content sharing services in aerial communications. To achieve this objective, the ABSs' trajectories and user associations are optimized jointly via an iterative algorithm.
The aforementioned studies agreed that lower latency could be achieved by an appropriate deployment of ABSs compared to baseline schemes with fixed deployment.

To improve the performance of CNPC links, several studies have integrated ultra-reliable and low-latency communications (URLLC) into LAP/HAP systems. She~\etal~\cite{she2019ultra} proposed improved distributed multiantenna systems to maximize horizontal communication distance (a.k.a. available range) to guarantee the end-to-end delay and total packet loss probability in URLLC-based LAP systems. This work emphasized that the number of antennas at each access point should be optimized to maximize the available range of the ABS. Ren~\etal~\cite{ren2019achievable} characterized the achievable data transmission rates of CNPC links. However, the analysis is complicated by the 3-D deployment of the aerial platforms and the design of short-packet transmission in URLLC. A notable observation is that the CNPC’s achievable link rate increases as block length increases, but this comes at the cost of greater latency. This work also indicated that leveraging the Shannon formula directly to compute the achievable rate is somewhat inefficient, especially when the block length is small. Another study on URLLC-based LAP communication was conducted in~\cite{ranjha2020quasi}, where an ABS was deployed to transfer URLLC packets among IoT devices and the block length is optimized. 

\begin{figure*}[t]
\centering
\includegraphics[width=.8\linewidth]{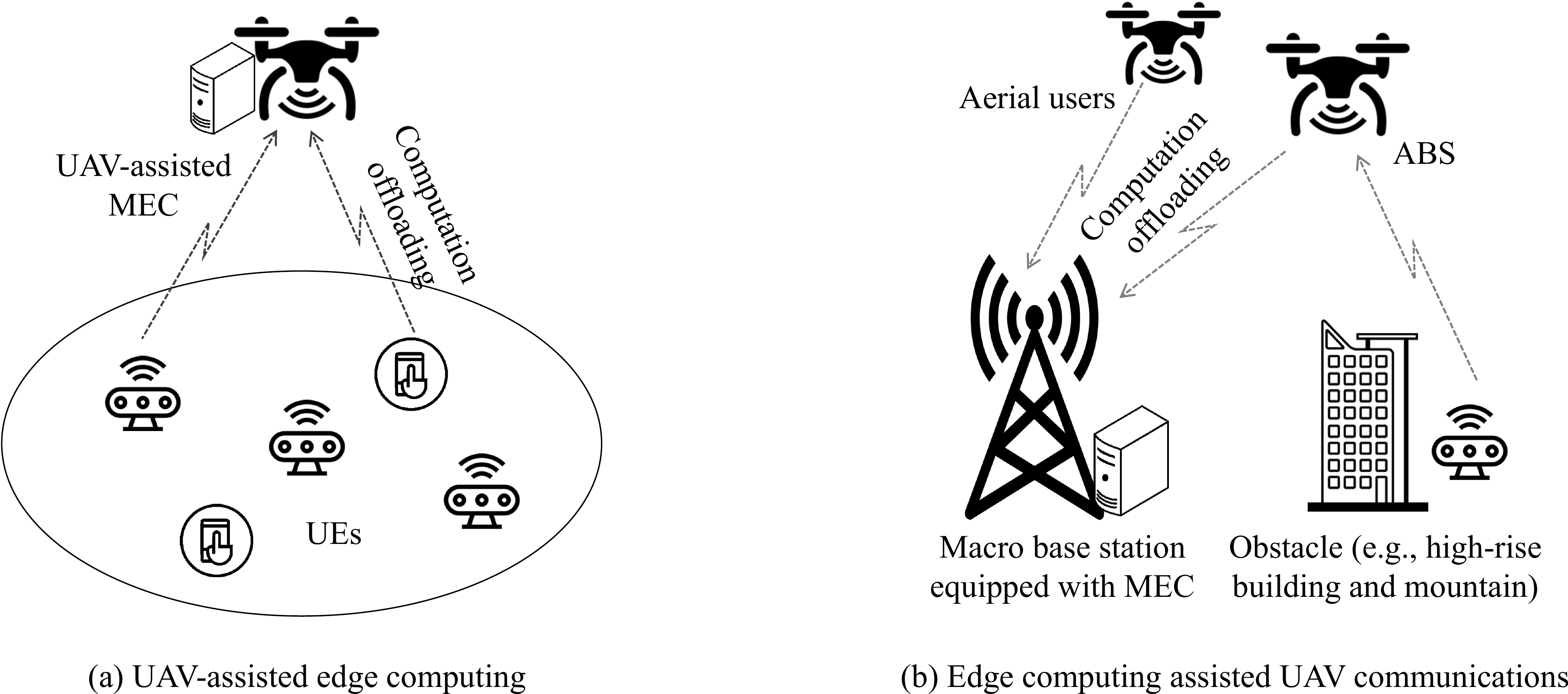}
\caption{Illustration of two UAV-MEC scenarios.} 
\label{fig:UAV_MEC_scenarios}
\end{figure*}

\subsubsection{Computation Latency} Considered as a key technology in B5G, edge computing (e.g., fog computing and multiaccess edge computing (MEC)) has recently received significant attention from both academia and industry. Edge computing markedly enables the emergence of delay-sensitive and computationally intensive applications (e.g., human activity recognition and autonomous driving) as computing resources are moved from the cloud to the network edges closest to end users~\cite{bellavista2019social}. MEC nodes can be deployed at various RAN locations, including both fixed points (e.g., macro base stations and WiFi access points) and mobile points (e.g., moving vehicles and smart phones)~\cite{bellavista2018human}. Corresponding to two types of aerial communications, there are also two MEC deployment scenarios, as shown in Fig.~\ref{fig:UAV_MEC_scenarios}. The first scenario considers aerial computing servers processing computational tasks migrated from ground users, whereas the second scenario considers aerial users, who may offload their computational tasks to terrestrial computing servers for remote processing. Different latency sources in MEC systems may include computing, transmission, queuing, and backhaul latencies~\cite{pham2019mobile,pham2020coalitional}. 

To minimize the weighted sum energy of the ABS and all the ground users under constraints on completion time, Hu~\etal~\cite{hu2019uav} considered an MEC system to process computational tasks of energy-limited devices and devised an iterative algorithm. This work reported a trade-off between energy consumption and task completion time, i.e., the network consumes less energy as the completion time increases and vice versa. 
In~\cite{han2020rate}, an aerial relay was deployed to collect and then forward data generated by IoT devices to an MEC server for remote processing. In this case, the challenges caused by limited computing capability at the MEC server and unstable transmission links were addressed jointly to enhance system endurance and task completion time. 
For IoT service provisioning in aerial communications, the work in~\cite{zhang2020latency} proposed that UAVs can serve as both computing servers and relays to minimize the average latency of all the IoT devices. The simulation results of this work revealed that completion time is reduced when the number of UAVs is increased, and the dual use of both aerial computing and aerial relaying significantly decreases the completion time compared to a case in which UAVs are used only for relaying purposes. 
In~\cite{zhang2019satellite}, integrating MEC into LEO communications was shown to be a promising solution to provide computing services for various scenarios, especially when ground users are sparsely distributed over different regions. An illustrative example of a satellite MEC system is shown in Fig.~\ref{fig:UAV_MEC_scenarios}, where UAVs can connect to MEC servers located at ground station via satellite connections. In such scenarios, cooperative computation offloading problems can be optimized to reduce both energy consumption and latency.

\subsection{Mobility Analysis} \label{sec3d}
While terrestrial infrastructures are constructed at predefined locations or follow constrained investment policies, ARANs can be designed for dynamic deployment to optimize benefits to infrastructure providers as well as to efficiently improve QoS for end users. 
Many studies have shown that the dynamic mobility and trajectory control of ARANs would offer significant advantages such as improving energy efficiency, supporting data collection applications, and powering energy-limited IoT devices. We explicate the benefits of and state-of-the-art studies pertaining to the optimization of mobility and trajectory in aerial systems as follows.

\subsubsection{System Mobility}
Aerial communications play an important role in the IoT, typically consisting of a large number of small, energy- and computation-limited devices. To reduce the power consumption of an IoT network and satisfy QoS requirements of IoT devices, in~\cite{mozaffari2017mobile}, multiple UAVs were deployed for IoT data collection. Since the locations of IoT devices vary over time and an IoT device may be under the service coverage of multiple UAVs, the joint problem of UAV deployment, IoT device association, and power control is extremely challenging. Simulation results demonstrated that mobile deployment of UAVs can reduce the average total transmission power by 45\% compared to that of the stationary ABSs. 
As analyzed by Chetlur~\etal~\cite{chetlur2017downlink}, increasing the number of UAVs can increase network spectral efficiency; however, this is reduced when the number of deployed UAVs is sufficiently large. Therefore, optimizing UAV placement is a critical problem and has been investigated in many research works. For example, Lyu~\etal~\cite{lyu2017placement} considered the UAV placement problem to guarantee that each ground user is served by at least one ABS.
In aerial relay communications, one significant challenge in optimizing UAV placement is that shadowing on the receiver side depends on the environment. Chen~\etal~\cite{chen2020efficient} addressed this dependence by developing a blockage-adaptive algorithm, which they found to be more suitable for propagation models than a statistical channel model for all environments in existing studies. 

Much research has focused on analyzing the performance of LAP/HAP systems with UAV mobility, e.g., coverage probability, secrecy rate, and outage probability. In~\cite{sharma2019random}, a set of ABSs was deployed to serve a ground user, and the coverage performance 
was analyzed in two scenarios. The first considers that the user randomly associates with one among multiple ABSs, whereas the user associates with the closest ABS in the second scenario. The use of UAVs for relaying satellite signals to ground users was considered~\cite{sharma2020outage}. In particular, aerial relays are to be deployed in the case in which direct communications links between the satellite and ground users are not well established (e.g., indoor users). This study reported that the deployment of 3-D aerial relays could achieve better performance in terms of outage probability. In~\cite{sharma2020secure}, the secrecy capacity and secrecy outage probability of a hybrid satellite-terrestrial network with 3-D aerial relays was analyzed. These metrics are necessary if eavesdroppers exist in the network. To generalize the secrecy performance, this work considered three criteria for selecting aerial relay: closest, random, and maximum signal-to-noise ratio. 

In the high tier of ARANs, the design of satellite constellations is important to ensure global coverage, that is, anywhere on Earth being covered by at least one satellite. Generally, constellation types and designs are dependent on several factors such as coverage requirements, propagation latency, launch cost, and designated missions. The reason for the first factor is that a GEO satellite has much greater coverage than a LEO satellite, and thus multiple LEO satellites are required to cover the same region of interest. Propagation latency is determined mainly by the satellites’ orbiting altitude and can be quite pronounced for satellite constellations. When the respective orbiting altitudes of LEO and GEO are set as 900 and 36,000 km, the propagation latency is 3 and 120 ms, respectively (the speed of light in space is $3 \times 10^{8}$ m/s). Launch cost, the third factor listed above, is related to the number of satellites to be deployed. As aforementioned, LEO satellites are attractive to the industry owing to their lower costs and deployment complexity compared with GEO and MEO satellites. Finally, the design of satellite constellations should be based on the designated mission, and there are multiple constellation designs for specific missions. These may include circular orbit constellations, elliptical orbit constellations, flower constellations, and Walker constellations. Several studies have investigated the design of satellite constellations from a communication perspective. For example, two classes (with and without inter-satellite links) of LEO constellation designs to support IoT applications were reviewed in~\cite{qu2017leo}. The former is suitable for delay-tolerant IoT applications, while the latter provides seamless connectivity. Papa~\etal~\cite{papa2020design} integrated software-defined networking (SDN) into LEO communication systems to separate the control plane from the data plane in satellite communications, and they further proposed a three-tier SDN-based LEO architecture in which SDN controllers are co-located on satellites. A technical comparison among three LEO satellite constellation systems (i.e., SpaceX, Telesat, and OneWeb) in terms of system throughput and satellite efficiency was conducted in~\cite{del2019technical}. 

\subsubsection{Trajectory Schedule}
Besides mobility, the trajectory of ABSs can be optimized for various purposes. For instance, in collecting IoT data in smart cities and agricultural scenarios, the deployment of ABSs can assure line-of-sight communication links for mobile users in emergency rescue operations, thus increasing data collection efficiency and saving more of the energy required by mobile devices to transmit data. In such cases, UAVs can be regarded as mobile hubs for data collection and processing. Another scenario may include situations where UAVs are deployed as energy sources to wirelessly power IoT devices. Recently, significant research efforts in the field of wireless communications have been dedicated to optimizing UAV trajectories. 

\begin{figure}[t]
\centering
\includegraphics[width=0.80\linewidth]{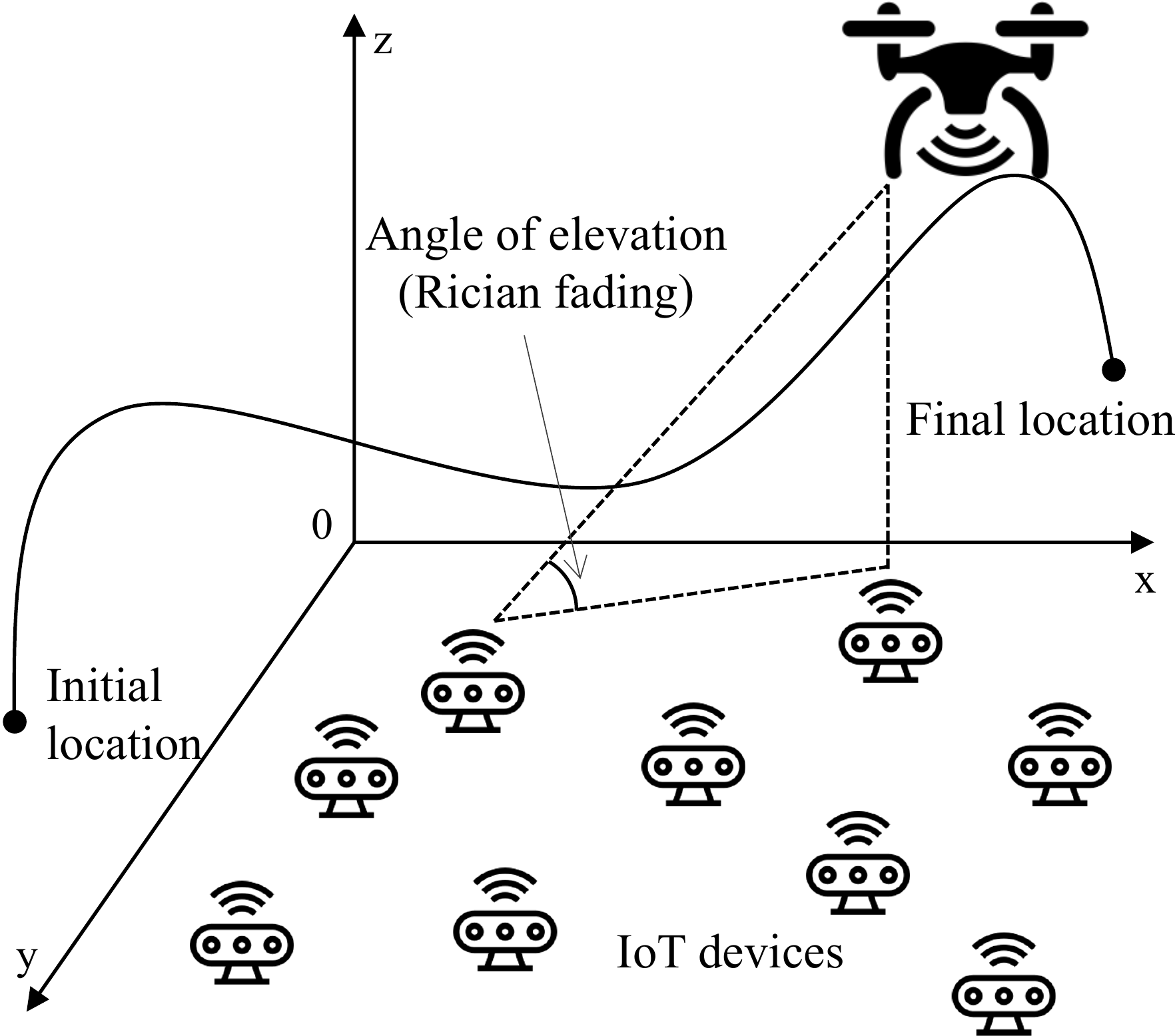}
\caption{Illustration of UAV-enabled data harvesting.} 
\label{fig:UAV_dataharvesting}
\end{figure}

UAV trajectory optimization was investigated in the seminal work~\cite{zeng2017energy}. After developing an energy consumption model, Zeng~\etal~considered two UAV trajectory scenarios: unconstrained and constrained. An interesting result for the unconstrained case is that the network was energy inefficient regardless of the design objective, thus demanding trajectory schemes with practical constraints such as initial and final locations, minimum and maximum velocity, and permitted altitude ranges. In contrast to~\cite{zeng2017energy}, You~\etal~\cite{you20193d} considered optimizing the joint UAV scheduling and 3-D trajectory problem in Rician fading channels, as illustrated in Fig.~\ref{fig:UAV_dataharvesting}. Particularly, at each time instance, one sensor device can be scheduled to transmit its data to a UAV traveling between two predefined locations to maximize the minimum data collection rate. The use of UAVs for wirelessly powering IoT devices and improving the freshness of IoT data was investigated in~\cite{hu2020aoi}. To address the non-convexity of the problem, a decomposition technique was applied, and a joint dynamic programming metaheuristic approach was adopted to solve the trajectory sub-problem. Numerical simulations in this work demonstrated that data freshness increases linearly with increments in UAVs' altitudes and the size of collected data. The application of AI for optimizing a UAV trajectory has been considered recently in various research studies. For instance, echo state networks and multiagent Q-learning were adopted to calculate effective UAVs' trajectories and predict ground users' locations~\cite{liu2019trajectory}, and a deep reinforcement learning approach was proposed to select UAVs' trajectories in UAV-enabled edge computing systems~\cite{liu2020path}. 

\begin{table*}[t]
	\caption{Summary of ARAN references on system model.}
	\label{tab_summary_system_model}
	\begin{tabular}{|c|c|p{5.40cm}|c|p{6.3cm}|}
		\hline
		\textbf{Aspect} & \textbf{Sub-aspect} & \textbf{Highlights} & \textbf{Ref.} & \textbf{Contributions} \\
		\hline
		\hline
		
		\multirow{9}{*}{\makecell{Transmission\\ Propagation}} 
		& \multirow{3}{*}{\makecell{Deterministic \\Models}} 
		& Deterministic models assume certain network layouts such as terrain topography and obstacles. & \multirow{3}{*}{\cite{al2014modeling}} & The path-loss is comprised  of power loss and free space path loss. The latter is dependent on the carrier frequency and transmission distance (urban areas). \\
		\cline{2-5} 
		
		& \multirow{3}{*}{\makecell{Stochastic \\Models}} 
		& Stochastic models consider multipath fading effects. & \multirow{3}{*}{\cite{rieth2019aircraft}} & Both small-scale and large-scale effects are considered for an air-to-ground channel. Respective max and min delay resolutions are 2184 ns and 52 ns. \\
		\cline{2-5}
		
		& \multirow{4}{*}{\makecell{Geometric-based \\Stochastic \\Models}} 
		& Geometric-based stochastic models consider spatio-temporal channel characteristics, and are thus suitable for deriving analytical performance metrics. & \multirow{4}{*}{\cite{cheng2019A3D}} & To cope with the non-stationarity of ABS in aerial communications, the channel model considers time-varying arrival and departure angles.\\
		\hline
		
		\multirow{15}{*}{\makecell{Energy\\ Consumption}} 
		& \multirow{8}{*}{\makecell{Hover and \\Vertical Moving}} 
		& Most of the existing studies model the total power in hover and vertical motion as a function of the total weight, including the frame weight, battery, and payload. & \multirow{8}{*}{\cite{dorling2017vehicle}} & The power consumption of a multirotor drone is 
		\begin{equation}
        P = (M + m)^{3/2}\sqrt{\frac{g^{3}}{2 \sigma A N}}, \notag
        \end{equation}
        where $M$ is the frame weight, $m$ is the battery and payload weight, $A$ is the rotor disc area, $N$ is the number of rotors, and $g$ is the gravitational acceleration. 
		\\
		\cline{2-5} 
		
		& \multirow{7}{*}{\makecell{Horizontal \\Moving}} 
		& Among the four principal forces (i.e., lift, weight, thrust, and drag), drag is mainly analyzed to compute the energy consumed by horizontal motion, focusing on lift-induced drag and parasitic drag. & \multirow{7}{*}{\cite{austin2011unmanned}} & The power needed for the drone to remain aloft is 
		\begin{equation}
		    P = \frac{2\kappa L^{2}}{\sigma \pi V b^{2}} + \frac{C_{D} S \sigma V^{3}}{2},\notag
		\end{equation}
		where main parameters are the wingspan, $b$, velocity, $V$, air density, $\sigma$, and wing area, $S$.  
		\\
		\hline

		\multirow{11}{*}{\makecell{Latency\\ Analysis}} 
		& \multirow{8}{*}{\makecell{Communication \\Latency}} 
		& $\bullet$ LEO communications typically have an average latency of a few tens of ms, e.g., 15--35 ms in Starlink and 32 ms in OneWeb. & \multirow{3}{*}{\cite{horani2018latency}} & The establishment of LoS links in UAV communications can reduce the queuing latency compared to terrestrial communications. \\
		\cline{4-5}
		&& $\bullet$ Low latency is of importance for CNPC links LAP/HAP communications. & \multirow{2}{*}{\cite{she2019ultra}} & A multiantenna system was proposed to guarantee latency and loss probability in URLLC LAP systems.  \\
		\cline{4-5}
		&& $\bullet$ In ARANs, the low tiers provide lower latency, while high tiers provide better coverage. & \multirow{2}{*}{\cite{ren2019achievable}} & The achievable rate of CNPC links, which becomes higher at the price of larger latency, was analyzed.  \\
		\cline{2-5} 
		
		& \multirow{4}{*}{\makecell{Computation \\Latency}} 
		& Computation latency would be a critical component in ARANs as many compute-intensive  & \multirow{2}{*}{\cite{hu2019uav}} & An aerial relay is deployed to facilitate the computation of end users, considering the energy-delay tradeoff.\\
		\cline{4-5}
		&& applications are emerging, e.g., virtual reality and data analytics in the sky. & \multirow{2}{*}{\cite{zhang2019satellite}} & MEC is integrated into satellite communications to better serve sparsely distributed users. \\
		\hline

		\multirow{10}{*}{\makecell{Mobility\\ Analysis}} 
		& \multirow{6}{*}{\makecell{System \\Mobility}} 
		& $\bullet$ The system mobility is mainly reflected by the 3-D deployment of ABSs and constellation designs in (LEO) satellite communications. & \multirow{3}{*}{\cite{mozaffari2017mobile}} & UAVs are 3-D deployed for IoT data collection, which can reduce the energy consumption by 45\% compared to fixed BSs in terrestrial communications.\\
		\cline{4-5}
		&& $\bullet$ ARANs can be dynamically configured and mobilized to improve the QoS of end users and to benefit service/infrastructure providers. & \multirow{3}{*}{\cite{papa2020design}} & The integration of SDN into LEO communications was proposed to reduce the (migration and reconfiguration) costs of constellation designs.\\
		\cline{2-5} 
		
		& \multirow{4}{*}{\makecell{Trajectory \\Optimization}} 
		& In the lower tier (LAP) of ARANs, the trajectory of ABSs can be optimized for various purposes, e.g., for reducing energy consumption and minimizing the completion latency.	& \multirow{4}{*}{\cite{you20193d}} & The UAV trajectory and the problem of user scheduling were considered to maximize the data collection rate. The problem was solved using a convex approximation algorithm and the decomposition technique. \\
		\hline
	\end{tabular}
\end{table*}

\subsection{Summary and Discussion} \label{sec3e}
This section presents four key theoretical aspects of an ARAN architecture, including transmission propagation, energy consumption, latency analysis, and system mobility. In Table~\ref{tab_summary_system_model}, we summarize key points of the ARAN system model as well as representative references and their contributions. In particular, three transmission propagation approaches for ARANs (i.e., deterministic, stochastic, and geometric-based stochastic models) are reviewed in Section~\ref{sec3a}. We observe that existing studies mainly focus on channel modelling in sub-6G GHz frequency bands, while almost ignoring consideration of THz and non-RF models (e.g., visible light, neural, and molecular links). As the first attempt, the THz channel model is investigated in~\cite{xu2020joint} assuming that the placement and power allocation of UAV are jointly optimized to minimize the total latency between the UAVs and ground users. Consequently, more studies should be conducted on ARAN channels to characterize potential scenarios and frequency bands in 6G wireless systems. In Section~\ref{sec3b}, we review energy consumption models for UAVs in two modes: hovering as well as vertical and horizontal movements. It is observed that these models are well-established for both fixed-wing and rotary-wing UAVs. However, because more potential technologies and network scenarios that can be integrated with ARAN infrastructures will be available in 6G, more studies should be conducted. For example, the energy consumption model may vary if UAVs are equipped with wireless power transfer and/or energy harvesting capabilities. Next, communication and computational latency analysis of ARANs is provided in Section~\ref{sec3c}. Similar to energy consumption models, latency analysis of ARANs is currently performed in typical 5G scenarios, and thus it should be carried out in potential network scenarios of future 6G wireless systems such as massive URLLC and zero-touch services. Finally, system mobility and trajectory scheduling of ARAN platforms are reviewed in Section~\ref{sec3d}. Owing to the fact that ARANs would be configured dynamically and controlled adaptively, system performance can be significantly improved when compared with conventional terrestrial/stationary RANs.

\section{Enabling Technologies} \label{sec4}
To realize ARANs in 6G networks, we dedicate this section to reviewing the enabling technologies, including energy refills, operational management, and data delivery.

\subsection{Energy Refills} \label{sec4a}

\begin{figure*}[t]
\centering
\includegraphics[width=0.85\linewidth]{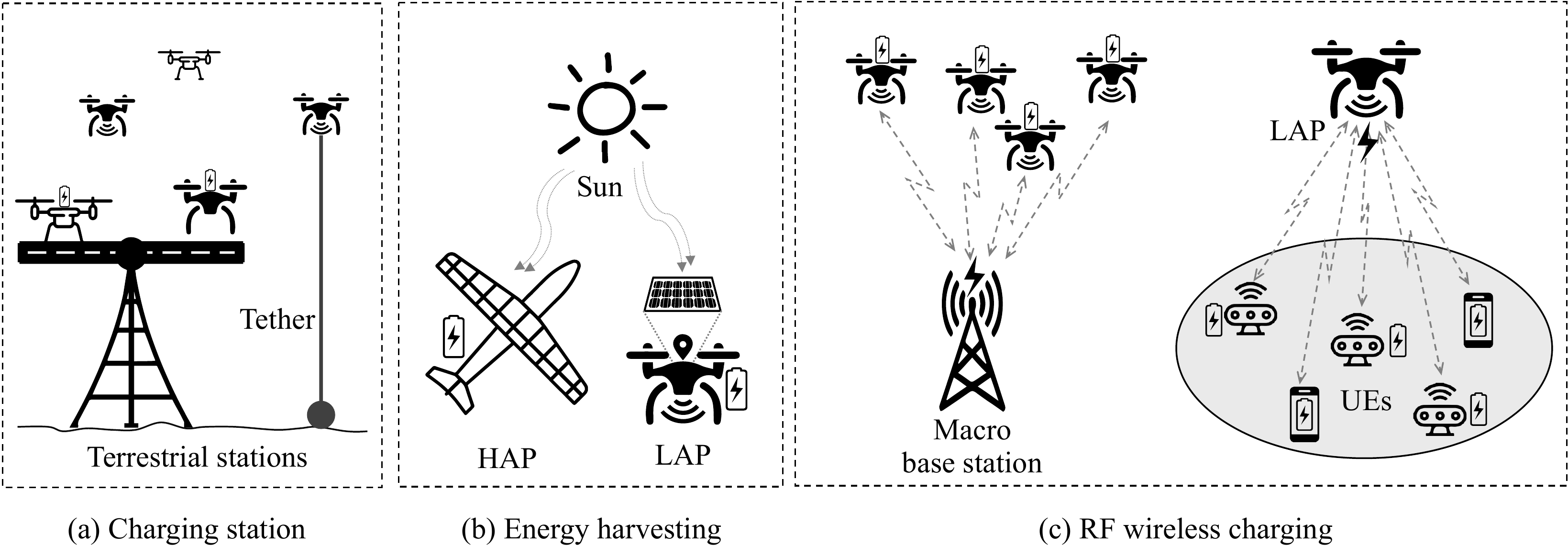}
\caption{Primary scenarios of energy refills in ARANs.} 
\label{fig:EnergyRefills}
\end{figure*}

As aforementioned in the ARAN access infrastructure's description, typical ABSs face battery storage capacity limitations during flight, which is the primary issue to be addressed.~\cite{youn2020aerodynamic}.
Therefore, research into energy replenishment strategies is essential. Aside from energy replenishment at charging stations, (self-)recharging through energy harvesting (EH) and RF wireless charging technologies are the most promising methods, with several advantages in terms of continuous provision of energy for ABSs working aloft.
Conversely, we can consider the capability of ABSs to use wireless power to serve massive low-power ground devices, especially in isolated and remote areas. An illustration of three primary scenarios for energy replenishment is shown in Fig.~\ref{fig:EnergyRefills}.

\subsubsection{Charging Station} \label{sec4a1}
This approach can be divided into three main categories of traditional charging stations, tethered technology, and near-field wireless charging.

\paragraph{Traditional Charging Stations} \label{sec4a1b}
For extended missions, terrestrial intermediate charging stations are the basic conventional approach to replenish ABSs’ energy.
Kantor~\etal~proposed a patent for a UAV-assisted recharging station~\cite{kantor2016unmanned}. 
In this work, a UAV can calculate a flight path and identify a terrestrial intermediate charging base station as the most suitable, considering the route characteristics to stop and recharge its batteries and then travel to the next mission location.
In addition to the conventional charging station model, Sharma~\etal~proposed the concept of charging stations where a UAV' exhausted batteries are replaced by fully charged ones from terrestrial macro base stations~\cite{sharma2017intelligent}.

To secure the charging process, several recent works have successfully incorporated blockchain technology to a system comprised of an ABS swarm and multiple charging stations as investigated in~\cite{dunphy2018first, rosa2018blockchain}. In an ARAN context, the blockchain technology allows secure peer-to-peer transactions among multiple networking components such as ABSs and user devices at different geographical locations. In the blockchain model, transaction information is stored in chained blocks. Every node in the committed chain can see all the transactions, and thus no node can fabricate or change the committed data in the chain~\cite{alladi2020applications}. In particular, the recent study in~\cite{hassija2020distributed} deployed a novel application for ABS systems and charging stations by implementing an advanced blockchain. The proposed solution is based on a tangled data structure that is equally secure, distributed as conventional blockchain, and simultaneously reduces power consumption and latency~\cite{popov2016tangle}. The numerical results in~\cite{hassija2020distributed} revealed that the solution can eventually provide an ABS swarm and charging stations with optimal cost and significantly outperform the conventional strategies. Although blockchain has been proven as a highly efficient technology for ARANs, these approaches suffer from several fundamental limitations, including a consensus mechanism consuming significant energy, considerably substantial latency from transaction confirmation, and constrained scalability~\cite{huang2019towards}. Nevertheless, traditional methods of energy transfer have several limitations such as service interruptions and small operational areas.

\paragraph{Tethered Technology}
Recently, tethered technology has been considered as a potential solution for ABS energy replenishment and system performance problems in ARANs~\cite{kishk2020,kishk2020aerial,alzidaneen2019resource,dicembrini2020modelling,bushnaq2020optimal}.
Assisted by the tethered connections, ABSs can not only replenish their limited battery onboard via a stable power supply from ground stations but can also extend the reliability of their backhaul links~\cite{kishk2020aerial}. Wrapped into a tether, two typical cables can be utilized separately, such as power and data transmission cables. In reality, various commercial products empowered by tethered technology have been launched by many companies around the world. Tethered UAV (T-UAV) systems are designed with diverse models, tether lengths, and fight time, as summarized in Table~\ref{Table:Summary_TUAV}.

 \begin{table*}[ht]
    
    \renewcommand{\arraystretch}{1.2}
 	\caption{Commercial tethered-UAV systems.}
	\label{Table:Summary_TUAV}
	\centering
	{\begin{tabular}{|c|p{8.3cm}|c|c|}
		\hline 
 		\textbf{Company} &
 		\centering\textbf{Series of products} &
		\textbf{Maximum tether length}&
		\textbf{Flight time}\\ 
		\hline
		\hline
        Lifeline~\cite{lifeline}&
        {Tethered Phantom 4, Mavic Pro, Mavic 2 Pro, Inspire 1 \& 2, etc.}&
        60 m&
        1.5--2.5 h	\\
		\hline
		
		\multirow{2}{*}{Acecore~\cite{acecore}} &
		\multirow{1}{*}{Neo Tethered System}&
		\multirow{2}{*}{60 m}&
		\multirow{2}{*}{$10+$ h}\\
		\cline{2-2}
		
		{} &
		\multirow{1}{*}{Zoe Tethered System}&
		{}&
		{}\\
		\hline
		
		{Ziyan~\cite{ziyan}}&
		{Ziyan's Tethering System}&
		{100 m}&
		{24 h}\\
		\hline
		
		{Tethered Drone System~\cite{TDS}}&
		{Tethered Drone System's T-UAV}&
		{120 m}&
		{24 h}\\
		\hline
		
		{Equinox Systems~\cite{equinox}}&
		{Falcon heavy, medium, and light}&
		{120 m}&
		{30 d}\\
		\hline
		
		\multirow{2}{*}{Hoverfly~\cite{hoverfly}}&
		{LiveSky$^{{\rm{TM}}}$ Sentry}&
		\multirow{2}{*}{120 m}&
		\multirow{2}{*}{20--30 d}\\
		\cline{2-2}
		
		{}&
		{BigSky}&
		{}&
		{}\\
		\hline
		
		\multirow{3}{*}{Elistair~\cite{elistair}}&
		{Orion 2}&
		{110 m}&
		{24 h}\\
		\cline{2-4}

		{}&
		{Safe-T 2}&
		{130 m}&
		{Unlimited}\\
		\cline{2-4}
		
		{}&
		{Light-T 4}&
		{70 m}&
		{Unlimited}\\
		\hline
		
		{Eagle Sky Light~\cite{eagleSkyLight}}&
		{Aquila 100}&
		{100 m}&
		{Unlimited}\\
		\hline

		\hline
	\end{tabular}
}
\end{table*}

In academic studies, the work in~\cite{bushnaq2020optimal} analyzed and compared performance between T-UAVs and conventional UAVs, referred to as untethered UAVs (U-UAVs). 
The numerical results in this research show that T-UAVs can achieve better performance for maximum cellular coverage than U-UAVs.
Similar observations have been revealed in~\cite{kishk2020aerial}, where T-UAVs with a 120-meter tether length can extend coverage probability up to 30\% compared to U-UAVs.
The study observed that the relative mission availability and recharging times of U-UAVs are approximately 70\% and 30\%, respectively, during operations. Meanwhile, 100\% available time is achieved by T-UAVs because of the usage of continuous power and data cables.
Nevertheless, it is necessary to further consider the mobility and endurance tradeoff to optimally select whether tethered or untethered solutions should be used because T-UAVs have several native drawbacks such as tether length limitation and strict inclination angle consideration to avoid tangling the tether on surrounding buildings or other obstacles.

\paragraph{Near-field Wireless Charging}
Near field wireless charging techniques are commonly categorized into two approaches, including inductive charging and magnetic resonance charging~\cite{kim2017review}. 
These approaches can be entirely integrated into a single charging station for ABSs. Owing to the short range of charging, it is harmless to the human body.
Although the boundary between the two strategies is not sharply delineated, they usually have distinctive characteristics that can be described as follows.

\begin{itemize}
    \item \textit{Inductive Charging:} Inductive charging is a wireless charging method in which a device uses magnetic induction to deliver electrical energy between two coils. The inductive charging method operates in the kHz frequency band with an effective charging distance generally within 20~cm~\cite{lu2016wireless}. Despite the limited transmission range, at a suitable distance, charging efficiency can be very high (up to 90\% within 17.5--26.5~cm~\cite{wu2012high}). In the context of ARANs, the technology is entirely suitable for deployment in charging stations to recharge the batteries of ABSs. For instance, a planar coil misalignment approach~\cite{campi2019wireless} was proposed for an inductive wireless charging system to recharge UAV batteries and reduce the weight of onboard components, leading to a cost reduction and providing a light-weight solution. Meanwhile, Obayashi~\etal~\cite{obayashi201985} elaborated a recharging prototype design with up to 450-W inductive power in the 85-kHz frequency band for a fast wireless charging port on a large airborne component like HAPs. 
    
    \item \textit{Magnetic Resonance Charging:} Magnetic resonance charging utilizes magnetic resonance between the transmitter and receiver antennas and operates in the MHz frequency bands.
    Because two resonant coils operate at the same resonant frequency, the energy transfer efficiency can be significantly high with only slight leakage and immunity to ambient frequency while meeting LoS transfer requirements.
    Magnetic resonance wireless charging has also proven its ability to transfer power over longer distances with higher energy efficiency than the inductive charging approach~\cite{lu2016wireless}. 
    In the context of ARANs, Wang~\etal~deployed a wireless magnetic resonant power transfer charging station for UAVs, solving the problems of endurance~\cite{wang2016design}. 
    Meanwhile, Qiong~\etal~\cite{qiong2018optimal} proposed an optimal design for magnetic resonance wireless charging in UAVs. 
    The experimental results in~\cite{qiong2018optimal} show that the energy transmission efficiency of their proposal can be kept stable despite the horizontal movement of the transmitting and receiving coils. 
    
\end{itemize}

\subsubsection{Energy Harvesting}
EH is another energy replenishment approach that is especially useful in the context of ARANs given their use of typical airborne components. The EH strategy is notably efficient for a swarm of ABSs instead of being limited to standalone devices as in the aforementioned traditional energy exchange methods. At present, EH is considered to be one of the best potential solutions for energy constraint problems, especially in self-powered systems like ARANs, because it allows the harvest of energy from external ambient sources such as solar, thermal, wind, vibration, and kinetic and then converts this energy into electrical current to support specific energy demands~\cite{sekander2020statistical}. Among these ambient sources, solar EH is the most popular approach, where LAP, HAP, and LEO airborne components can be powered by solar energy via photovoltaic (PV) cells.

The amount of available solar energy mainly depends on altitude, geographic location, radiation, the number of daylight hours, and the day of the year. The solar EH power resource under various conditions such as  cruising altitude, cloud types, and solar positions can be estimated as~\cite{zhang2020power}
\begin{equation}\label{eq:EH}
P_{EH}(y) =
    \begin{cases}
        SG\varepsilon(y) & \text{if } y \ge H_{up},
        \\
        SG{e^{ - {\beta_c}\left( {{H_{up}} - y} \right)}}\varepsilon \left( y \right) & \text{if } {H_{low}} < y < {H_{up}},
        \\
        SG{e^{ - {\beta_c}\left( {{H_{up}} - {H_{low}}} \right)}}\varepsilon \left( y \right) & \text{if } y \le {H_{low}},
    \end{cases}
\end{equation}
where $y$ is the cruising altitude of the ABS, $S$ and $G$ correspond to the  PV cell area onboard the solar-powered UAV, and the average solar radiation intensity on Earth. $\beta_c$ denotes an absorption coefficient modeling the optical characteristics of the surrounding clouds. While ${{H_{up}}}$ and ${{H_{low}}}$ are the upper and lower boundary altitudes of the clouds and ${\varepsilon \left( y \right)}$ is the energy harvesting efficiency as a function of $y$, which can be estimated as
\begin{align}
\label{eq:EHefficiency}
\varepsilon \left( y \right) = \varphi \left( y \right)\cos \theta ,
\end{align}
where $\varphi \left( y \right)$ and $\theta$ represent the atmospheric transmittance of the cloud at altitude $y$ depending on the cloud type, and the angle between the solar light rays and the PV cell depending on the solar position, respectively. 

Derived from \eqref{eq:EH}, it is seen that the altitude of an ABS influences the solar EH power significantly. Hence, it is necessary to consider a trade-off between EH and communication because more energy can be harvested at higher altitudes, whereas the communication path-loss is also larger as analyzed in Section \ref{sec3a}. There is a variety of research contributions regarding these issues. In particular, Jashnani~\etal~emphasized that the altitude, payload changes, and flight duration of solar-powered UAVs can be affected by its physical dimensions, weight, area of PV cells, the aspect ratio of the wing, and maximum battery recharging capacity~\cite{jashnani2013sizing}. The authors estimated the available solar power in two typical models with an altitude below approximately 2.5 km for one and a higher altitude for the other. The experimental results on charging capabilities for an engineering ground model with test UAVs in~\cite{jashnani2013sizing} showed excellent efficiency for  more than 24 h of continuous operation. Meanwhile, the solar-based UAV prototypes developed in~\cite{morton2015solar,oettershagen2016perpetual} demonstrated a continuous flight potential of up to 28 h. In~\cite{dwivedi2018maraal}, Dwivedi~\etal~experimented with a day flight for a solar-based UAV on April 1, 2017, where the test time was from 9:30 to 18:00. From these measurement results, it is recognized that the generated power is less than the power required by the UAV system from approximately 15:30 to 17:10. This can be explained by the throttle being cut and the UAV being directed into an extremely slow glide maneuver at 0.15~m/s while the available solar power was still being generated normally. In addition, the power generated during this period was utilized to charge the batteries. In~\cite{wang2018analysis}, Wang~\etal~studied a simulation model of solar cell behavior for a solar-powered UAV via the MATLAB/Simulink platform where the test time was from 6:00 to 18:00 over three days from July 10 to 13, 2017. This included light rain, cloudy, and sunny days as three typical weather types. The experimental results demonstrated the impact of light intensity and ambient temperature on solar EH efficiency. Meanwhile, Rajendran~\etal~investigated a maximum power point tracker for the optimal operation of solar-based UAVs by demonstrating the impact of temperature and solar irradiance intensity on various solar module angles~\cite{rajendran2018experimental}. The results showed that approximately 45$^{\circ}$C is the optimal operating temperature, and the solar power rose almost linearly along the tilt angle of this solar module.

From the standpoint of optimization problems, Hosseini~\etal~considered optimal path scheduling and power allocation for UAVs, where the UAVs' wings are equipped with PV cells to harvest solar energy and recharge the batteries~\cite{hosseini2013optimal}. The crucial objectives in~\cite{hosseini2013optimal} were to optimize energy storage coupled with a flight path trajectory based on the allocation of available power. However, a sophisticated path-scheduling algorithm was called for by the constraints of strict moving trajectories and flight altitudes. To this end, Sun~\etal~examined a design with optimal joint resource allocation including 3-D position, power, and subcarrier for a solar-powered UAV~\cite{sun2018resource}. Unfortunately, the assumption of constant aerodynamic power-consumption in~\cite{sun2018resource} is unrealistic because it still depends on flight velocity. For this issue, they further investigated a communication system with a multicarrier solar-based UAV by jointly optimizing the amount of solar EH, aerodynamic power consumption, capability of the onboard energy storage, and QoS requirements of ground users~\cite{sun2019optimal}. The experimental results revealed that EH efficiency may be higher when flying above the clouds. Furthermore, Zhang~\etal~\cite{zhang2020power} exploited an intelligent power approach by combining a reinforcement learning mechanism with the impact of a solar-based UAV’s flight altitude. Simulation results demonstrated that this approach could simultaneously improve both communication performance and EH efficiency.

Conversely, EH strategies from other ambient sources including thermal, wind, vibration, and kinetic, can be seen as complementary ways of utilizing more energy. For instance, Wang~\etal~published a patent of a multirotor aerial drone with thermal energy scavenging capability in~\cite{wang2019multi}. The patent presented a thermoelectric generator that harvests the waste thermal energy from a processor in a UAV, and then the UAV batteries are recharged to prolong flight time. Bonnin~\etal~in~\cite{bonnin2015energy} described a comprehensive understanding of principles and mathematical formulae for a dynamic soaring technique. This technique is inspired by albatross flight behavior, with which they can fly against the flow without flapping their wings, i.e., without having to waste energy. Meanwhile, Anton is a pioneer in investigating and designing novel piezoelectric elements based on UAV platforms~\cite{anton2008vibration}. Koszewnik~\etal~in~\cite{koszewnik2019performance} proposed a common system for vibrational energy scavenging using a piezoelectric harvester integrated into UAV platforms. Experiment results showed that it is possible to extend flight duration using integrated piezo patch elements with flexible batteries for each UAV wing. Moreover, Anton~\etal~in~\cite{anton2011performance} developed an onboard hybrid model using solar and vibrational EH. This hybrid approach takes advantage of both EH methods and supports higher energy efficiency.

\subsubsection{RF Wireless Charging}
RF wireless charging (a.k.a. far-field wireless charging) operates in various frequency bands from 300~MHz to 300~GHz~\cite{lu2016wireless}. From the standpoint of estimation, the amount of RF power, $P_r$, received from this RF wireless charging strategy is dependent on several typical parameters such as transmission power, wavelength, and the distance between transmitter and receiver and can be estimated following the Friis formula as~\cite{belo2019selective}
\begin{align}
\label{eq:RFWirelessCharging}
P_r = {P_t}{G_t}{G_r}{\left( {\frac{\lambda }{{4\pi d}}} \right)^2},
\end{align}
where ${P_t}$ is the total power of the transmitter,
$\lambda$ is the wavelength of the RF signals,
$d$ denotes the distance between transmitter and receiver, while
${G_t}$ and ${G_r}$ are the transmitter and receiver gains, respectively.

The charging distance of this approach can be much further than that of near-field wireless charging, i.e., up to several tens of km.
Furthermore, the power inversion efficiency obtained is significantly high at up to 84\% at 5.8~dBm cumulative received RF power~\cite{kuhn2015multi}. In particular, the energy efficiency parameter, $\delta$, can be estimated by the ratio of the output of usable electrical power, $P_{DC}$, and the received RF power, $P_r$, as~\cite{belo2019selective}
\begin{equation}
\label{eq:RFPowerConversionEfficiecy}
\delta  = \frac{P_{DC}}{{P_r}}.
\end{equation}

Next, we invoke two ARAN scenarios applying RF wireless charging strategies, where ABSs act as either (\textit{i}) powered devices recharged by terrestrial macro base stations or (\textit{ii}) wireless power supply to user devices.

\paragraph{ABSs as Powered Devices}
This category includes the works wherein ABSs are powered by terrestrial macro base stations using RF wireless charging technology. For instance, the work in~\cite{dunbar2015wireless} investigated a wireless charging method for a micro-UAV operating at a frequency of 5.9~GHz, where the batteries were replenished via an integrated rectifier antenna with approximately 5~W of transmit power. 
A design for the power receiving side of a wireless charging system for ABS applications was proposed in~\cite{muharam2017design}. 
The results reported in~\cite{muharam2017design} indicated that the proposed method achieved an energy conversion efficiency of more than 77\%. 
Long~\etal~\cite{long2018energy} developed a framework to energize ABSs using a hybrid of the EH and RF wireless charging methods. The framework consists of communication and networking architectures as well as protocols designed for realizing multidimensional objectives of a charging system to significantly prolong the continuous operating lifetime of ABSs. The authors in~\cite{chen2019resonant} conducted research that investigated the possibility of using resonant beam charging to replenish the energy for ABSs and achieved the anticipated results.
Moreover, they considered the joint optimization of the ABSs' trajectory and the recharging station’s power. 
Lahmeri~\etal~in~\cite{lahmeri2020stochastic} proved that at least 6 laser beaming directors per 10~km$^2$ are required to ensure probable energy coverage of above 0.9. 
Meanwhile, Ouyang~\etal~solved the problem of jointly optimizing the transmit power allocation and trajectory of ABSs~\cite{ouyang2018throughput}. 
Chen~\etal~in~\cite{chen2015design} proved the feasibility of this laser approach with an energy conversion efficiency of up to 17.55\%.
However, this laser beaming method requires safety measures under Federal Communications Commission (FCC) regulations in residential and industrial areas where human health must be protected and managed.

\paragraph{ABSs as Power Supply}
In this category, once ABSs have more available energy, they can be considered as an aerial power supply to simultaneously energize massive numbers of low-power user devices. For instance, the work in~\cite{marano2018resource} examined optimal energy allocation in the context of a wireless sensor network energized by dynamic ABSs using RF wireless charging technology. By exploiting the ABSs' trajectory approach, Xu~\etal~proposed the first work on characterizing the energy obtainable by terrestrial users through ABS-enabled wireless power transfer~\cite{xu2018uav}. The authors deployed their approach among multiple terrestrial users, where the optimal trajectory subject to flying velocity constraints was considered to maximize the energy received by users and minimize energy consumption. Furthermore, Su~\etal~\cite{su2020uav} recently introduced a dynamic bipartite matching mechanism, where ABSs act as wireless power chargers to replenish the energy of low-power terrestrial devices with maximal charging efficiency. To jointly optimize charging efficiency and communication throughput maximization,~\cite{wang2018resource} applied a time and power optimization algorithm to maximize average throughput, with ABSs functioning as an original energy source to supply multiple terrestrial devices. To this end, they designed a harvest-transmit-store strategy.

A summary of energy replenishment contributions is presented in Table.~\ref{Table:Summary_EnergyRefills}.

 \begin{table*}[h]
     \renewcommand{\arraystretch}{1.2}
 	\caption{Summary of contributions to energy replenishment for ABSs.}
	\label{Table:Summary_EnergyRefills}
	\centering
	\begin{tabular}{|c|c|p{11.9cm}|}
		\hline 
 		\centering\multirow{1}{*}{\textbf{Ref.}} &
 		\centering\multirow{1}{*}{\textbf{Technology}} &
		\hspace{4.5cm}\multirow{1}{*}{\textbf{Main contributions}}  \\ \hline
		\hline 
		\multirow{1}{*}{\cite{kantor2016unmanned}} &
		\multirow{3}{*}{Conventional charging station} &
		\multirow{1}{*}{Electrical charging for UAVs from stations}\\
		
		\cline{1-1}	\cline{3-3}		\multirow{1}{*}{\cite{sharma2017intelligent}} &
		\multirow{1}{*}{} &
		\multirow{1}{*}{Replacing UAV's exhausted batteries from charging stations}\\
		
		\cline{1-1}	\cline{3-3}	
		\multirow{1}{*}{\cite{hassija2020distributed}} &
		\multirow{1}{*}{} &
		\multirow{1}{*}{Optimized costs, security, power-consumption, and low latency for an ABS swarm}\\
		\hline

		\multirow{2}{*}{\cite{lifeline,acecore,ziyan,TDS,equinox,hoverfly,elistair,eagleSkyLight}} &
		\multirow{4}{*}{Tethered technology} &
		Commercial T-UAVs with the various supported maximum tethered lengths up to 130 m as well as the capability of unlimited flight duration\\

		\cline{1-1}	\cline{3-3}	
		\multirow{2}{*}{\cite{bushnaq2020optimal,kishk2020aerial}} &
		\multirow{1}{*}{} &
		The significant out-performance is achieved when using the help from T-UAVs' energy replenishment perspectives compared to U-UAVs\\
		\hline

		\multirow{1}{*}{\cite{campi2019wireless}} &
		\multirow{2}{*}{Inductive charging} &
		\multirow{1}{*}{A planar coil misalignment to recharge UAV batteries and reduce weight of UAV onboard components}\\
		
		\cline{1-1}	\cline{3-3}	
		\multirow{1}{*}{\cite{obayashi201985}} &
		\multirow{1}{*}{} &
		\multirow{1}{*}{Fast wireless charging prototype for HAP components at 450-W power recharging in the 85-kHz band}\\
		\hline
		
		\multirow{1}{*}{\cite{wang2016design}} &
		\multirow{2}{*}{Magnetic resonance charging} &
		\multirow{1}{*}{Powering UAV batteries and solving the problems of endurance}\\
		
		\cline{1-1}	\cline{3-3}
		\multirow{1}{*}{\cite{qiong2018optimal}} &
		\multirow{1}{*}{} &
		\multirow{1}{*}{An optimal design for UAVs to maintain stable energy transmission efficiency}\\
		\hline
		
		\multirow{1}{*}{\cite{jashnani2013sizing}} &
		\multirow{9}{*}{Solar EH} &
		\multirow{1}{*}{More than 24 h of continuous operation for UAVs}\\
		
		\cline{1-1}	\cline{3-3}
		\multirow{1}{*}{\cite{oettershagen2016perpetual}} &
		\multirow{1}{*}{} &
		\multirow{1}{*}{Up to 28 h of continuous UAV operation}\\
		
		\cline{1-1}	\cline{3-3}
		\multirow{1}{*}{\cite{dwivedi2018maraal}} &
		\multirow{1}{*}{} &
		\multirow{1}{*}{The impact of daytime hours of flight for a solar-powered UAV}\\
		
		\cline{1-1}	\cline{3-3}
		\multirow{1}{*}{\cite{wang2018analysis}} &
		\multirow{1}{*}{} &
		\multirow{1}{*}{The impact of light intensity and ambient temperature on solar EH efficiency of UAVs}\\
		
		\cline{1-1}	\cline{3-3}
		\multirow{1}{*}{\cite{rajendran2018experimental}} &
		\multirow{1}{*}{} &
		\multirow{1}{*}{The impact of temperature and solar irradiance intensity on solar-based UAVs for optimal operation}\\
		
		\cline{1-1}	\cline{3-3}
		\multirow{1}{*}{\cite{hosseini2013optimal}} &
		\multirow{1}{*}{} &
		\multirow{1}{*}{Joint optimal path scheduling and power allocation for a UAV with PV cells}\\
		
		\cline{1-1}	\cline{3-3}
		\multirow{1}{*}{\cite{sun2018resource}} &
		\multirow{1}{*}{} &
		\multirow{1}{*}{Joint optimal 3-D position, power, and subcarrier for a solar-powered UAV}\\
		
		\cline{1-1}	\cline{3-3}
		\multirow{1}{*}{\cite{sun2019optimal}} &
		\multirow{1}{*}{} &
		\multirow{1}{*}{Joint optimal solar EH, power consumption, onboard energy storage, and users' QoS requirements}\\
		
		\cline{1-1}	\cline{3-3}
		\multirow{1}{*}{\cite{zhang2020power}} &
		\multirow{1}{*}{} &
		\multirow{1}{*}{The intelligent power mechanism using ML regarding the flight altitude of solar-based UAVs}\\
		\hline
		
		\multirow{1}{*}{\cite{wang2019multi}} &
		\multirow{1}{*}{Thermal EH} &
		\multirow{1}{*}{Harvesting the waste thermal energy from a UAV processor to prolong flight time}\\
		\hline

		\multirow{1}{*}{\cite{bonnin2015energy}} &
		\multirow{1}{*}{Wind EH} &
		\multirow{1}{*}{A dynamic soaring technique to harvest wind energy and reduce the UAVs energy consumption}\\
		\hline
		
		\multirow{1}{*}{\cite{koszewnik2019performance}} &
		\multirow{1}{*}{Vibration EH} &
		\multirow{1}{*}{A piezoelectric harvester integrated into UAV platforms to extend the flight duration}\\
		\hline

		\multirow{1}{*}{\cite{anton2011performance}} &
		\multirow{1}{*}{Solar and vibration EH} &
		\multirow{1}{*}{An onboard UAV hybrid model to support higher energy efficiency}\\
		\hline
		
		\multirow{1}{*}{\cite{long2018energy}} &
		\multirow{1}{*}{EH and RF wireless charging} &
		\multirow{1}{*}{Prolongation of the continuous operating lifetime of ABSs via combination frameworks}
		\\
		
		\hline
		\multirow{1}{*}{\cite{dunbar2015wireless}} &
		\multirow{11}{*}{RF wireless charging} &
		\multirow{1}{*}{Replenishing batteries via an integrated rectifier antenna with approximately 5-W transmission power}\\
		\cline{1-1}	\cline{3-3}

		\multirow{1}{*}{\cite{muharam2017design}} &
		\multirow{1}{*}{} &
		\multirow{1}{*}{More than 77\% energy conversion efficiency via power on the receiving side of ABSs}\\
		\cline{1-1}	\cline{3-3}
		
		\multirow{1}{*}{\cite{chen2019resonant}} &
		\multirow{1}{*}{} &
		\multirow{1}{*}{Joint optimal ABSs’ trajectory and the recharging station power}\\
		\cline{1-1}	\cline{3-3}

		\multirow{1}{*}{\cite{lahmeri2020stochastic}} &
		\multirow{1}{*}{} &
		\multirow{1}{*}{At least six laser beaming directors are required to ensure the ABSs coverage probability per 10~km${^2}$}\\
		\cline{1-1}	\cline{3-3}

		\multirow{1}{*}{\cite{ouyang2018throughput}} &
		\multirow{1}{*}{} &
		\multirow{1}{*}{Joint optimal transmission power allocation and ABSs' trajectory}\\
		\cline{1-1}	\cline{3-3}

		\multirow{1}{*}{\cite{chen2015design}} &
		\multirow{1}{*}{} &
		\multirow{1}{*}{Up to 17.55\% of energy conversion efficiency for UAVs via the laser beaming approach}\\
		\cline{1-1}	\cline{3-3}

		\multirow{1}{*}{\cite{marano2018resource}} &
		\multirow{1}{*}{} &
		\multirow{1}{*}{Optimal energy allocation for dynamic ABS networks}\\
		\cline{1-1}	\cline{3-3}

		\multirow{1}{*}{\cite{xu2018uav}} &
		\multirow{1}{*}{} &
		\multirow{1}{*}{Optimal UAV's trajectory subject to flying velocity constraint for optimal energy efficiency}\\
		\cline{1-1}	\cline{3-3}

		\multirow{1}{*}{\cite{su2020uav}} &
		\multirow{1}{*}{} &
		\multirow{1}{*}{A dynamic bipartite matching mechanism with maximal charging efficiency for ABS networks}\\
		\cline{1-1}	\cline{3-3}

		\multirow{1}{*}{\cite{wang2018resource}} &
		\multirow{1}{*}{} &
		\multirow{1}{*}{Jointly-optimized charging efficiency and communication throughput maximization}\\
		\hline
	\end{tabular}
\end{table*}

\subsection{Operational Management} \label{sec4b}
This subsection addresses three foundational pillars relating to operational management planes: network softwarization, mobile cloudization, and data mining.

\subsubsection{Network Softwarization}
Network softwarization plays a crucial role in harmonizing network and computational resources across multiple tiers in ARANs~\cite{chen2015software}, a.k.a. software-defined networking (SDN). To this end, network softwarization improves programmable network management control, enabling global visibility through a central orchestrator~\cite{mamushiane2018comparative,oubbati2020softwarization}.
As a result, ARANs are empowered with abilities to (\textit{i}) provide network operators with enhanced control, situational awareness, and flexibility, (\textit{ii}) coordinate interference avoidance, and (\textit{iii}) facilitate interoperability between network nodes.

Typical publications that applied the network softwarization architecture to ARAN were perused. In~\cite{zhao2019software}, Zhao~\etal~proposed an SDN-UAV architecture where SDN-integrated UAV-based radio networks are deployed with separate data and control planes, and the behaviors of UAVs are controlled by providing network programmability. The results in~\cite{zhao2019software} show that a controller can successfully consider circumstantial information related to a global UAV group to select the most suitable flight path route trajectory and avoid collisions among multiple UAVs. The work in~\cite{zhang2018sdn} incorporates SDN deployment in the LAP/HAP tiers, where an SDN controller based on a monitoring platform can make intelligent optimization decisions owing to its ability to learn and synthesize network information gathered by ABSs.  In considerations of the ABSs' power limitation, a load-balancing algorithm was proposed. Xiong \textit{at al.}~\cite{xiong2019sdn} studied an SDN structure for ABS ad hoc topologies, where the network supports flexible data transmission among payloads, adapts to frequent change, and improves the security of the network topology. In~\cite{hermosilla2020security}, the security management of automatic orchestration, deployment, and configuration in the MEC-LAP/HAP networks was considered. In this context, an integration of SDN orchestration and network function virtualization (NFV) considered several contextual virtual and physical conditions and metrics for coordination in ABSs.

A large-scale implementation of conventional SDNs, known as Loon SDN, has been exploited to optimally interoperate and coordinate complex networks, especially aerospace networks~\cite{barritt2018loon}. Here integrated terrestrial and aerial segments, corresponding to ground and LAP/HAP/LEO platforms in the ARAN network architecture, are considered in view of the packet routing and physical wireless topology. Regarding LEO communication systems, the work in~\cite{papa2018dynamic} deployed a use-case study of an SDN-enabled LEO satellite space segment that considers dynamically changing traffic demands based on geographical position and end users’ time zones. Recently, this research group has also further investigated an SDN with LEO constellation based on programmable and reconfigurable concepts~\cite{papa2020design}. Here, programmable SDN controllers that update the forwarding rules of data plane devices are integrated with the control logic. Additionally, the authors in~\cite{jia2020virtual} devised a virtual network based on an orchestration model to realize virtual resource management for LEO satellite networks.

\subsubsection{Mobile Cloudization}
Mobile cloudization encompasses computing resources at all tiers of the ARAN and responds on demand to the flexibility of network resource allocation~\cite{dao2020ic,abbas2020fog,mohamed2017uavfog}. As investigated in Section~\ref{sec3c}, the application of mobile cloudization includes mutual assistance scenarios between ABSs and computing platforms; prime examples are shown in Fig.~\ref{fig:UAV_MEC_scenarios}. In these contexts, ARANs benefit from mobile cloudization capabilities for computational offloading in terms of energy efficiency and latency reduction.

For instance, the issue of offloading highly intensive computational tasks to LAP/HAP tiers with MEC capability to reduce energy overhead and execution delay is reviewed in~\cite{messous2017computation}. Meanwhile, the work in~\cite{pinto2019framework} studied fog-cloud computing cooperation that develops the ability to reduce latency and power consumption as well as improve the scalability and efficiency of end user information exchange in an ARAN. A similar approach with excellent results between edge and cloud computing and one more feature at high long term performance in online environments for ABS swarms is also examined in~\cite{liu2020online}.
Furthermore, a joint optimization in data allocation, trajectory, and energy consumption is achieved for UAV-assisted MEC systems in~\cite{diao2019fair}, where a UAV plays the role of a computing server to provide computational offloading from the multiple mobile users' tasks.
One more valuable contribution is found in~\cite{yang2020offloading}, where the proposed system model belongs to the category of Fig.~\ref{fig:UAV_MEC_scenarios}~(b).
In this work, the authors envisioned a visual target tracking experiment test using deep learning (DL) for a trained convolutional neural network (CNN) model.
The lower layers of a CNN deployed on a UAV can provide sufficient tracking performance in good image quality conditions, whereas the higher layers at the MEC server are used in conditions of poor image quality.
In addition, other constraints such as resource sharing among multiple UAVs and bandwidth for communicating between UAVs and MEC servers are considered as well, showing significant performance benefits from an MEC implementation.

Regarding cloudization in LEO satellite systems, the authors in~\cite{ilchenko2019solution} investigated an LEO satellite constellation with nano-satellites and CubeSats using fog computing. To realize fog computing, satellite computers were included in each of the distributed satellites~\cite{qiu2019deep}. This achieved several advantages including (\textit{i}) delay value of round-trip IoT data transmission time, (\textit{ii}) data processing time of IoT devices, (\textit{iii}) orbital distributed database transmission time, and (\textit{iv}) computational load balancing through the LEO satellite IoT System's orbital distributed computing network. An edge computing implementation in LEO satellite networks is reviewed in~\cite{zhang2019satellite}, where a user device without an edge server nearby also enjoyed edge computing services via satellite links. Moreover, to achieve parallel computation in LEO--terrestrial networks, a cooperative computation offloading model was designed and simultaneously integrated with network resources via a dynamic NFV technique.

\subsubsection{Data Mining}
This subsection presents the results of our investigation of data mining techniques as another key enabler for achieving optimal targets such as reducing energy consumption, improving network security, sharing workload across an entire network, increasing bandwidth, self-organizing networking configurations, achieving autonomous training operations,  handling mobile big data, and analyzing mobility~\cite{qian2017survey, zhang2019deep}.
Data mining is defined as the process of applying specific algorithms to extract useful patterns from data to feed further activities such as feature prediction and optimal action decision making~\cite{han2011data}. 
In ARAN architecture models, the data is an entire information set located on network nodes such as LAP/HAP/LEO components and terrestrial devices.
By exploiting the value of data mining, machine learning (ML) can provide solutions for simultaneous massive numbers of user connections in a dynamic, heterogeneous, and unpredictable network resource such as ARANs in the context of 6G~\cite{bithas2019survey}.

The work in~\cite{carrio2017review} reviewed applications of DL for LAP/HAP components and services where the main application scenarios were feature extraction, planning, situational awareness, and motion control. Conversely, the work in~\cite{luong2019applications}  surveyed the application of deep reinforcement learning to address issues emerging in multitier LAP/HAP systems as well as communications such as data rate control, dynamic network access, wireless caching, network security, data offloading, connectivity preservation, traffic routing, resource sharing, and data collection.
In~\cite{mukherjee2020distributed}, a deployment of DL for the UAV-enabled MEC network was exploited to minimize the energy consumption and the weighted sum of latency. Meanwhile, the work in~\cite{fotouhi2019survey} provided a comprehensive survey on using ABS-assisted cellular communications, where ML algorithms were used for data popularity analysis for trajectory and placement purposes. From the standpoint of cybersecurity, the work in~\cite{mozaffari2019tutorial} provided a comprehensive research on LAP/HAP systems and communications from a cyber physical security perspective. Furthermore, the authors in~\cite{shakeri2019design} provided a comprehensive survey of multi-UAV systems, where ML is applied for fine-grained cyberphysical applications, highlighting key aspects such as coverage spanning, tracking of targets and infrastructure objects, energy efficient navigation, and image analysis assessment.

Regarding LEO tiers, Ferreira~\etal~\cite{ferreira2019reinforcement} devised and deployed reinforcement learning for LEO satellite communications. ML decisions based on reinforcement learning were used to configure a satellite link from an LEO constellation to a ground station to promote high throughput, low bit error rate, bandwidth optimization, and reduced power consumption. From another perspective, the work in~\cite{werth2020silo} described the use of a LEO satellite dataset recorded at a terrestrial optical observatory, used along with ML strategies to produce higher resolution image recovery on degraded image sets and perform image interpretability assessment.
Subsequently, Chen~\etal~introduced a novel ML application for estimating precipitation based on LEO satellite observations, where an ML platform was used to enhance estimation accuracy~\cite{chen2020machine}.
Especially in~\cite{kato2019optimizing}, multitier LAP/HAP/LEO communication systems in ARANs and ground stations were used to investigate DL algorithms and training method implementation to improve various aspects of network performance.

\subsection{Data Delivery} \label{sec4c}
This subsection provides readers with the enabling technologies that support data delivery in ARANs. In this regard, key technologies for addressing particular issues such as frequency spectrum, communication protocol, and multiaccess approaches are introduced.

\subsubsection{Frequency Spectrum}
As mentioned regarding the ARAN architecture above, mobile wireless technologies such as 5G NR and Wi-Fi are considered to support links among ABSs at LAP/HAP tiers and between ABSs and (terrestrial and aerial) users. 5G NR and Wi-Fi operate at various frequency bands including sub-6 and THz frequencies in both the licensed and unlicensed spectra~\cite{parkvall2017nr}.
Motivated by this, the work in~\cite{mozaffari2019tutorial} demonstrated that ultra-high-speed wireless backhaul can be achieved using 3-D beamforming for ABSs in the mmWave bands, significantly improving flexibility and reducing the comparative cost of wired backhaul in terrestrial networks.
The author in~\cite{zhou2018beam} proposed a mmWave UAV mesh network using a fast beam-tracking mechanism to provide ultra-high-speed wireless backhaul between UAVs and relays for terrestrial base stations.
Meanwhile, an efficient channel tracking strategy was proposed for mmWave UAV communications, where the complexity of the downlink channel tracking was significantly decreased by improving both angle and Doppler reciprocities~\cite{zhao2018efficient}.
With regard to LEO satellite systems, seamless integration of ultra-broadband short-range wireless
network segments into satellite constellations was devised to provide multigigabit multimedia to nomadic users via mmWave satellite links~\cite{mudonhi2018sdn}.
The study achieved advantages of increased throughout and reduced latency in the proposed network.
The authors in~\cite{artiga2018shared} introduced two technological enablers applying to mmWave satellite-terrestrial communications, which were a smart antenna approach and programmable intelligent management to enhance mobile broadband access in denser conditions.

Furthermore, THz frequency spectrum bands (i.e., 0.1--10~THz) higher than mmWave are discussed with regard to LAP/HAP/LEO communication links.
These THz bands are utilized for LEO satellite links~\cite{tekbiyik2020reconfigurable}, using reconfigurable intelligent surface (RIS) technology to compensate for the high path loss at high carrier frequencies, leading to improvements in signal-to-noise ratio, i.e., it significantly improves system performance.
Meanwhile,~\cite{wang2019inter} explored the use of THz bands between micro-satellites and ground stations, in a scenario where THz quantum entanglement distribution and THz quantum key distribution are suitable for deployment to achieve desirable high performance.
In the context of LAP/HAP communications, the authors in~\cite{mendrzik2018error} presented MIMO links in a UAV swarm using orthogonal frequency division multiplexing in the THz band, reaching millimeter-scale positioning accuracy with respect to separation of dimension, bandwidth, and transmitter-receiver
array.

\subsubsection{Communication Protocol}
As particularly analyzed in Section~\ref{sec3c}, URLLC communications are considered one of the most promising protocols to accommodate ARANs’ broad coverage, low latency, and ultra-reliability towards 6G requirements. Although ARANs and URLLC have a mutual supportive relation, this section is dedicated to reviewing URLLC communication as an enabler from the perspective of ARANs.

For instance, the authors in~\cite{chu2019uav} provided an assessment of short packet communications in a UAV-IoT network, where spectrum sharing based on intelligent radio strategies programs and dynamically configures the use of the optimized wireless channels to avoid user interference and congestion. This work invokes the concept of deploying blocklengths as short as possible to reduce latency and meet URLLC service expectations for reliable communications between UAVs and IoT devices.
In~\cite{han2019uav}, non-orthogonal multiple access (NOMA)-based URLLC communications in ARANs without any assistance from terrestrial base stations was analyzed.
The authors minimized a block error rate by appropriately narrowing the beam width to improve transmission distance and communication reliability on access links. The beam width reduction was adopted via a user grouping approach to reduce coverage of areas with no users.
Additionally, the work in~\cite{she2019ultra} investigated a framework for enabling URLLC in the context of LAP/HAP systems with control and non-payload communications, where a modified distributed multiantenna system was adopted to judiciously optimize UAVs' altitude, the uplink and downlink duration, and the antenna configuration. From the standpoint of LEO satellite segments, the work in~\cite{cho2019cross} proposed a power-efficient control link algorithm to minimize total power consumption and response reliability by adopting URLLC communications for the satellite links in the context of ARANs.
In addition, the literature review in~\cite{leyva2020leo} showed that URLLC communications can support ultra-reliability as well as low latency requirements within tens of milliseconds for LEO constellations in ARANs.

\subsubsection{Multiaccess}
From a multiaccess perspective, massive MIMO, NOMA, and RIS are considered the main technological enablers to further enhance an ARAN network’s performance.
\paragraph{Massive MIMO}
It is well known that a large volume of data signals can be transmitted and received simultaneously in the spatial domain by equipping many independently controlled antennas to participate in network nodes, known as massive MIMO~\cite{jameel2017massive}.
Additionally, beamforming, based on a highly directional antenna beam, can be considered as one of the effective solutions to achieve sufficient antenna gain to overcome path loss, handle multipath and interference phenomena, and ensure a high signal-to-noise ratio at output~\cite{molisch2017hybrid}.

Returning to ARAN networks, Chandhar~\etal~introduced a massive MIMO deployment to a LAP-based (e.g., UAV/drone swarms) network~\cite{chandhar2017massive}, where multiple single-antenna UAVs simultaneously communicate with a multiantenna ground station. 
This work established the optimal antenna distance and demonstrated that the ergodic rate per UAV reached a maximal value because UAVs are spherically evenly distributed around a ground station with an antenna distance respecting an integer multiple number of half a wavelength.
The authors in~\cite{geraci2018supporting} provided support for multiuser massive MIMO systems by UAVs, where the UAV downlink control and command channel was considered under realistic 3GPP assumptions. 
The results from~\cite{geraci2018supporting} hold that the massive MIMO utilized with UAV communications can significantly improve system performance across several typical parameters such as spatial multiplexing gain, interference mitigation, and carrier signal strength.
Similarly, a UAV network assisted by a cell-free massive MIMO architecture where Rician fading is examined to obtain a closed-form expression of spectral efficiency was studied in~\cite{d2019cell}.
In addition, user scheduling and power allocation strategies were proposed and shown to provide superior performance.

 A massive MIMO channel model for LEO satellite systems was established in~\cite{you2020leo}, where Doppler and time delay compensation techniques and user grouping algorithms were proposed.
The study presented a user grouping algorithm aiming to schedule access for terrestrial end users using the same time and frequency resources for various groups.
In brief, the data rate of LEO satellite systems was significantly improved by massive MIMO model.
The work in~\cite{Li2020downlink} designed a massive MIMO with downlink transmission for LEO satellite communications, where all participating nodes are equipped with uniform planar arrays.
The maximization of the ergodic sum-rate was achieved via a low-complexity algorithm for Lagrangian multiplier optimization and provided the expected performance improvement.

\paragraph{NOMA}
The NOMA concept has been widely utilized in various communication schemes as being advantageous for enhancing network performance by improving spectrum efficiency, increasing capacity, decreasing transmission latency, heightening throughput, and enabling data transmission among a vast array of UEs that simultaneously request access~\cite{makki2020survey,fang2020energy}.
There are several publications concerned with the advantages of the NOMA concept and its integration with LAP/HAP/LEO communications.
Chu~\etal~\cite{chu2020robust} investigated an LEO-terrestrial network of satellites equipped with multibeam technology, considering massive access using the NOMA scheme.
Two robust beamforming algorithms were proposed to minimize total power consumption.
Using a similar approach, Gao~\etal~\cite{gao2020performance} contributed to the NOMA technique used in LEO satellite communications, achieving a significantly enhanced performance relating to ergodic capacity, outage probability, and mutual information.

In the manner of LAP/HAP communications, Nasir~\etal~\cite{nasir2019uav} deployed a UAV-assisted network to provide a large number of terrestrial UEs with multiconnection capability using the NOMA technique, where a path-following algorithm was proposed to reach the optimal max-min rate. 
The work in~\cite{botsinis2018air} employed beamforming based on the arriving signals' angle and an array of antennas using the NOMA strategy for HAP networks to improve performance by reducing the bit error rate.
A contribution to the integration of MIMO and NOMA for UAV-assisted networks can be found in~\cite{hou2019multiple}, in which, as expected, the outage probability and ergodic rate improved. The diversity order and high signal-to-noise ratio slope were also confirmed. More recently, the integration of NOMA into UAV-assisted visible light communications was investigated in~\cite{pham2020sum} to maximize the total sum-rate under QoS requirements and various constraints. 

\paragraph{RIS}
The novel concept of RIS has been recently devised~\cite{gong2020towards}, where a joint combination of phase control, angles of incident RF signals, and reflecting phases can be arbitrarily adjusted to create a desirable multipath effect to improve the received signal power or mitigate interference.
As mentioned in~\cite{tekbiyik2020reconfigurable} in the high-frequency spectrum subsection, RIS has been integrated into LEO satellite communications at THz bands with good results.
Moreover, an RIS approximately the size of two large billboards could significantly improve the signal-to-noise ratio of LEO-terrestrial satellite links as demonstrated in~\cite{matthiesen2020intelligient}.
Meanwhile, Hua~\etal~\cite{hua2020uav} studied a UAV-assisted RIS radio system, where a passive beamforming technique was applied by the RIS to enhance UAV transmission.
The joint optimization of a UAV’s trajectory, RIS scheduling, and RIS phase shift matrix was also considered in this contribution.
In \cite{ge2020joint}, the joint optimization of active beamforming toward a UAV, passive beamforming toward the RIS, and the trajectory of the UAV were achieved in RIS-assisted UAV communications.
The results in this publication have demonstrated that this scheme outperforms other benchmark schemes regarding feasibility and efficacy.
Furthermore, the integrating concept among IRS, mmWave beamforming, and ML approaches was invoked in the supporting context of a UAV network \cite{zhang2019reflections}.
The preliminary results in \cite{zhang2019reflections} has revealed that the reflection coefficient of an IRS-based UAV can be optimized to achieve maximum downlink capacity, and the RL approach can further significantly improve system performance compared to a scheme without RL deployment.

\subsection{Summary and Discussion} \label{sec4d}
In brief, three key technological pillars for emerging ARANs including energy replenishment, operational management, and data delivery have been reviewed.
In particular, Section~\ref{sec4a} investigates energy replenishment, which is the foremost limiting aspect in ARAN realization to overcome battery constraints.
Three primary categories of energy replenishment in ARANs including charging stations, EH, and RF wireless charging are summarized in Table~\ref{Table:Summary_EnergyRefills}.
It is observed that the conventional energy trading techniques such as a charging station and ABS swapping are inefficient and less dynamic. Although the energy replenishment techniques of renewable energy sources and RF wireless charging are promising solutions for ARANs enabling sustainable services, joint optimization of the charging efficiency and communication objectives (e.g., throughput maximization and service availability) considering user mobility and QoS requires further investigation.
Next, Section~\ref{sec4b} presents recent advances in operational management, which can be listed as network softwarization, mobile cloudization, and data mining.
While network softwarization is beneficial for ARANs with the abilities to (\textit{i}) provide greater control, situational awareness, and flexibility, (\textit{ii}) coordinate interference avoidance, and (\textit{iii}) facilitate interoperability among network nodes, ARANs are further enabled by mobile cloudization with computational offloading in terms of energy efficiency and latency reduction. Additionally, another key enabler for achieving some optimal performance perspectives in ARANs' scopes towards 6G is presented in the data mining subsection.
Nevertheless, cooperative resource management in 3-D space is needed for efficient utilization of the limited computation, communication, storage, and battery capacity of ABSs. More specifically, important challenges such as optimal power control, placement of content on the ABSs, user device association, ABS trajectory optimization, and computation resource allocation need further investigation. 
Ultimately, data delivery perspectives including (\textit{i}) prospective employment of mmWave and THz frequency spectrums, (\textit{ii}) URLLC protocol utilization, and (\textit{iii}) advanced multiaccess based on massive MIMO, NOMA, and RIS are within the review of Section~\ref{sec4c}. As 6G is expected to exploit higher frequency bands, designing an effective and efficient mobility management scheme for these frequency bands deserves to be investigated. Furthermore, an adaptive bandwidth resource allocation scheme considering the distribution of terrestrial users’ traffic, line-of-sight interference, mobility, and positioning of UAVs' is an important avenue of future research.

\section{Application Scenarios} \label{sec5}
As aforementioned in Section~\ref{sec2}, hierarchical ARAN architectures provide heterogeneous communication capabilities from multiple tiers with distinct functionality and features. As a result, ARAN supports a broad range of emerging applications and services such as wireless coverage expansion, aerial surveillance, precision agriculture, and commercial delivery. Depending on the networking requirements, we classify applications into three categories: event-based, scheduled, and permanent communications. The details are described below.

\subsection{Event-based Communications} \label{sec5a}
Event-based communications define application scenarios in which networking infrastructures are temporarily required to provide and/or boost communication services for short duration events. As prime examples in this category, disaster and SAR scenarios are investigated.
\subsubsection{Disaster}
Emergency management systems currently depend on the wireless communication infrastructure. When large-scale disasters or catastrophes occur in an area, the communication infrastructures are often severely damaged. Despite some communication components remaining functional, the communication systems have to deal with network congestion. This leads to various problems for emergency service providers, such as difficulty in acquiring real-time information from users. To this end, the work in~\cite{bupe2015relief} demonstrated that UAVs/drones are suitable for deployment, where these networking devices are configured to form an impermanent cluster providing base station functions.
The proposed cluster including multiple UAVs forming hexagonal cells was investigated. 
Regardless of the initial deployment topology, the UAV clusters self-organize and operate as a temporary access infrastructure to provide reliable airborne communication links.
Furthermore, to enhance the ability to detect and secure emergency services in disasters, proposals on photogrammetry and geocomputing based on UAV-assisted functions are investigated in~\cite{gomez2016uav}.
In these proposals, the UAVs’ main uses are given as (\textit{i}) regular scheduling to monitor and map land features and their transformation over time to anticipate potential hazards and disasters, (\textit{ii}) observing human activities during an emergency, (\textit{iii}) altering telecommunications components when they are damaged in disasters, and (\textit{iv}) delivering essential materials to disaster-stricken isolated areas.
For instance, the authors in~\cite{ejaz2020energy} presented several platforms (e.g., UAVs/IoT-based or inter-integration strategies) for disaster management, where the information harvested by UAVs in disaster-affected areas can be analyzed to make timely and correct decisions on the assistance for people in these localities. 
A schedule for data collection that especially considers the energy-efficiency of UAVs is also considered in this contribution.
Additionally, UAV applications for assessing the damage to cultural heritage sites after earthquakes have been also considered in~\cite{baranwal2019application}.
As an example, the Sulamani Pagoda in Bagan, Myanmar was severely damaged by a strong earthquake on August 24, 2016.
This work revealed the classification of the point cloud strategies for UAV-based functions intended to distinguish damaged parts of the pagoda as well as analyze and quantify the level of damage at this architectural heritage site.

With the help of LEO satellite systems, satellite image analysis enhances disaster detection accuracy improvement capabilities to undertake rescue missions, coordinate relief efforts, and respond to disasters, etc. in a rapid, timely, and efficient manner as well as to identify the extent of the damage.
Amit~\etal~\cite{amit2016analysis} exploited an automatic disaster detection system based on DL techniques to analyze images collected from LEO satellites.
The experimental results mainly focus on two disasters (e.g., flood and landslide) in Japan and Thailand and demonstrated an accuracy of 80\%--90\% in disaster detection for both.
Meanwhile, a framework for change detection in flood and fire disasters was introduced in~\cite{doshi2018satellite}.
Based on DL algorithms using satellite images, 81.2\% and 83.5\% detection accuracy are obtained respectively for flood and fire disasters.
Moreover, the work in~\cite{deepak2019overview} provided an overview of several communication paradigms (e.g., device-to-device, IoT, UAVs, cellular, mobile ad-hoc, and satellite networks) inter-linked for post-disaster emergency communication services. 
These cooperative networks support recovery and ensure failure detection, failure recovery, and the localization of failures that occur owing to disasters.

\subsubsection{SAR}
Regarding SAR application scenarios, the authors in~\cite{hayat2017multi} proposed a multiobjective optimization algorithm for a UAV swarm aiming to find a targeted object and establish continuous communication between the target and terrestrial personnel as rapidly as possible.
It is recognized that overall mission completion times have been significantly reduced prosegmentally to the number of deployed UAVs.
Another work in~\cite{alotaibi2019lsar} similarly elaborated on emergencies where multiple UAVs are assigned for SAR missions as rapidly as possible such that the maximum number of people are saved.
The proposed algorithm’s performance has been compared to other algorithms (e.g., multi-UAV task allocation and opportunistic task allocation) in terms of mission duration and survivor rescue rate, and has demonstrated superior performance.
Furthermore, the work in~\cite{nakadai2017development} described several useful mechanisms for sound source enhancement, sound source localization, and robust communication, where these approaches were applied to a UAV conducting SAR missions with a microphone array to enhance rescue performance in outdoor conditions. 

The proposed model in~\cite{bracciale2016smartsos} was designed to utilize LEO satellite systems as follows. (\textit{i}) Life vests are equipped with a global positioning system (GPS) receiver to simplify location services and send information at low cost with high data integrity to remote locations, and with health sensors to quickly detect any health problems as well as maximize QoS and delivery probability. (\textit{ii}) These life vests can be considered an ad-hoc sensor network communicating with an LEO satellite system via a satellite master node. (\textit{iii}) LEO satellite terminals are used to provide long-range connectivity between a satellite master node and a terrestrial satellite gateway, and (\textit{iv}) distress signals are forwarded to a mission control center to assist in planning rescue missions accordingly.
The proposed system’s experimental results demonstrated its capability to improve performance and ensure the high reliability of data transmissions for maritime emergency communications.

\subsection{Scheduled Communications} \label{sec5b}
Scheduled communications cover scenarios where the ARAN components fly on a predefined path to provide users with networking services of a given duration. Two primary examples of this category, aerial surveillance and smart agriculture, are discussed in this subsection. 

\subsubsection{Aerial Surveillance}
Airborne surveillance using UAVs/drones is well known as a thoroughly complete, flexible, and unique solution for several special use cases such as ensuring territorial integrity and national security, resource exploration in dangerous areas, wildfire and oil spill detection, and environmental surveillance.
In~\cite{7369958.2016}, Rossi~\etal~proposed a modular embedded sensing system in which UAVs carried sensors used for measuring reducing gases. Furthermore, a monitoring algorithm for gas leakage localization was proposed, which considers optimizing UAV flight duration and energy consumption.
A practical model for an aerial surveillance system using drones was proposed in~\cite{7759885.2016}. The advantages of this proposed system are real-time monitoring with audio, extended operating time, and multimode operation at an affordable cost.
Inspired by this work, many reports investigated various interesting aspects under considerations including surveillance in remote areas~\cite{8234482.2017}, extended coverage with an energy-efficient routing path scheme under resolution constraints~\cite{dai2018quality}, and mobility-aware control mechanisms for cooperative drones via heterogeneous Wi-Fi and satellite networks~\cite{8700598.2019}.

Specifically, the UAV-based model proposed in~\cite{8234482.2017} can be used where dense foresting obstructs the detection of risks and hidden activities, e.g., guerilla activities. By using thermal cameras, UAVs can monitor and report minute details of these hidden activities. The proposed system is suitable for reconnaissance by army forces or for surveillance in remote areas that cannot be reached by human beings. In addition, a UAV-based network with the operating scheme proposed in~\cite{dai2018quality} can be deployed in a geometrically complex target area. It can capture images of a specified spatial resolution. The authors of this contribution also investigated the solutions to a routing problem to reduce a UAV cluster’s maximum energy consumption to attain better coverage.
Meanwhile, a control problem in a heterogeneous network was examined under three scenarios in~\cite{8700598.2019}. 
In particular, a cooperative UAVs-based surveillance system operates under the control of a server via ground base stations (access points) through Wi-Fi and satellite communications. 
Based on the function of UAVs in a cluster under multiple constraints, the control mechanism automatically adjusts the distance between the cluster and access points in terms of Multipath TCP. 
The results reveal that the designed system can operate stably and support robust bandwidth. 

\subsubsection{Smart Agriculture}
In smart agriculture, farmers use UAVs not only for crop spraying but also for many functions including crop monitoring and disease detection.
The primary feature of smart agriculture is to integrate the latest information and communication technologies (ICT) such as IoT, big data, ML, and UAVs into farming operations~\cite{garg2018uav,Bacco2018smart, Islam2019bigdata, boursianis2020internet}. 
In~\cite{maddikunta2020unmanned} many kinds of UAVs equipped with suitable sensors, and cases of UAVs being used in smart farming scenarios were thoroughly reviewed. In particular, multitier ARANs can be exploited for many potential agricultural applications, such as weed management, crop health monitoring, plant counting and numbering, pest management, and assessing plant quality. It is clear that an integrated system with the aforementioned functions would be very useful to farmers.

Furthermore, smart agriculture applications based on accurate land monitoring for timely support actions to ensure safe food production need remote sensing and satellite data for cheap and timely paddy mapping.
With this as motivation, the work in~\cite{tam2020monitoring} introduced a contribution to the development of an autonomous and intelligent agriculture system built using available LEO satellite data.
A novel multitemporal high-spatial-resolution classification method was applied to a testbed based on a case study of the Landsat 8 data.
Landsat 8’s data has been widely utilized in various publications including satellite field support for smart agriculture because of its high spatial resolution.
Particularly, in this contribution, the tested model for land monitoring attained an accuracy of up to 93\%.
Other works~\cite{murugan2016fusion,murugan2017development} were conducted based on this Landsat 8 data (i.e., satellite-based data). It further encompassed these approaches to LAP/HAP components as well.
Along with the linkage between satellite and drone/UAV data, larger scale precision monitoring in agriculture has been achieved.

\subsection{Permanent Communications} \label{sec5c}
Permanent communications involve application scenarios where networking services are required continuously over a long period.
Prime examples in this category are smart city applications such as urban monitoring, advanced healthcare services, and intelligent transportation systems (ITSs) as well as networking infrastructure for remote and isolated areas~\cite{Chowdhury20206G}.

\subsubsection{Urban Monitoring}
In contrast to UAV-based aerial surveillance applications mentioned in the scheduled communication category, which focus on control algorithms as well as detection of unusual movement, research on urban monitoring typically focuses on solutions for collecting real-time multimedia data with surveillance systems.
A design for a UAV assisted urban monitoring system under tactile Internet constraints was examined in~\cite{8460024.2018}. This work proposed tactile Internet architecture to determine system elements’ compliance with tactile Internet requirements in a 5G ecosystem.
In addition, Jin~\etal~\cite{9040423.2020} proposed four algorithms: (\textit{i}) UAV placement, (\textit{ii}) addressing the bi-objective fragile bin packing problem, (\textit{iii}) obtaining circles for a complete graph, and (\textit{iv}) refining UAV candidate routes. In this network, a UAV cluster functioning as ABSs forms a rectangular boundary in optimal positions to provide reliable, real-time, and feasible services while operating in a heterogeneous communication environment in a smart city. The authors established that their algorithms improve throughput, video quality, and delay in video surveillance systems.

\subsubsection{Health Care}
Applying advanced networking technologies for healthcare can significantly reduce healthcare expenditure for both government and citizens, in particular to the elderly who need daily assistance. Recently, electronic healthcare (e-health) services provided electronically via the Internet are being gradually formed to provide ubiquitous, continuous, and personalized medical assistance.  The authors in~\cite{ullah2019uav} proposed a smart UAV-assisted healthcare architecture to solve the limitations on the Internet of Medical Things in body area networks (BANs). Particularly, a UAV connects to multiple BANs via wake-up radio-based communication in a star topology. The UAVs function as data collectors with efficient power consumption.
In~\cite{9148581.2020}, the UAVs' positioning in a service area was examined to ensure the minimum number of serving UAVs in the network for each patient to be within their coverage area. Besides, the positioning problem, UAV services are also considered under the constraint that the collected health data must be efficiently processed. The authors proved that their particle-swarm optimization-based algorithm could significantly reduce the number of UAVs used for e-healthcare services under data-driven constraints.
Subsequently, for decentralized secure and reliable data linking between the UAVs and other entities, a scheme that applies blockchain-based outdoor medical delivery, namely VAHAK, was proposed in~\cite{9162738.2020}. Performance analysis parameters relating to scalability, bandwidth, latency, and data storage cost were investigated to prove the efficacy of using VAHAK in health care services. 
Conversely, the research work in~\cite{8355261.2018} proposed a framework for exchanging medical information when medical devices are distributed in many scattered locations. Specifically, medical data from patients  in remote/rural areas are directly transmitted via secure and robust satellite links.
Similarly, to improve global access to e-health, especially for remote rural communities, Anema~\etal~\cite{anema2020shaping} reviewed Canadian space technology applications using LEO connections for e-healthcare.

\subsubsection{ITSs}
According to development trends in ITSs, vehicle components with autonomous properties will be integrated into systems forming a large-scale ITS. 
This leads to many opportunities for new services as well as new applications to be introduced with the support of 6G wireless networks, e.g., flying taxi service. 
In the next-generation ITSs, transport automation needs to be fully implemented, not only automating the vehicles but also including the automation of remaining components such as roads and end-to-end transport systems, e.g., terminal communications, field support teams, traffic police, road surveys, and rescue teams. This can be realized by the large-scale use of intelligent and reliable ARAN components~\cite{menouar2017uav}. In some congestion prediction problems, the proposed algorithms must count the number of vehicles circulating in many places in a large region \cite{8326145.2018}.
UAVs with cameras used for traffic monitoring as well as for data collection produce better results than fixed recording detectors permanently installed at traffic intersections because the targeting mechanism of UAV is based on the true density. 

Furthermore, UAVs can act as intermediate agents to interconnect either among vehicles or to roadside units in the ITSs to improve the overall quality of experience \cite{9076661.2020}. In particular, a joint optimization of trajectory planning and resource allocation was investigated for critical data delivery in UAV-assisted vehicular networks \cite{samir2019trajectory}. Adapting to situational changes, a low-complexity trajectory design with minimum communication resources is achieved to fully serve all vehicles. By contrast, the work in \cite{wu2020optimal} considered data caching and trajectory design of UAVs to support ITSs. A convolutional neural network-based approach was utilized to maximize network throughput for reliable content delivery. In addition, a dynamic positioning technique for 3-D UAV positions was investigated in \cite{hadiwardoyo2020three} to optimize vehicular communications' QoS. Next, the minimum number of power-limited UAVs in a swarm involved in the optimal routing scheme with strict considerations of communication delay and energy consumption was addressed in densely crowded vehicular environments \cite{9110438.2020}.

Additionally, to meet the stringent requirements for ITS services with improved reliability, positioning accuracy, and continuous location updates, more complex navigation strategies must be adopted.
Global navigation satellite systems are the potential candidates that demonstrated two major aspects of ITSs and location-based applications to provide accurate global position, velocity, and time data where they are supported by LEO satellite systems.
The authors in~\cite{dovis2020recent} introduced critical problems that can be expressed as follows.
In addition to classical inertial measurement devices (e.g., accelerometers or high/low-grade gyroscopes), there are several other types of signals and sensors such as barometers, magnetometers, cameras, mobile network signals, and signals of opportunity to be considered.
Such signals and sensors that are integrated into navigation systems can be assessed and used for autonomous mobility to deploy ITS applications appropriately.

\subsubsection{Networking in Underserved Areas}
In~\cite{7451189.2016}, UAVs were assigned as intermediate aerial nodes to improve coverage as well as boost the system capacity. In this system, UAVs create multiple intermediate links to connect users in macro cells and small cells. The proposed system is evaluated according to several performance parameters: network delay, throughput coverage, and spectral efficiency. Compared to systems without UAV assistance, it can improve efficiency by up to 38\% and reduce delays by up to 37.5\%. Therefore, this model is suitable for areas with high demand for connections.
Similarly, the foundation of an airborne wireless cellular network was first introduced in~\cite{8533634.2019}. By cooperating with local base stations and LEO satellite communications, the network establishes a large coverage to provide Internet services for sparsely populated communities (in forests, deserts, at sea, etc.). 
In such circumstances, the LEO satellites help to transfer data between these localities and the Internet~\cite{7289337.2016}.

\subsection{Summary and Discussion} 
This section presents three emerging application scenarios based on networking requirements including event-based, scheduled, and permanent communications. In particular, Section~\ref{sec5a} represents existing studies primarily focused on approaches to allow clusters of UAVs to self-organize and automatically form network replacements when the current network needs to be supported in providing and/or boosting communication services for short duration events. In addition, UAVs/IoT-based or inter-integration strategies are proposed for post-disaster management. Moreover, in this model, LEO satellite systems implementing advanced deep learning techniques are taken into account in detecting and managing natural disasters. For SAR application scenarios, related works focus on optimization algorithms to control a UAV swarm in seeking targeted objects and in establishing transparent and immediate communication links between the targets and terrestrial personnel. In most event-based application scenarios, LEO satellite systems are utilized as an auxiliary support system to enhance the accuracy of the UAVs assigned to SAR missions. With scheduled communications mentioned in Section~\ref{sec5b}, the proposed works mainly focus on detecting unusual activities in cases of aerial surveillance and methods for utilizing UAVs to autonomously perform smart farming jobs for the other sub-group application. Next, Section~\ref{sec5c} provides researches relevant to smart city applications that require permanent communications to be deployed. In contrast to applications classified as the scheduled communications application group (Section~\ref{sec5b}), smart city applications (Section~\ref{sec5c}) must deal with tremendous amount of data for storage and analysis. Consequently, related works in this Section~\ref{sec5c} are primarily designed for UAV-assisted models under tactile Internet constraints, UAV position trajectory optimization approaches, algorithms to improve throughput and video quality, methods to allow UAVs to function as sufficient data collection systems to decrease operational costs. Finally, some articles in underserved areas are mentioned to prove that the multi-tier ARAN networks can improve coverage and reduce delay in data transmission, especially in sparsely populated areas.
Besides the significant advantages of exploiting ARAN in cost-efficiency 6G scenarios as discussed in \cite{9034074.2020}, optimal trade-offs between cost efficiency and technological facilitation still need further realistic studies. In addition, telecommunication subscribers may have concerns regarding their privacy (e.g., urban surveillance), personal information leakage (UAVs often collect data by design), and high-tech education (e.g., doctors or farmers need to be well-equipped with high-tech knowledge to efficiently use complicated 6G-driven applications). These concerns can be used as guideline for future investigations on 6G networks.

\section{Research Challenges} \label{sec7}
This section specifies and discusses the challenges and open research issues to spur further investigation of ARANs in 6G contexts.

\subsection{Intelligent Radio}
The emerging AI chip revolutions empower communication devices with high computing capabilities for adaptive reaction to environmental changes in real time. In this circumstance, the 6G envisions RANs growth from the current NR (of 5G) to the next generation: intelligent radio (IR)~\cite{letaief2019roadmap}. IR defines RAN capabilities of exploiting advanced AI chip powers at both user and network devices to intelligently select the most appropriate algorithms for radio frequency planning, spectrum sharing, channel modulation/coding/estimation, and multiaccess schemes. Considered as a native component of comprehensive 6G access infrastructures, ARANs have to be designed to embed a flexible federated learning model to orchestrate networking behaviors between the ABSs and multiple user devices interactively. In addition, the variety of AI chip classes in user devices is a significant challenge to building effective learning models according to different service requirements. Moreover, the locality of user traffic and behaviors such as service type, volume, and reliability as well as mobility should to be studied and exploited owing to their significant impact on channel communication between ABSs and user devices.

\subsection{Extremely High Spectrum Exploitation}
Continuing the success of 5G in exploiting high spectrum for Gbps data rates, 6G access networks will expand broader and higher spectrum utilization to achieve Tbps connections. In this regard, THz and visible light communications are the most promising candidates~\cite{huang2019survey,yang20196g}. Although literature has witnessed some breakthroughs in this extremely high spectrum exploitation for intra-tier and inter-tier communications in ARAN, there exist inevitable challenges ahead to realize effective and optimal communications. First, high spectrum communications are sensitive to attenuation because of environmental condition changes, especially within the 3-D space between ABSs and end users. Therefore, an adaptive channel propagation model and estimation should be carefully considered to react to any environmental changes in real time. In addition, super-narrow beamforming techniques with high direction degree and fast transformation should be further studied to improve transmission efficiency. Moreover, even though ARANs have advantages in high probability of LoS signal propagation, a relay solution through RIS cannot be ignorable in the future research.

\subsection{Network Stability}
In ARANs, ABSs interconnect with each other using aerial ad hoc technologies and have hierarchical networking overlays among the LAP, HAP, and LEO communication tiers. Although the high mobility of ABSs and hierarchical networking overlays assist ARANs to adapt flexibly to the requirements of airborne and terrestrial end users globally, guaranteeing network stability for such a highly dynamic and resource-constrained platform is difficult. More specifically, the highly dynamic network topology and intermittent connectivity due to the high mobility of ABSs in 3-D space, limited resources, and varying QoS requirements of applications are challenging for the network organization design of both intra-tier and inter-tier communications in ARANs. Thus, the routing protocols and network organization designs for ARANs should consider different mobility patterns in 3-D space, traffic characteristics, available resources, and load balancing among networking infrastructures as well as multiple backup routes for fault tolerance and reliable packet delivery.

\subsection{Security and Privacy Issues}
Security and privacy are the most critical issues to be resolved for ARANs. The shared wireless links, potential line-of-sight links among real platforms and aerial platforms and ground-users, limited resources (energy, computation, etc.) at some typical aerial nodes, mobility features, a massive device communications, among other factors, contribute to the challenge of meeting ARAN security requirements. Cryptographic solutions are mainly considered to combat network security problems at the higher communication protocol layers. However, conventional cryptographic techniques, for instance, public-key cryptography involving high computation and delay costs in the encryption and decryption processes may not be feasible for dynamic ARANs using resource-constrained aerial platforms. The high mobility and deployment of UAVs in open areas can also render achieving physical layer security solutions utilizing the inherent features of wireless channels more challenging. In addition, the potential line-of-sight links and high mobility features of UAVs in ARANs can be another threat to ground communication networks as they can be easily jammed or eavesdropped on by malicious UAVs once they are part of the network. Thus, designing a secure communication protocol for ARANs that considers the limited resource and mobility features of typical ABSs is crucial and worth further investigation. UAVs fitted with video cameras in the LAP/HAP tiers of ARAN are perceived to be privacy hazards owing to their ability to capture videos from unanticipated angles or areas. Hence, a privacy-preserving mechanism may be required for UAVs that capture videos. 

\subsection{Simulation Tools}
The performance of the proposed solutions for ARANs can be evaluated using either real experiments or software-based simulations. Real experiments enable the evaluation and analysis of the proposed algorithms, systems, and protocols in real environments. It is, however, very difficult to conduct experiments with aerial platforms in many cases because of the high cost, large space, strict regulations, difficulties related to scenario repetition, and the complexity of building large-scale networks with varying topologies. Consequently, simulation-based performance evaluation is considered to be a viable option by most researchers owing to its flexibility and lower cost. However, the existing simulation-based evaluations are often conducted with the assumption that aerial platforms such as UAVs or drones can move in any direction at any time with no mechanical and aerodynamic constraints due to different UAV types and/or any specific constraints occasioned by environmental obstacles. Hence, the simulation results may not accurately reflect the real environments in many specific scenarios. Thus, it would be highly beneficial to design simulation tools for aerial platforms that can fully characterize mechanical and aerodynamic constraints and capture environmental factors, allowing researchers to precisely simulate various types of UAVs based on their hardware specifications and subject to environmental factors.

\section{Conclusion} \label{sec8}
This paper has presented a thorough survey of publications on network design, system models, enabling technologies, and application of ARANs toward a comprehensive 6G access infrastructure. First, ARANs are positioned in the 6G architecture to demonstrate their roles, features, and relationships with other elements of the networks. Thereafter, an ARAN reference model is derived from recent ETSI and 3GPP standards released for 5G networks and STINs. Subsequently, we analyzed key technical aspects of the systems regarding transmission propagation, energy consumption, communication latency, and network mobility, followed by enabling technologies for improving the performance of these aspects. Finally, research challenges are detailed to offer readers the expected future trends in ARAN studies in the context of 6G networks.

\balance

\begin{thebibliography}{100}
	\providecommand{\url}[1]{#1}
	\csname url@samestyle\endcsname
	\providecommand{\newblock}{\relax}
	\providecommand{\bibinfo}[2]{#2}
	\providecommand{\BIBentrySTDinterwordspacing}{\spaceskip=0pt\relax}
	\providecommand{\BIBentryALTinterwordstretchfactor}{4}
	\providecommand{\BIBentryALTinterwordspacing}{\spaceskip=\fontdimen2\font plus
		\BIBentryALTinterwordstretchfactor\fontdimen3\font minus
		\fontdimen4\font\relax}
	\providecommand{\BIBforeignlanguage}[2]{{%
			\expandafter\ifx\csname l@#1\endcsname\relax
			\typeout{** WARNING: IEEEtran.bst: No hyphenation pattern has been}%
			\typeout{** loaded for the language `#1'. Using the pattern for}%
			\typeout{** the default language instead.}%
			\else
			\language=\csname l@#1\endcsname
			\fi
			#2}}
	\providecommand{\BIBdecl}{\relax}
	\BIBdecl
	
	\bibitem{david20186g}
	K.~David and H.~Berndt, ``{6G} vision and requirements: Is there any need for
	beyond {5G?}'' \emph{IEEE Vehicular Technology Magazine}, vol.~13, no.~3, pp.
	72--80, Jul. 2018.
	
	\bibitem{dao2020ic}
	N.-N. Dao, W.~Na, and S.~Cho, ``Mobile cloudization storytelling: Current
	issues from an optimization perspective,'' \emph{IEEE Internet Computing},
	vol.~24, no.~1, pp. 39--47, Jan. 2020.
	
	\bibitem{dao2017adaptive}
	N.-N. Dao, J.~Lee, D.-N. Vu, J.~Paek, J.~Kim, S.~Cho, K.-S. Chung, and C.~Keum,
	``Adaptive resource balancing for serviceability maximization in fog radio
	access networks,'' \emph{IEEE Access}, vol.~5, pp. 14\,548--14\,559, Aug.
	2017.
	
	\bibitem{network2030}
	{ITU-T Rec. Y.3172}, ``New services and capabilities for {Network 2030}:
	Description, technical gap and performance target analysis,'' Oct. 2019.
	
	\bibitem{bi2019ten}
	Q.~Bi, ``Ten trends in the cellular industry and an outlook on 6g,'' \emph{IEEE
		Communications Magazine}, vol.~57, no.~12, pp. 31--36, Dec. 2019.
	
	\bibitem{chen2020vision}
	S.~Chen, Y.-C. Liang, S.~Sun, S.~Kang, W.~Cheng, and M.~Peng, ``Vision,
	requirements, and technology trend of {6G}: how to tackle the challenges of
	system coverage, capacity, user data-rate and movement speed,'' \emph{IEEE
		Wireless Communications}, vol.~27, no.~2, pp. 218--228, Feb. 2020.
	
	\bibitem{samdanis2020road}
	K.~Samdanis and T.~Taleb, ``The road beyond {5G}: A vision and insight of the
	key technologies,'' \emph{IEEE Network}, vol.~34, no.~2, pp. 135--141, Feb.
	2020.
	
	\bibitem{imt2020}
	{ITU-R Report ITU-R M.2410}, ``Minimum requirements related to technical
	performance for {IMT-2020} radio interface(s),'' Nov. 2017.
	
	\bibitem{zhang2020uav}
	H.~Zhang, L.~Song, and Z.~Han, ``{UAV} assisted cellular communications,'' in
	\emph{Unmanned Aerial Vehicle Applications over Cellular Networks for 5G and
		Beyond}.\hskip 1em plus 0.5em minus 0.4em\relax Springer, 2020, pp. 61--100.
	
	\bibitem{abo2019survey}
	M.~Abo-Zeed, J.~B. Din, I.~Shayea, and M.~Ergen, ``Survey on land mobile
	satellite system: Challenges and future research trends,'' \emph{IEEE
		Access}, vol.~7, pp. 137\,291--137\,304, Sep. 2019.
	
	\bibitem{Saad2020AVision}
	W.~{Saad}, M.~{Bennis}, and M.~{Chen}, ``A vision of {6G} wireless systems:
	Applications, trends, technologies, and open research problems,'' \emph{IEEE
		Network}, vol.~34, no.~3, pp. 134--142, May/Jun. 2020.
	
	\bibitem{giordani2020toward}
	M.~Giordani, M.~Polese, M.~Mezzavilla, S.~Rangan, and M.~Zorzi, ``Toward {6G}
	networks: Use cases and technologies,'' \emph{IEEE Communications Magazine},
	vol.~58, no.~3, pp. 55--61, Mar. 2020.
	
	\bibitem{Yaacoub2020AKey6G}
	E.~{Yaacoub} and M.~{Alouini}, ``A key {6G} challenge and
	opportunity—connecting the base of the pyramid: A survey on rural
	connectivity,'' \emph{Proceedings of the IEEE}, vol. 108, no.~4, pp.
	533--582, Apr. 2020.
	
	\bibitem{Shafin2020Artificial}
	R.~{Shafin}, L.~{Liu}, V.~{Chandrasekhar}, H.~{Chen}, J.~{Reed}, and J.~C.
	{Zhang}, ``Artificial intelligence-enabled cellular networks: A critical path
	to beyond-{5G} and {6G},'' \emph{IEEE Wireless Communications}, vol.~27,
	no.~2, pp. 212--217, Apr. 2020.
	
	\bibitem{Song2020Artificial}
	H.~{Song}, J.~{Bai}, Y.~{Yi}, J.~{Wu}, and L.~{Liu}, ``Artificial intelligence
	enabled {Internet} of things: Network architecture and spectrum access,''
	\emph{IEEE Computational Intelligence Magazine}, vol.~15, no.~1, pp. 44--51,
	Feb. 2020.
	
	\bibitem{zhang20196g}
	Z.~Zhang, Y.~Xiao, Z.~Ma, M.~Xiao, Z.~Ding, X.~Lei, G.~K. Karagiannidis, and
	P.~Fan, ``{6G} wireless networks: Vision, requirements, architecture, and key
	technologies,'' \emph{IEEE Vehicular Technology Magazine}, vol.~14, no.~3,
	pp. 28--41, Jul. 2019.
	
	\bibitem{ali20206g}
	S.~Ali, W.~Saad, N.~Rajatheva, K.~Chang, D.~Steinbach, B.~Sliwa, C.~Wietfeld,
	K.~Mei, H.~Shiri, H.-J. Zepernick \emph{et~al.}, ``{6G} white paper on
	machine learning in wireless communication networks,'' \emph{arXiv preprint
		arXiv:2004.13875}, 2020.
	
	\bibitem{peltonen20206g}
	E.~Peltonen, M.~Bennis, M.~Capobianco, M.~Debbah, A.~Ding,
	F.~Gil-Casti{\~n}eira, M.~Jurmu, T.~Karvonen, M.~Kelanti, A.~Kliks
	\emph{et~al.}, ``{6G} white paper on edge intelligence,'' \emph{arXiv
		preprint arXiv:2004.14850}, 2020.
	
	\bibitem{rajatheva2020white}
	N.~Rajatheva, I.~Atzeni, E.~Bjornson, A.~Bourdoux, S.~Buzzi, J.-B. Dore,
	S.~Erkucuk, M.~Fuentes, K.~Guan, Y.~Hu \emph{et~al.}, ``White paper on
	broadband connectivity in {6G},'' \emph{arXiv preprint arXiv:2004.14247},
	2020.
	
	\bibitem{mahmood2020white}
	N.~H. Mahmood, S.~B{\"o}cker, A.~Munari, F.~Clazzer, I.~Moerman, K.~Mikhaylov,
	O.~Lopez, O.-S. Park, E.~Mercier, H.~Bartz \emph{et~al.}, ``White paper on
	critical and massive machine type communication towards {6G},'' \emph{arXiv
		preprint arXiv:2004.14146}, 2020.
	
	\bibitem{you2021towards}
	X.~You, C.-X. Wang, J.~Huang, X.~Gao, Z.~Zhang, M.~Wang, Y.~Huang, C.~Zhang,
	Y.~Jiang, J.~Wang \emph{et~al.}, ``Towards {6G} wireless communication
	networks: Vision, enabling technologies, and new paradigm shifts,''
	\emph{Science China Information Sciences}, vol.~64, no.~1, pp. 1--74, Jan.
	2021.
	
	\bibitem{huang2019survey}
	T.~Huang, W.~Yang, J.~Wu, J.~Ma, X.~Zhang, and D.~Zhang, ``A survey on green
	{6G} network: Architecture and technologies,'' \emph{IEEE Access}, vol.~7,
	pp. 175\,758--175\,768, Dec. 2019.
	
	\bibitem{olwal2016survey}
	T.~O. Olwal, K.~Djouani, and A.~M. Kurien, ``A survey of resource management
	toward {5G} radio access networks,'' \emph{IEEE Communications Surveys \&
		Tutorials}, vol.~18, no.~3, pp. 1656--1686, Thirdquarter 2016.
	
	\bibitem{mohamed2015control}
	A.~Mohamed, O.~Onireti, M.~A. Imran, A.~Imran, and R.~Tafazolli, ``Control-data
	separation architecture for cellular radio access networks: A survey and
	outlook,'' \emph{IEEE Communications Surveys \& Tutorials}, vol.~18, no.~1,
	pp. 446--465, Firstquarter 2015.
	
	\bibitem{Peng2016Recent}
	M.~{Peng}, Y.~{Sun}, X.~{Li}, Z.~{Mao}, and C.~{Wang}, ``Recent advances in
	cloud radio access networks: System architectures, key techniques, and open
	issues,'' \emph{IEEE Communications Surveys \& Tutorials}, vol.~18, no.~3,
	pp. 2282--2308, Thirdquarter 2016.
	
	\bibitem{Alimi2018Toward}
	I.~A. {Alimi}, A.~L. {Teixeira}, and P.~P. {Monteiro}, ``Toward an efficient
	{C-RAN} optical fronthaul for the future networks: A tutorial on
	technologies, requirements, challenges, and solutions,'' \emph{IEEE
		Communications Surveys \& Tutorials}, vol.~20, no.~1, pp. 708--769,
	Firstquarter 2018.
	
	\bibitem{habibi2019comprehensive}
	M.~A. Habibi, M.~Nasimi, B.~Han, and H.~D. Schotten, ``A comprehensive survey
	of {RAN} architectures toward {5G} mobile communication system,'' \emph{IEEE
		Access}, vol.~7, pp. 70\,371--70\,421, May 2019.
	
	\bibitem{pham2020survey}
	Q.-V. Pham, F.~Fang, V.~N. Ha, M.~J. Piran, M.~Le, L.~B. Le, W.-J. Hwang, and
	Z.~Ding, ``A survey of multi-access edge computing in {5G} and beyond:
	Fundamentals, technology integration, and state-of-the-art,'' \emph{IEEE
		Access}, vol.~8, pp. 116\,974--117\,017, 2020.
	
	\bibitem{peng2016fog}
	M.~Peng, S.~Yan, K.~Zhang, and C.~Wang, ``Fog-computing-based radio access
	networks: Issues and challenges,'' \emph{IEEE Network}, vol.~30, no.~4, pp.
	46--53, Jul.-Aug. 2016.
	
	\bibitem{reznik2018cloud}
	A.~Reznik, L.~M.~C. Murillo, Y.~Fang, W.~Featherstone, M.~Filippou, F.~Fontes,
	F.~Giust, Q.~Huang, A.~Li, C.~Turyagyenda \emph{et~al.}, ``Cloud {RAN} and
	{MEC}: A perfect pairing,'' \emph{ETSI MEC}, no.~23, p.~25, Feb. 2018.
	
	\bibitem{vu2015energy}
	Q.-D. Vu, L.-N. Tran, M.~Juntti, and E.-K. Hong, ``Energy-efficient bandwidth
	and power allocation for multi-homing networks,'' \emph{IEEE Transactions on
		Signal Processing}, vol.~63, no.~7, pp. 1684--1699, Apr. 2015.
	
	\bibitem{wang2018millimeter}
	X.~Wang, L.~Kong, F.~Kong, F.~Qiu, M.~Xia, S.~Arnon, and G.~Chen, ``Millimeter
	wave communication: A comprehensive survey,'' \emph{IEEE Communications
		Surveys \& Tutorials}, vol.~20, no.~3, pp. 1616--1653, Thirdquarter 2018.
	
	\bibitem{han2019terahertz}
	C.~Han, Y.~Wu, Z.~Chen, and X.~Wang, ``Terahertz communications ({TeraCom}):
	Challenges and impact on {6G} wireless systems,'' \emph{arXiv preprint
		arXiv:1912.06040}, 2019.
	
	\bibitem{Lagen2020NewRadio}
	S.~{Lagen}, L.~{Giupponi}, S.~{Goyal}, N.~{Patriciello}, B.~{Bojović},
	A.~{Demir}, and M.~{Beluri}, ``New radio beam-based access to unlicensed
	spectrum: Design challenges and solutions,'' \emph{IEEE Communications
		Surveys \& Tutorials}, vol.~22, no.~1, pp. 8--37, Firstquarter 2020.
	
	\bibitem{ling2019blockchain}
	X.~Ling, J.~Wang, T.~Bouchoucha, B.~C. Levy, and Z.~Ding, ``Blockchain radio
	access network ({B-RAN}): Towards decentralized secure radio access
	paradigm,'' \emph{IEEE Access}, vol.~7, pp. 9714--9723, 2019.
	
	\bibitem{yao2019artificial}
	M.~Yao, M.~Sohul, V.~Marojevic, and J.~H. Reed, ``Artificial intelligence
	defined {5G} radio access networks,'' \emph{IEEE Communications Magazine},
	vol.~57, no.~3, pp. 14--20, Mar. 2019.
	
	\bibitem{kodheli2020satellite}
	O.~Kodheli, E.~Lagunas, N.~Maturo, S.~K. Sharma, B.~Shankar, J.~F.~M. Montoya,
	J.~C.~M. Duncan, D.~Spano, S.~Chatzinotas, S.~Kisseleff \emph{et~al.},
	``Satellite communications in the new space era: A survey and future
	challenges,'' \emph{IEEE Communications Surveys \& Tutorials}, 2021.
	
	\bibitem{saeed2020cubesat}
	N.~Saeed, A.~Elzanaty, H.~Almorad, H.~Dahrouj, T.~Y. Al-Naffouri, and M.-S.
	Alouini, ``{CubeSat} communications: Recent advances and future challenges,''
	\emph{IEEE Communications Surveys \& Tutorials}, 2020, in press.
	
	\bibitem{wang2019convergence}
	P.~Wang, J.~Zhang, X.~Zhang, Z.~Yan, B.~G. Evans, and W.~Wang, ``Convergence of
	satellite and terrestrial networks: A comprehensive survey,'' \emph{IEEE
		Access}, vol.~8, pp. 5550--5588, 2019.
	
	\bibitem{fotouhi2019survey}
	A.~Fotouhi, H.~Qiang, M.~Ding, M.~Hassan, L.~G. Giordano, A.~Garcia-Rodriguez,
	and J.~Yuan, ``Survey on {UAV} cellular communications: Practical aspects,
	standardization advancements, regulation, and security challenges,''
	\emph{IEEE Communications Surveys \& Tutorials}, vol.~21, no.~4, pp.
	3417--3442, Fourthquarter 2019.
	
	\bibitem{cao2018airborne}
	X.~Cao, P.~Yang, M.~Alzenad, X.~Xi, D.~Wu, and H.~Yanikomeroglu, ``Airborne
	communication networks: A survey,'' \emph{IEEE Journal on Selected Areas in
		Communications}, vol.~36, no.~9, pp. 1907--1926, Sep. 2018.
	
	\bibitem{he2017drone}
	D.~He, S.~Chan, and M.~Guizani, ``Drone-assisted public safety networks: The
	security aspect,'' \emph{IEEE Communications Magazine}, vol.~55, no.~8, pp.
	218--223, Apr. 2017.
	
	\bibitem{shi2018drone}
	W.~Shi, H.~Zhou, J.~Li, W.~Xu, N.~Zhang, and X.~Shen, ``Drone assisted
	vehicular networks: Architecture, challenges and opportunities,'' \emph{IEEE
		Network}, vol.~32, no.~3, pp. 130--137, Jan. 2018.
	
	\bibitem{zhao2019uav}
	N.~Zhao, W.~Lu, M.~Sheng, Y.~Chen, J.~Tang, F.~R. Yu, and K.-K. Wong,
	``{UAV}-assisted emergency networks in disasters,'' \emph{IEEE Wireless
		Communications}, vol.~26, no.~1, pp. 45--51, Feb. 2019.
	
	\bibitem{shah2017association}
	S.~A. Shah, T.~Khattab, M.~Z. Shakir, and M.~O. Hasna, ``Association of
	networked flying platforms with small cells for network centric {5G+
		C-RAN},'' in \emph{IEEE 28th Annual International Symposium on Personal,
		Indoor, and Mobile Radio Communications (PIMRC)}, Montreal, Quebec, Canada,
	2017, pp. 1--7.
	
	\bibitem{huang2019airplane}
	X.~Huang, J.~A. Zhang, R.~P. Liu, Y.~J. Guo, and L.~Hanzo, ``Airplane-aided
	integrated networking for {6G} wireless: Will it work?'' \emph{IEEE Vehicular
		Technology Magazine}, vol.~14, no.~3, pp. 84--91, Jul. 2019.
	
	\bibitem{wang2019deployment}
	H.~Wang, H.~Zhao, W.~Wu, J.~Xiong, D.~Ma, and J.~Wei, ``Deployment algorithms
	of flying base stations: {5G} and beyond with {UAVs},'' \emph{IEEE Internet
		of Things Journal}, vol.~6, no.~6, pp. 10\,009--10\,027, Aug. 2019.
	
	\bibitem{lee2020path}
	J.~Lee and V.~Friderikos, ``Path optimization for flying base stations in
	multi-cell networks,'' in \emph{2020 IEEE Wireless Communications and
		Networking Conference (WCNC)}, Seoul, South Korea, 2020, pp. 1--6.
	
	\bibitem{chriki2019fanet}
	A.~Chriki, H.~Touati, H.~Snoussi, and F.~Kamoun, ``{FANET}: Communication,
	mobility models and security issues,'' \emph{Computer Networks}, vol. 163, p.
	106877, Nov. 2019.
	
	\bibitem{lakew2020routing}
	D.~S. Lakew, U.~Sa’ad, N.-N. Dao, W.~Na, and S.~Cho, ``Routing in flying ad
	hoc networks: A comprehensive survey,'' \emph{IEEE Communications Surveys \&
		Tutorials}, vol.~22, no.~2, pp. 1071--1120, Mar. 2020.
	
	\bibitem{lin2019joint}
	Z.~Lin, M.~Lin, J.-B. Wang, T.~de~Cola, and J.~Wang, ``Joint beamforming and
	power allocation for satellite-terrestrial integrated networks with
	non-orthogonal multiple access,'' \emph{IEEE Journal of Selected Topics in
		Signal Processing}, vol.~13, no.~3, pp. 657--670, Feb. 2019.
	
	\bibitem{guidotti2019integrated}
	A.~Guidotti, B.~Evans, and M.~Di~Renzo, ``Integrated satellite-terrestrial
	networks in future wireless systems,'' \emph{International Journal of
		Satellite Communications and Networking}, vol.~37, no.~2, pp. 73--75, Mar.
	2019.
	
	\bibitem{zhu2019cooperative}
	X.~Zhu, C.~Jiang, L.~Kuang, N.~Ge, S.~Guo, and J.~Lu, ``Cooperative
	transmission in integrated terrestrial-satellite networks,'' \emph{IEEE
		network}, vol.~33, no.~3, pp. 204--210, Jan. 2019.
	
	\bibitem{li2020energy}
	J.~Li, K.~Xue, D.~S. Wei, J.~Liu, and Y.~Zhang, ``Energy efficiency and traffic
	offloading optimization in integrated satellite/terrestrial radio access
	networks,'' \emph{IEEE Transactions on Wireless Communications}, vol.~19,
	no.~4, pp. 2367--2381, Jan. 2020.
	
	\bibitem{song2020icc}
	L.~Song, Z.~Han, and B.~Di, ``Aerial access networks for {6G}: From {UAV, HAP},
	to satellite communication networks,'' in \emph{IEEE International Conference
		on Communications (ICC)}, Virtual Program, 2020, pp. 1--1.
	
	\bibitem{dao2021aran}
	N.-N. Dao, Q.-V. Pham, D.-T. Do, and S.~Dustdar, ``The sky is the
	edge--{T}oward mobile coverage from the sky,'' \emph{IEEE Internet
		Computing}, 2021.
	
	\bibitem{uavclassification}
	{United States. Department of the Army}, ``'{E}yes of the army': {U.S.} army
	roadmap for unmanned systems, 2010--2035,'' Apr. 2010.
	
	\bibitem{kishk2020}
	M.~A. Kishk, A.~Bader, and M.-S. Alouini, ``On the {3-D} placement of airborne
	base stations using tethered {UAVs},'' \emph{IEEE Transactions on
		Communications}, vol.~68, no.~8, pp. 5202--5215, Aug. 2020.
	
	\bibitem{kishk2020aerial}
	M.~Kishk, A.~Bader, and M.-S. Alouini, ``Aerial base station deployment in {6G}
	cellular networks using tethered drones: The mobility and endurance
	tradeoff,'' \emph{IEEE Vehicular Technology Magazine}, vol.~15, no.~4, pp.
	103--111, Dec. 2020.
	
	\bibitem{alzidaneen2019resource}
	A.~Alzidaneen, A.~Alsharoa, and M.-S. Alouini, ``Resource and placement
	optimization for multiple {UAVs} using backhaul tethered balloons,''
	\emph{IEEE Wireless Communications Letters}, vol.~9, no.~4, pp. 543--547,
	Apr. 2020.
	
	\bibitem{pelton2020high}
	J.~Pelton, ``High altitude platform systems ({HAPS}) and unmanned aerial
	vehicles ({UAV}) as an alternative to small satellites,'' \emph{Handbook of
		Small Satellites: Technology, Design, Manufacture, Applications, Economics
		and Regulation}, pp. 1--16, 2020.
	
	\bibitem{arum2020review}
	S.~C. Arum, D.~Grace, and P.~D. Mitchell, ``A review of wireless communication
	using high-altitude platforms for extended coverage and capacity,''
	\emph{Computer Communications}, vol. 157, no.~1, pp. 232--256, May 2020.
	
	\bibitem{su2019broadband}
	Y.~Su, Y.~Liu, Y.~Zhou, J.~Yuan, H.~Cao, and J.~Shi, ``Broadband {LEO}
	satellite communications: Architectures and key technologies,'' \emph{IEEE
		Wireless Communications}, vol.~26, no.~2, pp. 55--61, Apr. 2019.
	
	\bibitem{davoli2019small}
	F.~Davoli, C.~Kourogiorgas, M.~Marchese, A.~Panagopoulos, and F.~Patrone,
	``Small satellites and {CubeSats}: Survey of structures, architectures, and
	protocols,'' \emph{International Journal of Satellite Communications and
		Networking}, vol.~37, no.~4, pp. 343--359, Jul. 2019.
	
	\bibitem{3gpp2020}
	{3GPP TS 23.501 V16.5.0 Release 16}, ``{5G; System} architecture for the {5G}
	system ({5GS}),'' Jul. 2020.
	
	\bibitem{etsi2020}
	{ETSI TR 103 611 V1.1.1}, ``Satellite {E}arth stations and systems ({SES});
	{S}eamless integration of satellite and/or {HAPS} (high altitude platform
	station) systems into {5G} and related architecture options,'' Jun. 2020.
	
	\bibitem{rinaldi2020non}
	F.~Rinaldi, H.-L. Maattanen, J.~Torsner, S.~Pizzi, S.~Andreev, A.~Iera,
	Y.~Koucheryavy, and G.~Araniti, ``Non-terrestrial networks in 5g \& beyond: A
	survey,'' \emph{IEEE Access}, vol.~8, pp. 165\,178--165\,200, Sep. 2020.
	
	\bibitem{3gpp2020nr}
	{3GPP TS 38.300 V16.2.0 Release 16}, ``{5G; NR; NR and NG-RAN} overall
	description; stage-2,'' Jul. 2020.
	
	\bibitem{khawaja2019survey}
	W.~Khawaja, I.~Guvenc, D.~W. Matolak, U.-C. Fiebig, and N.~Schneckenburger, ``A
	survey of air-to-ground propagation channel modeling for unmanned aerial
	vehicles,'' \emph{IEEE Communications Surveys \& Tutorials}, vol.~21, no.~3,
	pp. 2361--2391, Thirdquarter 2019.
	
	\bibitem{panagopoulos2004satellite}
	A.~D. Panagopoulos, P.-D.~M. Arapoglou, and P.~G. Cottis, ``Satellite
	communications at {Ku}, {Ka}, and {V} bands: Propagation impairments and
	mitigation techniques,'' \emph{IEEE Communications Surveys \& Tutorials},
	vol.~6, no.~3, pp. 2--14, Thirdquarter 2004.
	
	\bibitem{dovis2002small}
	F.~Dovis, R.~Fantini, M.~Mondin, and P.~Savi, ``Small-scale fading for
	high-altitude platform ({HAP}) propagation channels,'' \emph{IEEE Journal on
		Selected areas in Communications}, vol.~20, no.~3, pp. 641--647, Apr. 2002.
	
	\bibitem{karapantazis2005broadband}
	S.~Karapantazis and F.~Pavlidou, ``Broadband communications via high-altitude
	platforms: a survey,'' \emph{IEEE Communications Surveys \& Tutorials},
	vol.~7, no.~1, pp. 2--31, Firstquarter 2005.
	
	\bibitem{feng2006path}
	Q.~Feng, J.~McGeehan, E.~K. Tameh, and A.~R. Nix, ``Path loss models for
	air-to-ground radio channels in urban environments,'' in \emph{2006 IEEE 63rd
		Vehicular Technology Conference}, vol.~6, Melbourne, Vic., Australia, 2006,
	pp. 2901--2905.
	
	\bibitem{al2014modeling}
	A.~Al-Hourani, S.~Kandeepan, and A.~Jamalipour, ``Modeling air-to-ground path
	loss for low altitude platforms in urban environments,'' in \emph{2014 IEEE
		Global Communications Conference (GLOBECOM)}, Austin, TX, USA, 2014, pp.
	2898--2904.
	
	\bibitem{bor2017environment}
	I.~Bor-Yaliniz, S.~S. Szyszkowicz, and H.~Yanikomeroglu, ``Environment-aware
	drone-base-station placements in modern metropolitans,'' \emph{IEEE Wireless
		Communications Letters}, vol.~7, no.~3, pp. 372--375, Jun. 2017.
	
	\bibitem{al2018modeling}
	A.~Al-Hourani and K.~Gomez, ``Modeling cellular-to-{UAV} path-loss for suburban
	environments,'' \emph{IEEE Wireless Communications Letters}, vol.~7, no.~1,
	pp. 82--85, 2018.
	
	\bibitem{cui2019low}
	Z.~Cui, C.~Briso, K.~Guan, D.~W. Matolak, C.~Calvo-Ram{\'\i}rez, B.~Ai, and
	Z.~Zhong, ``Low-altitude {UAV} air-ground propagation channel measurement and
	analysis in a suburban environment at 3.9 {GHz},'' \emph{IET Microwaves,
		Antennas \& Propagation}, vol.~13, no.~9, pp. 1503--1508, Apr. 2019.
	
	\bibitem{matolak2017airI}
	D.~W. Matolak and R.~Sun, ``Air--ground channel characterization for unmanned
	aircraft systems—part {I}: Methods, measurements, and models for over-water
	settings,'' \emph{IEEE Transactions on Vehicular Technology}, vol.~66, no.~1,
	pp. 26--44, Jan. 2017.
	
	\bibitem{sun2017airII}
	R.~Sun and D.~W. Matolak, ``Air--ground channel characterization for unmanned
	aircraft systems part {II}: Hilly and mountainous settings,'' \emph{IEEE
		Transactions on Vehicular Technology}, vol.~66, no.~3, pp. 1913--1925, Mar.
	2016.
	
	\bibitem{matolak2017airIII}
	D.~W. Matolak and R.~Sun, ``Air--ground channel characterization for unmanned
	aircraft systems—part {III}: The suburban and near-urban environments,''
	\emph{IEEE Transactions on Vehicular Technology}, vol.~66, no.~8, pp.
	6607--6618, Aug. 2017.
	
	\bibitem{sun2017airIV}
	R.~Sun, D.~W. Matolak, and W.~Rayess, ``Air-ground channel characterization for
	unmanned aircraft systems—part {IV}: Airframe shadowing,'' \emph{IEEE
		Transactions on Vehicular Technology}, vol.~66, no.~9, pp. 7643--7652, Sep.
	2017.
	
	\bibitem{rieth2019aircraft}
	D.~Rieth, C.~Heller, and G.~Ascheid, ``Aircraft to ground-station {C}-band
	channel—small airport scenario,'' \emph{IEEE Transactions on Vehicular
		Technology}, vol.~68, no.~5, pp. 4306--4315, May 2019.
	
	\bibitem{gutierrez2019comparison}
	R.~M. Gutierrez, H.~Yu, Y.~Rong, and D.~W. Bliss, ``Comparison of
	{UAS}-to-ground small-scale fading in residential and mountainous desert
	terrains,'' \emph{IEEE Transactions on Vehicular Technology}, vol.~68,
	no.~10, pp. 9348--9358, Oct. 2019.
	
	\bibitem{cheng2019A3D}
	X.~Cheng and Y.~Li, ``A {3-D} geometry-based stochastic model for {UAV-MIMO}
	wideband nonstationary channels,'' \emph{IEEE Internet of Things Journal},
	vol.~6, no.~2, pp. 1654--1662, Apr. 2019.
	
	\bibitem{jiang2020novel}
	H.~Jiang, Z.~Zhang, C.-X. Wang, J.~Zhang, J.~Dang, L.~Wu, and H.~Zhang, ``A
	novel {3D UAV} channel model for {A2G} communication environments using {AoD}
	and {AoA} estimation algorithms,'' \emph{IEEE Transactions on
		Communications}, 2020, in press.
	
	\bibitem{ma2020wideband}
	Z.~Ma, B.~Ai, R.~He, G.~Wang, Y.~Niu, and Z.~Zhong, ``A wideband non-stationary
	air-to-air channel model for {UAV} communications,'' \emph{IEEE Transactions
		on Vehicular Technology}, vol.~69, no.~2, pp. 1214--1226, Feb. 2020.
	
	\bibitem{stolaroff2018energy}
	J.~K. Stolaroff, C.~Samaras, E.~R. O’Neill, A.~Lubers, A.~S. Mitchell, and
	D.~Ceperley, ``Energy use and life cycle greenhouse gas emissions of drones
	for commercial package delivery,'' \emph{Nature communications}, vol.~9,
	no.~1, pp. 1--13, Feb. 2018.
	
	\bibitem{tseng2017autonomous}
	C.-M. Tseng, C.-K. Chau, K.~Elbassioni, and M.~Khonji, ``Autonomous recharging
	and flight mission planning for battery-operated autonomous drones,''
	\emph{arXiv preprint arXiv:1703.10049}, 2017.
	
	\bibitem{thibbotuwawa2018energy}
	A.~Thibbotuwawa, P.~Nielsen, B.~Zbigniew, and G.~Bocewicz, ``Energy consumption
	in unmanned aerial vehicles: a review of energy consumption models and their
	relation to the {UAV} routing,'' in \emph{International Conference on
		Information Systems Architecture and Technology}, Nysa, Poland, 2018, pp.
	173--184.
	
	\bibitem{dorling2017vehicle}
	K.~Dorling, J.~Heinrichs, G.~G. Messier, and S.~Magierowski, ``Vehicle routing
	problems for drone delivery,'' \emph{IEEE Transactions on Systems, Man, and
		Cybernetics: Systems}, vol.~47, no.~1, pp. 70--85, Jan. 2017.
	
	\bibitem{leishman2006principles}
	G.~J. Leishman, \emph{Principles of helicopter aerodynamics}.\hskip 1em plus
	0.5em minus 0.4em\relax Cambridge university press, 2006.
	
	\bibitem{Marins2018AClosed}
	J.~L. {Marins}, T.~M. {Cabreira}, K.~S. {Kappel}, and P.~R. {Ferreira}, ``A
	closed-form energy model for multi-rotors based on the dynamic of the
	movement,'' in \emph{VIII Brazilian Symposium on Computing Systems
		Engineering (SBESC)}, Salvador, Brazil, 2018, pp. 256--261.
	
	\bibitem{Michel2019Multiphysical}
	N.~{Michel}, A.~K. {Sinha}, Z.~{Kong}, and X.~{Lin}, ``Multiphysical modeling
	of energy dynamics for multirotor unmanned aerial vehicles,'' in
	\emph{International Conference on Unmanned Aircraft Systems (ICUAS)},
	Atlanta, GA, USA, 2019, pp. 738--747.
	
	\bibitem{austin2011unmanned}
	R.~Austin, \emph{Unmanned aircraft systems: {UAVs} design, development and
		deployment}.\hskip 1em plus 0.5em minus 0.4em\relax John Wiley \& Sons, 2011,
	vol.~54.
	
	\bibitem{zeng2017energy}
	Y.~Zeng and R.~Zhang, ``Energy-efficient {UAV} communication with trajectory
	optimization,'' \emph{IEEE Transactions on Wireless Communications}, vol.~16,
	no.~6, pp. 3747--3760, Jun. 2017.
	
	\bibitem{zeng2019energy}
	Y.~Zeng, J.~Xu, and R.~Zhang, ``Energy minimization for wireless communication
	with rotary-wing {UAV},'' \emph{IEEE Transactions on Wireless
		Communications}, vol.~18, no.~4, pp. 2329--2345, Apr. 2019.
	
	\bibitem{horani2018latency}
	M.~Horani and M.~O. Hasna, ``Latency analysis of {UAV} based communication
	networks,'' in \emph{International Conference on Information and
		Communication Technology Convergence (ICTC)}, Jeju, South Korea, 2018, pp.
	385--390.
	
	\bibitem{liao2020end}
	K.-M. Liao, G.-Y. Chen, Y.-J. Chen, and Y.-F. Chen, ``End-to-end delay analysis
	in {mmWave} {UAV}-assisted wireless caching networks,'' in \emph{2020 IEEE
		Wireless Communications and Networking Conference Workshops (WCNCW)}, Seoul,
	Korea, 2020, pp. 1--6.
	
	\bibitem{wu2020latency}
	X.~Wu, Q.~Li, Y.~Lu, H.~V. Poor, V.~C. Leung, and P.~Ching, ``Latency-minimized
	design of secure transmissions in {UAV}-aided communications,'' in \emph{IEEE
		International Conference on Acoustics, Speech and Signal Processing
		(ICASSP)}, Barcelona, Spain, 2020, pp. 8753--8757.
	
	\bibitem{she2019ultra}
	C.~She, C.~Liu, T.~Q. Quek, C.~Yang, and Y.~Li, ``Ultra-reliable and
	low-latency communications in unmanned aerial vehicle communication
	systems,'' \emph{IEEE Transactions on Communications}, vol.~67, no.~5, pp.
	3768--3781, May 2019.
	
	\bibitem{ren2019achievable}
	H.~Ren, C.~Pan, K.~Wang, Y.~Deng, M.~Elkashlan, and A.~Nallanathan,
	``Achievable data rate for {URLLC}-enabled {UAV} systems with {3-D} channel
	model,'' \emph{IEEE Wireless Communications Letters}, vol.~8, no.~6, pp.
	1587--1590, Dec. 2019.
	
	\bibitem{ranjha2020quasi}
	A.~Ranjha and G.~Kaddoum, ``Quasi-optimization of distance and blocklength in
	{URLLC} aided multi-hop {UAV} relay links,'' \emph{IEEE Wireless
		Communications Letters}, vol.~9, no.~3, pp. 306--310, Mar. 2020.
	
	\bibitem{bellavista2019social}
	P.~Bellavista, D.~Belli, S.~Chessa, and L.~Foschini, ``A social-driven edge
	computing architecture for mobile crowd sensing management,'' \emph{IEEE
		Communications Magazine}, vol.~57, no.~4, pp. 68--73, Apr. 2019.
	
	\bibitem{bellavista2018human}
	P.~Bellavista, S.~Chessa, L.~Foschini, L.~Gioia, and M.~Girolami,
	``Human-enabled edge computing: Exploiting the crowd as a dynamic extension
	of mobile edge computing,'' \emph{IEEE Communications Magazine}, vol.~56,
	no.~1, pp. 145--155, Jan. 2018.
	
	\bibitem{pham2019mobile}
	Q.-V. Pham, L.~B. Le, S.-H. Chung, and W.-J. Hwang, ``Mobile edge computing
	with wireless backhaul: Joint task offloading and resource allocation,''
	\emph{IEEE Access}, vol.~7, pp. 16\,444--16\,459, 2019.
	
	\bibitem{pham2020coalitional}
	Q.-V. Pham, H.~T. Nguyen, Z.~Han, and W.-J. Hwang, ``Coalitional games for
	computation offloading in {NOMA}-enabled multi-access edge computing,''
	\emph{IEEE Transactions on Vehicular Technology}, vol.~69, no.~2, pp.
	1982--1993, Feb. 2020.
	
	\bibitem{hu2019uav}
	X.~Hu, K.-K. Wong, K.~Yang, and Z.~Zheng, ``{UAV}-assisted relaying and edge
	computing: Scheduling and trajectory optimization,'' \emph{IEEE Transactions
		on Wireless Communications}, vol.~18, no.~10, pp. 4738--4752, Oct. 2019.
	
	\bibitem{han2020rate}
	R.~Han, Y.~Wen, L.~Bai, J.~Liu, and J.~Choi, ``Rate splitting on mobile edge
	computing for {UAV}-aided {IoT} systems,'' \emph{IEEE Transactions on
		Cognitive Communications and Networking}, 2020, in press.
	
	\bibitem{zhang2020latency}
	L.~Zhang and N.~Ansari, ``Latency-aware {IoT} service provisioning in
	{UAV}-aided mobile edge computing networks,'' \emph{IEEE Internet of Things
		Journal}, 2020, in press.
	
	\bibitem{zhang2019satellite}
	Z.~Zhang, W.~Zhang, and F.-H. Tseng, ``Satellite mobile edge computing:
	Improving {QoS} of high-speed satellite-terrestrial networks using edge
	computing techniques,'' \emph{IEEE network}, vol.~33, no.~1, pp. 70--76,
	Jan./Feb. 2019.
	
	\bibitem{mozaffari2017mobile}
	M.~Mozaffari, W.~Saad, M.~Bennis, and M.~Debbah, ``Mobile unmanned aerial
	vehicles ({UAVs}) for energy-efficient internet of things communications,''
	\emph{IEEE Transactions on Wireless Communications}, vol.~16, no.~11, pp.
	7574--7589, Nov. 2017.
	
	\bibitem{chetlur2017downlink}
	V.~V. Chetlur and H.~S. Dhillon, ``Downlink coverage analysis for a finite
	{3-D} wireless network of unmanned aerial vehicles,'' \emph{IEEE Transactions
		on Communications}, vol.~65, no.~10, pp. 4543--4558, Oct. 2017.
	
	\bibitem{lyu2017placement}
	J.~Lyu, Y.~Zeng, R.~Zhang, and T.~J. Lim, ``Placement optimization of
	{UAV}-mounted mobile base stations,'' \emph{IEEE Communications Letters},
	vol.~21, no.~3, pp. 604--607, Mar. 2017.
	
	\bibitem{chen2020efficient}
	J.~Chen and D.~Gesbert, ``Efficient local map search algorithms for the
	placement of flying relays,'' \emph{IEEE Transactions on Wireless
		Communications}, vol.~19, no.~2, pp. 1305--1319, Feb. 2020.
	
	\bibitem{sharma2019random}
	P.~K. Sharma and D.~I. Kim, ``Random {3D} mobile {UAV} networks: Mobility
	modeling and coverage probability,'' \emph{IEEE Transactions on Wireless
		Communications}, vol.~18, no.~5, pp. 2527--2538, May 2019.
	
	\bibitem{sharma2020outage}
	P.~K. Sharma, D.~Deepthi, and D.~I. Kim, ``Outage probability of {3-D} mobile
	{UAV} relaying for hybrid satellite-terrestrial networks,'' \emph{IEEE
		Communications Letters}, vol.~24, no.~2, pp. 418--422, Feb. 2020.
	
	\bibitem{sharma2020secure}
	P.~K. Sharma and D.~I. Kim, ``Secure {3D} mobile {UAV} relaying for hybrid
	satellite-terrestrial networks,'' \emph{IEEE Transactions on Wireless
		Communications}, vol.~19, no.~4, pp. 2770--2784, Apr. 2020.
	
	\bibitem{qu2017leo}
	Z.~Qu, G.~Zhang, H.~Cao, and J.~Xie, ``{LEO} satellite constellation for
	{Internet} of things,'' \emph{IEEE Access}, vol.~5, pp. 18\,391--18\,401,
	2017.
	
	\bibitem{papa2020design}
	A.~Papa, T.~de~Cola, P.~Vizarreta, M.~He, C.~Mas-Machuca, and W.~Kellerer,
	``Design and evaluation of reconfigurable {SDN LEO} constellations,''
	\emph{IEEE Transactions on Network and Service Management}, May 2020.
	
	\bibitem{del2019technical}
	I.~Del~Portillo, B.~G. Cameron, and E.~F. Crawley, ``A technical comparison of
	three low earth orbit satellite constellation systems to provide global
	broadband,'' \emph{Acta Astronautica}, vol. 159, pp. 123--135, Jun. 2019.
	
	\bibitem{you20193d}
	C.~You and R.~Zhang, ``{3D} trajectory optimization in {Rician} fading for
	{UAV}-enabled data harvesting,'' \emph{IEEE Transactions on Wireless
		Communications}, vol.~18, no.~6, pp. 3192--3207, Jun. 2019.
	
	\bibitem{hu2020aoi}
	H.~Hu, K.~Xiong, G.~Qu, Q.~Ni, P.~Fan, and K.~B. Letaief, ``{AoI}-minimal
	trajectory planning and data collection in {UAV}-assisted wireless powered
	{IoT} networks,'' \emph{IEEE Internet of Things Journal}, 2020, in press.
	
	\bibitem{liu2019trajectory}
	X.~Liu, Y.~Liu, Y.~Chen, and L.~Hanzo, ``Trajectory design and power control
	for multi-{UAV} assisted wireless networks: A machine learning approach,''
	\emph{IEEE Transactions on Vehicular Technology}, vol.~68, no.~8, pp.
	7957--7969, Aug. 2019.
	
	\bibitem{liu2020path}
	Q.~Liu, L.~Shi, L.~Sun, J.~Li, M.~Ding, and F.~Shu, ``Path planning for
	{UAV}-mounted mobile edge computing with deep reinforcement learning,''
	\emph{IEEE Transactions on Vehicular Technology}, vol.~69, no.~5, pp.
	5723--5728, May 2020.
	
	\bibitem{xu2020joint}
	L.~Xu, M.~Chen, M.~Chen, Z.~Yang, C.~Chaccour, W.~Saad, and C.~S. Hong, ``Joint
	location and power optimization for {THz}-enabled {UAV} communications,''
	\emph{arXiv preprint arXiv:2011.01065}, 2020.
	
	\bibitem{youn2020aerodynamic}
	W.~Youn, H.~Choi, A.~Cho, S.~Kim, and M.~B. Rhudy, ``Aerodynamic
	{M}odel-{A}ided estimation of {A}ttitude, {3D} {W}ind, {A}irspeed, {AOA}, and
	{SSA} for {H}igh-{A}ltitude {L}ong-{E}ndurance {UAV},'' \emph{IEEE
		Transactions on Aerospace and Electronic Systems}, Apr. 2020.
	
	\bibitem{kantor2016unmanned}
	I.~Kantor, A.~N. Srivastava, D.~M. Pasko, H.~Batla, and G.~Ubhi, ``Unmanned
	aerial vehicle network-based recharging,'' Aug.~9 2016, {US} Patent
	9,412,279.
	
	\bibitem{sharma2017intelligent}
	V.~Sharma, K.~Srinivasan, H.-C. Chao, K.-L. Hua, and W.-H. Cheng, ``Intelligent
	deployment of {UAVs} in {5G} heterogeneous communication environment for
	improved coverage,'' \emph{Journal of Network and Computer Applications},
	vol.~85, pp. 94--105, May 2017.
	
	\bibitem{dunphy2018first}
	P.~Dunphy and F.~A. Petitcolas, ``A first look at identity management schemes
	on the blockchain,'' \emph{IEEE Security \& Privacy}, vol.~16, no.~4, pp.
	20--29, Aug. 2018.
	
	\bibitem{rosa2018blockchain}
	R.~Rosa and C.~E. Rothenberg, ``Blockchain-based decentralized applications for
	multiple administrative domain networking,'' \emph{IEEE Communications
		Standards Magazine}, vol.~2, no.~3, pp. 29--37, Sep. 2018.
	
	\bibitem{alladi2020applications}
	T.~Alladi, V.~Chamola, N.~Sahu, and M.~Guizani, ``Applications of blockchain in
	unmanned aerial vehicles: A review,'' \emph{Vehicular Communications}, p.
	100249, Jan. 2020.
	
	\bibitem{hassija2020distributed}
	V.~Hassija, V.~Chamola, D.~N.~G. Krishna, and M.~Guizani, ``A distributed
	framework for energy trading between {UAVs} and charging stations for
	critical applications,'' \emph{IEEE Transactions on Vehicular Technology},
	vol.~69, no.~5, pp. 5391--5402, Mar. 2020.
	
	\bibitem{popov2016tangle}
	\BIBentryALTinterwordspacing
	S.~Popov, ``The tangle,'' \emph{cit. on}, p. 131, Jun. 2016. [Online].
	Available: \url{https://www.iota.org/foundation/research-papers.}
	\BIBentrySTDinterwordspacing
	
	\bibitem{huang2019towards}
	J.~Huang, L.~Kong, G.~Chen, M.-Y. Wu, X.~Liu, and P.~Zeng, ``Towards secure
	industrial {IoT}: Blockchain system with credit-based consensus mechanism,''
	\emph{IEEE Transactions on Industrial Informatics}, vol.~15, no.~6, pp.
	3680--3689, Mar. 2019.
	
	\bibitem{dicembrini2020modelling}
	E.~Dicembrini, M.~Scanavino, F.~Dabbene, and G.~Guglieri, ``Modelling and
	simulation of a tethered {UAS},'' in \emph{2020 International Conference on
		Unmanned Aircraft Systems (ICUAS)}.\hskip 1em plus 0.5em minus 0.4em\relax
	IEEE, Oct. 2020, pp. 1801--1808.
	
	\bibitem{bushnaq2020optimal}
	O.~M. Bushnaq, M.~A. Kishk, A.~{\c{C}}elik, M.-S. Alouini, and T.~Y.
	Al-Naffouri, ``Optimal deployment of tethered drones for maximum cellular
	coverage in user clusters,'' \emph{IEEE Transactions on Wireless
		Communications}, Nov. 2020.
	
	\bibitem{lifeline}
	\BIBentryALTinterwordspacing
	``{Lifeline Tethered Drone},'' {A}ccessed: Dec. 22, 2020. [Online]. Available:
	\url{https://www.lifeline-drone.com/}
	\BIBentrySTDinterwordspacing
	
	\bibitem{acecore}
	\BIBentryALTinterwordspacing
	``{Acecore Crafting Elevating Drones},'' {A}ccessed: Dec. 22, 2020. [Online].
	Available: \url{https://acecoretechnologies.com/}
	\BIBentrySTDinterwordspacing
	
	\bibitem{ziyan}
	\BIBentryALTinterwordspacing
	``{Ziyan UAS},'' {A}ccessed: Dec. 22, 2020. [Online]. Available:
	\url{http://ziyanuav.com/}
	\BIBentrySTDinterwordspacing
	
	\bibitem{TDS}
	\BIBentryALTinterwordspacing
	``{Tethered Drone Systems: The Future of Tethered UAV Technology},''
	{A}ccessed: Dec. 22, 2020. [Online]. Available:
	\url{https://tethereddronesystems.co.uk/}
	\BIBentrySTDinterwordspacing
	
	\bibitem{equinox}
	\BIBentryALTinterwordspacing
	``{Equinox Innovatice Systems: Ultra High Bandwidth Drone Services and
		Solutions},'' {A}ccessed: Dec. 22, 2020. [Online]. Available:
	\url{https://equinoxinnovativesystems.com/}
	\BIBentrySTDinterwordspacing
	
	\bibitem{hoverfly}
	\BIBentryALTinterwordspacing
	``{Hoverfly Technology},'' {A}ccessed: Dec. 22, 2020. [Online]. Available:
	\url{https://hoverflytech.com/}
	\BIBentrySTDinterwordspacing
	
	\bibitem{elistair}
	\BIBentryALTinterwordspacing
	``{Elistair: The Tethered Drone Company},'' {A}ccessed: Dec. 22, 2020.
	[Online]. Available: \url{https://elistair.com/}
	\BIBentrySTDinterwordspacing
	
	\bibitem{eagleSkyLight}
	\BIBentryALTinterwordspacing
	``{Eagle Sky Light},'' {A}ccessed: Dec. 22, 2020. [Online]. Available:
	\url{https://eagleskylight.it/en/}
	\BIBentrySTDinterwordspacing
	
	\bibitem{kim2017review}
	H.-J. Kim, H.~Hirayama, S.~Kim, K.~J. Han, R.~Zhang, and J.-W. Choi, ``Review
	of near-field wireless power and communication for biomedical applications,''
	\emph{IEEE Access}, vol.~5, pp. 21\,264--21\,285, 2017.
	
	\bibitem{lu2016wireless}
	X.~Lu, P.~Wang, D.~Niyato, D.~I. Kim, and Z.~Han, ``Wireless charging
	technologies: Fundamentals, standards, and network applications,'' \emph{IEEE
		Communications Surveys \& Tutorials}, vol.~18, no.~2, pp. 1413--1452,
	Secondquarter 2016.
	
	\bibitem{wu2012high}
	H.~H. Wu, A.~Gilchrist, K.~D. Sealy, and D.~Bronson, ``A high efficiency {5 kW}
	inductive charger for {EVs} using dual side control,'' \emph{IEEE
		Transactions on Industrial Informatics}, vol.~8, no.~3, pp. 585--595, Aug.
	2012.
	
	\bibitem{campi2019wireless}
	T.~Campi, S.~Cruciani, F.~Maradei, and M.~Feliziani, ``Wireless charging system
	integrated in a small unmanned aerial vehicle ({UAV}) with high tolerance to
	planar coil misalignment,'' in \emph{2019 Joint International Symposium on
		Electromagnetic Compatibility, Sapporo and Asia-Pacific International
		Symposium on Electromagnetic Compatibility (EMC Sapporo/APEMC)}.\hskip 1em
	plus 0.5em minus 0.4em\relax Sapporo, Japan: IEEE, 2019, pp. 601--604.
	
	\bibitem{obayashi201985}
	S.~Obayashi, Y.~Kanekiyo, K.~Nishizawa, and H.~Kusada, ``85-{kHz} band 450-{W}
	{I}nductive {P}ower {T}ransfer for {U}nmanned {A}erial {V}ehicle {W}ireless
	{C}harging {P}ort,'' in \emph{2019 IEEE Wireless Power Transfer Conference
		(WPTC)}, London, United Kingdom, 2019, pp. 80--84.
	
	\bibitem{wang2016design}
	C.~Wang and Z.~Ma, ``Design of wireless power transfer device for {UAV},'' in
	\emph{2016 IEEE International Conference on Mechatronics and Automation},
	Harbin, China, 2016, pp. 2449--2454.
	
	\bibitem{qiong2018optimal}
	L.~Qiong, W.~Huayun, W.~Wenbin, M.~Tianqi, and W.~Yongyue, ``Optimal {D}esign
	of {M}agnetic {R}esonance {W}ireless {C}harging {C}oil for {U}nmanned
	{A}erial {V}ehicles,'' in \emph{2018 3rd International Conference on Smart
		City and Systems Engineering (ICSCSE)}.\hskip 1em plus 0.5em minus
	0.4em\relax Xiamen, China, China: IEEE, 2018, pp. 469--472.
	
	\bibitem{sekander2020statistical}
	S.~Sekander, H.~Tabassum, and E.~Hossain, ``Statistical performance modeling of
	solar and wind-powered {UAV} communications,'' \emph{IEEE Transactions on
		Mobile Computing}, pp. 1--1, Apr. 2020.
	
	\bibitem{zhang2020power}
	J.~Zhang, M.~Lou, L.~Xiang, and L.~Hu, ``Power cognition: Enabling intelligent
	energy harvesting and resource allocation for solar-powered {UAVs},''
	\emph{Future Generation Computer Systems}, vol. 110, pp. 658--664, Sep. 2020.
	
	\bibitem{jashnani2013sizing}
	S.~Jashnani, T.~Nada, M.~Ishfaq, A.~Khamker, and P.~Shaholia, ``Sizing and
	preliminary hardware testing of solar powered {UAV},'' \emph{The Egyptian
		Journal of Remote Sensing and Space Science}, vol.~16, no.~2, pp. 189--198,
	Dec. 2013.
	
	\bibitem{morton2015solar}
	S.~Morton, R.~D'Sa, and N.~Papanikolopoulos, ``Solar powered {UAV}: Design and
	experiments,'' in \emph{2015 IEEE/RSJ International Conference on Intelligent
		Robots and Systems (IROS)}, Hamburg, Germany, 2015, pp. 2460--2466.
	
	\bibitem{oettershagen2016perpetual}
	P.~Oettershagen, A.~Melzer, T.~Mantel, K.~Rudin, T.~Stastny, B.~Wawrzacz,
	T.~Hinzmann, K.~Alexis, and R.~Siegwart, ``Perpetual flight with a small
	solar-powered {UAV}: Flight results, performance analysis and model
	validation,'' in \emph{2016 IEEE Aerospace Conference}, Big Sky, MT, USA,
	2016, pp. 1--8.
	
	\bibitem{dwivedi2018maraal}
	V.~S. Dwivedi, J.~Patrikar, A.~Addamane, and A.~Ghosh, ``{MARAAL}: A low
	altitude long endurance solar powered {UAV} for surveillance and mapping
	applications,'' in \emph{2018 23rd International Conference on Methods \&
		Models in Automation \& Robotics (MMAR)}.\hskip 1em plus 0.5em minus
	0.4em\relax Miedzyzdroje, Poland: IEEE, 2018, pp. 449--454.
	
	\bibitem{wang2018analysis}
	H.~Wang and J.~Shen, ``Analysis of the characteristics of solar cell array
	based on {MATLAB}/{S}imulink in solar unmanned aerial vehicle,'' \emph{IEEE
		Access}, vol.~6, pp. 21\,195--21\,201, Apr. 2018.
	
	\bibitem{rajendran2018experimental}
	P.~Rajendran and H.~Smith, ``Experimental study of solar module \& maximum
	power point tracking system under controlled temperature conditions,''
	\emph{Internation Journal on Advanced Science Engineering Information
		Technology}, vol.~8, no.~4, pp. 1147--1153, 2018.
	
	\bibitem{hosseini2013optimal}
	S.~Hosseini, R.~Dai, and M.~Mesbahi, ``Optimal path planning and power
	allocation for a long endurance solar-powered {UAV},'' in \emph{2013 American
		Control Conference}.\hskip 1em plus 0.5em minus 0.4em\relax Washington, DC,
	USA: IEEE, 2013, pp. 2588--2593.
	
	\bibitem{sun2018resource}
	Y.~Sun, D.~W.~K. Ng, D.~Xu, L.~Dai, and R.~Schober, ``Resource allocation for
	solar powered {UAV} communication systems,'' in \emph{2018 IEEE 19th
		International Workshop on Signal Processing Advances in Wireless
		Communications (SPAWC)}, Kalamata, Greece, 2018, pp. 1--5.
	
	\bibitem{sun2019optimal}
	Y.~Sun, D.~Xu, D.~W.~K. Ng, L.~Dai, and R.~Schober, ``Optimal {3D}-trajectory
	design and resource allocation for solar-powered {UAV} communication
	systems,'' \emph{IEEE Transactions on Communications}, vol.~67, no.~6, pp.
	4281--4298, Feb. 2019.
	
	\bibitem{wang2019multi}
	P.~Wang, J.~L. Rosales, and V.~A. Chiriac, ``Multi-rotor aerial drone with
	thermal energy harvesting,'' Feb.~28 2019, {US} Patent App. 15/689,994.
	
	\bibitem{bonnin2015energy}
	V.~Bonnin, E.~B{\'e}nard, J.-M. Moschetta, and C.~Toomer, ``Energy-harvesting
	mechanisms for {UAV} flight by dynamic soaring,'' \emph{International Journal
		of Micro Air Vehicles}, vol.~7, no.~3, pp. 213--229, Sep. 2015.
	
	\bibitem{anton2008vibration}
	S.~R. Anton and D.~J. Inman, ``Vibration energy harvesting for unmanned aerial
	vehicles,'' in \emph{Active and Passive Smart Structures and Integrated
		Systems 2008}, vol. 6928.\hskip 1em plus 0.5em minus 0.4em\relax San Diego,
	California, United States: International Society for Optics and Photonics,
	2008, p. 692824.
	
	\bibitem{koszewnik2019performance}
	A.~Koszewnik and D.~Oldziej, ``Performance assessment of an energy harvesting
	system located on a copter,'' \emph{The European Physical Journal Special
		Topics}, vol. 228, no.~7, pp. 1677--1692, Aug. 2019.
	
	\bibitem{anton2011performance}
	S.~R. Anton and D.~J. Inman, ``Performance modeling of unmanned aerial vehicles
	with on-board energy harvesting,'' in \emph{Active and Passive Smart
		Structures and Integrated Systems 2011}, vol. 7977.\hskip 1em plus 0.5em
	minus 0.4em\relax Diego, California, United States: International Society for
	Optics and Photonics, 2011, p. 79771H.
	
	\bibitem{belo2019selective}
	D.~Belo, D.~C. Ribeiro, P.~Pinho, and N.~B. Carvalho, ``A selective, tracking,
	and power adaptive far-field wireless power transfer system,'' \emph{IEEE
		Transactions on Microwave Theory and Techniques}, vol.~67, no.~9, pp.
	3856--3866, Sep. 2019.
	
	\bibitem{kuhn2015multi}
	V.~Kuhn, C.~Lahuec, F.~Seguin, and C.~Person, ``A multi-band stacked {RF}
	energy harvester with {RF}-to-{DC} efficiency up to 84\%,'' \emph{IEEE
		Transactions on Microwave Theory and Techniques}, vol.~63, no.~5, pp.
	1768--1778, May 2015.
	
	\bibitem{dunbar2015wireless}
	S.~Dunbar, F.~Wenzl, C.~Hack, R.~Hafeza, H.~Esfeer, F.~Defay, S.~Prothin,
	D.~Bajon, and Z.~Popovic, ``Wireless far-field charging of a micro-{UAV},''
	in \emph{2015 IEEE Wireless Power Transfer Conference (WPTC)}, Boulder, CO,
	USA, 2015, pp. 1--4.
	
	\bibitem{muharam2017design}
	A.~Muharam, T.~M. Mostafa, and R.~Hattori, ``Design of power receiving side in
	wireless charging system for {UAV} application,'' in \emph{2017 International
		Conference on Sustainable Energy Engineering and Application (ICSEEA)}.\hskip
	1em plus 0.5em minus 0.4em\relax Jakarta, Indonesia: IEEE, 2017, pp.
	133--139.
	
	\bibitem{long2018energy}
	T.~Long, M.~Ozger, O.~Cetinkaya, and O.~B. Akan, ``Energy neutral internet of
	drones,'' \emph{IEEE Communications Magazine}, vol.~56, no.~1, pp. 22--28,
	Jan. 2018.
	
	\bibitem{chen2019resonant}
	W.~Chen, S.~Zhao, Q.~Shi, and R.~Zhang, ``Resonant {B}eam {C}harging-{P}owered
	{UAV}-{A}ssisted {S}ensing {D}ata {C}ollection,'' \emph{IEEE Transactions on
		Vehicular Technology}, vol.~69, no.~1, pp. 1086--1090, Oct. 2019.
	
	\bibitem{lahmeri2020stochastic}
	M.-A. Lahmeri, M.~A. Kishk, and M.-S. Alouini, ``Stochastic geometry-based
	analysis of airborne base stations with laser-powered {UAVs},'' \emph{IEEE
		Communications Letters}, vol.~24, no.~1, pp. 173--177, Feb. 2020.
	
	\bibitem{ouyang2018throughput}
	J.~Ouyang, Y.~Che, J.~Xu, and K.~Wu, ``Throughput maximization for
	laser-powered {UAV} wireless communication systems,'' in \emph{2018 IEEE
		International Conference on Communications Workshops (ICC Workshops)}.\hskip
	1em plus 0.5em minus 0.4em\relax Kansas City, MO, USA: IEEE, 2018, pp. 1--6.
	
	\bibitem{chen2015design}
	Q.~Chen, D.~Zhang, D.~Zhu, Q.~Shi, J.~Gu, and Y.~Ai, ``Design and experiment
	for realization of laser wireless power transmission for small unmanned
	aerial vehicles,'' in \emph{AOPC 2015: Advances in Laser Technology and
		Applications}, vol. 9671.\hskip 1em plus 0.5em minus 0.4em\relax Beijing,
	China: International Society for Optics and Photonics, 2015, p. 96710N.
	
	\bibitem{marano2018resource}
	S.~Marano and P.~Willett, ``Resource allocation in energy-harvesting sensor
	networks,'' \emph{IEEE Transactions on Signal and Information Processing over
		Networks}, vol.~4, no.~3, pp. 585--598, Jan. 2018.
	
	\bibitem{xu2018uav}
	J.~Xu, Y.~Zeng, and R.~Zhang, ``{UAV}-enabled wireless power transfer:
	{T}rajectory design and energy optimization,'' \emph{IEEE Transactions on
		Wireless Communications}, vol.~17, no.~8, pp. 5092--5106, May 2018.
	
	\bibitem{su2020uav}
	C.~Su, F.~Ye, L.-C. Wang, L.~Wang, Y.~Tian, and Z.~Han, ``{UAV}-assisted
	wireless charging for energy-constrained {IoT} devices using dynamic
	matching,'' \emph{IEEE Internet of Things Journal}, Jan. 2020.
	
	\bibitem{wang2018resource}
	H.~Wang, J.~Wang, G.~Ding, L.~Wang, T.~A. Tsiftsis, and P.~K. Sharma,
	``Resource allocation for energy harvesting-powered {D2D} communication
	underlaying {UAV}-assisted networks,'' \emph{IEEE Transactions on Green
		Communications and Networking}, vol.~2, no.~1, pp. 14--24, Oct. 2018.
	
	\bibitem{chen2015software}
	T.~Chen, M.~Matinmikko, X.~Chen, X.~Zhou, and P.~Ahokangas, ``Software defined
	mobile networks: concept, survey, and research directions,'' \emph{IEEE
		Communications Magazine}, vol.~53, no.~11, pp. 126--133, Nov. 2015.
	
	\bibitem{mamushiane2018comparative}
	L.~Mamushiane, A.~Lysko, and S.~Dlamini, ``A comparative evaluation of the
	performance of popular {SDN} controllers,'' in \emph{2018 Wireless Days
		(WD)}.\hskip 1em plus 0.5em minus 0.4em\relax Dubai, United Arab Emirates:
	IEEE, 2018, pp. 54--59.
	
	\bibitem{oubbati2020softwarization}
	O.~S. Oubbati, M.~Atiquzzaman, T.~A. Ahanger, and A.~Ibrahim, ``Softwarization
	of {UAV} networks: A survey of applications and future trends,'' \emph{IEEE
		Access}, vol.~8, pp. 98\,073--98\,125, May 2020.
	
	\bibitem{zhao2019software}
	Z.~Zhao, P.~Cumino, A.~Souza, D.~Rosario, T.~Braun, E.~Cerqueira, and M.~Gerla,
	``Software-defined unmanned aerial vehicles networking for video
	dissemination services,'' \emph{Ad Hoc Networks}, vol.~83, pp. 68--77, Feb.
	2019.
	
	\bibitem{zhang2018sdn}
	X.~Zhang, H.~Wang, and H.~Zhao, ``An {SDN} framework for {UAV} backbone network
	towards knowledge centric networking,'' in \emph{IEEE Conference on Computer
		Communications Workshops (INFOCOM WKSHPS)}, Honolulu, HI, USA, Apr. 2018, pp.
	456--461.
	
	\bibitem{xiong2019sdn}
	F.~Xiong, A.~Li, H.~Wang, and L.~Tang, ``An {SDN-MQTT} based communication
	system for battlefield {UAV} swarms,'' \emph{IEEE Communications Magazine},
	vol.~57, no.~8, pp. 41--47, Aug. 2019.
	
	\bibitem{hermosilla2020security}
	A.~Hermosilla, A.~M. Zarca, J.~B. Bernabe, J.~Ortiz, and A.~Skarmeta,
	``Security orchestration and enforcement in {NFV/SDN-aware UAV}
	deployments,'' \emph{IEEE Access}, vol.~8, pp. 131\,779--131\,795, Jul. 2020.
	
	\bibitem{barritt2018loon}
	B.~Barritt and V.~Cerf, ``Loon {SDN}: Applicability to {NASA}'s next-generation
	space communications architecture,'' in \emph{2018 IEEE Aerospace
		Conference}, Big Sky, MT, USA, Mar. 2018, pp. 1--9.
	
	\bibitem{papa2018dynamic}
	A.~Papa, T.~De~Cola, P.~Vizarreta, M.~He, C.~M. Machuca, and W.~Kellerer,
	``Dynamic {SDN} controller placement in a {LEO} constellation satellite
	network,'' in \emph{IEEE Global Communications Conference (GLOBECOM)}, Abu
	Dhabi, United Arab Emirates, United Arab Emirates, Dec. 2018, pp. 206--212.
	
	\bibitem{jia2020virtual}
	Z.~Jia, M.~Sheng, J.~Li, Y.~Zhu, W.~Bai, and Z.~Han, ``Virtual network
	functions orchestration in software defined {LEO} small satellite networks,''
	in \emph{IEEE International Conference on Communications (ICC)}, Dublin,
	Ireland, Jun. 2020, pp. 1--6.
	
	\bibitem{abbas2020fog}
	A.~Abbas, S.~U. Khan, and A.~Y. Zomaya, \emph{Fog {C}omputing: Theory and
		{P}ractice}.\hskip 1em plus 0.5em minus 0.4em\relax John Wiley \& Sons, 2020.
	
	\bibitem{mohamed2017uavfog}
	N.~Mohamed, J.~Al-Jaroodi, I.~Jawhar, H.~Noura, and S.~Mahmoud, ``{UAVFog}: A
	{UAV}-based fog computing for {I}nternet of {T}hings,'' in \emph{2017 IEEE
		SmartWorld, Ubiquitous Intelligence \& Computing, Advanced \& Trusted
		Computed, Scalable Computing \& Communications, Cloud \& Big Data Computing,
		Internet of People and Smart City Innovation
		(SmartWorld/SCALCOM/UIC/ATC/CBDCom/IOP/SCI)}, San Francisco, CA, USA, 2017,
	pp. 1--8.
	
	\bibitem{messous2017computation}
	M.-A. Messous, H.~Sedjelmaci, N.~Houari, and S.-M. Senouci, ``Computation
	offloading game for an {UAV} network in mobile edge computing,'' in
	\emph{2017 IEEE International Conference on Communications (ICC)}, Paris,
	France, 2017, pp. 1--6.
	
	\bibitem{pinto2019framework}
	M.~F. Pinto, A.~L. Marcato, A.~G. Melo, L.~M. Hon{\'o}rio, and C.~Urdiales, ``A
	framework for analyzing fog-cloud computing cooperation applied to
	information processing of {UAVs},'' \emph{Wireless Communications and Mobile
		Computing}, vol. 2019, 2019.
	
	\bibitem{liu2020online}
	B.~Liu, W.~Zhang, W.~Chen, H.~Huang, and S.~Guo, ``Online {C}omputation
	{O}ffloading and {T}raffic {R}outing for {UAV} {S}warms in {E}dge-{C}loud
	{C}omputing,'' \emph{IEEE Transactions on Vehicular Technology}, May 2020.
	
	\bibitem{diao2019fair}
	X.~Diao, J.~Zheng, Y.~Cai, Y.~Wu, and A.~Anpalagan, ``Fair data allocation and
	trajectory optimization for {UAV}-assisted mobile edge computing,''
	\emph{IEEE Communications Letters}, vol.~23, no.~12, pp. 2357--2361, Dec.
	2019.
	
	\bibitem{yang2020offloading}
	B.~Yang, X.~Cao, C.~Yuen, and L.~Qian, ``Offloading optimization in edge
	computing for deep learning enabled target tracking by internet-of-{UAVs},''
	\emph{IEEE Internet of Things Journal}, Aug. 2020.
	
	\bibitem{ilchenko2019solution}
	M.~Ilchenko, T.~Narytnik, V.~Prisyazhny, S.~Kapshtyk, and S.~Matvienko, ``The
	solution of the problem of the delay determination in the information
	transmission and processing in the {LEO} satellite internet of things
	system,'' in \emph{2019 IEEE International Scientific-Practical Conference
		Problems of Infocommunications, Science and Technology (PIC S\&T)}.\hskip 1em
	plus 0.5em minus 0.4em\relax Kyiv, Ukraine, Ukraine: IEEE, Oct. 2019, pp.
	419--425.
	
	\bibitem{qiu2019deep}
	C.~Qiu, H.~Yao, F.~R. Yu, F.~Xu, and C.~Zhao, ``Deep {Q-learning} aided
	networking, caching, and computing resources allocation in software-defined
	satellite-terrestrial networks,'' \emph{IEEE Transactions on Vehicular
		Technology}, vol.~68, no.~6, pp. 5871--5883, Jun. 2019.
	
	\bibitem{qian2017survey}
	L.~{Qian}, J.~{Zhu}, and S.~{Zhang}, ``Survey of {W}ireless {B}ig {D}ata,''
	\emph{Journal of Communications and Information Networks}, vol.~2, no.~1, pp.
	1--18, Mar. 2017.
	
	\bibitem{zhang2019deep}
	C.~Zhang, P.~Patras, and H.~Haddadi, ``Deep learning in mobile and wireless
	networking: A survey,'' \emph{IEEE Communications Surveys \& Tutorials},
	vol.~21, no.~3, pp. 2224--2287, Mar. 2019.
	
	\bibitem{han2011data}
	J.~Han, J.~Pei, and M.~Kamber, \emph{Data mining: concepts and
		techniques}.\hskip 1em plus 0.5em minus 0.4em\relax Elsevier, 2011.
	
	\bibitem{bithas2019survey}
	P.~S. Bithas, E.~T. Michailidis, N.~Nomikos, D.~Vouyioukas, and A.~G. Kanatas,
	``A survey on machine-learning techniques for {UAV}-based communications,''
	\emph{Sensors}, vol.~19, no.~23, p. 5170, Nov. 2019.
	
	\bibitem{carrio2017review}
	A.~Carrio, C.~Sampedro, A.~Rodriguez-Ramos, and P.~Campoy, ``A review of deep
	learning methods and applications for unmanned aerial vehicles,''
	\emph{Journal of Sensors}, vol. 2017, Aug. 2017.
	
	\bibitem{luong2019applications}
	N.~C. Luong, D.~T. Hoang, S.~Gong, D.~Niyato, P.~Wang, Y.-C. Liang, and D.~I.
	Kim, ``Applications of deep reinforcement learning in communications and
	networking: A survey,'' \emph{IEEE Communications Surveys \& Tutorials},
	vol.~21, no.~4, pp. 3133--3174, Fourthquarter 2019.
	
	\bibitem{mukherjee2020distributed}
	M.~Mukherjee, V.~Kumar, A.~Lat, M.~Guo, R.~Matam, and Y.~Lv, ``Distributed
	{D}eep {L}earning-based {T}ask {O}ffloading for {UAV}-enabled {M}obile {E}dge
	{C}omputing,'' in \emph{IEEE INFOCOM 2020-IEEE Conference on Computer
		Communications Workshops (INFOCOM WKSHPS)}, Toronto, ON, Canada, 2020, pp.
	1208--1212.
	
	\bibitem{mozaffari2019tutorial}
	M.~Mozaffari, W.~Saad, M.~Bennis, Y.-H. Nam, and M.~Debbah, ``A tutorial on
	{UAVs} for wireless networks: Applications, challenges, and open problems,''
	\emph{IEEE Communications Surveys \& Tutorials}, vol.~21, no.~3, pp.
	2334--2360, Thirdquarter 2019.
	
	\bibitem{shakeri2019design}
	R.~Shakeri, M.~A. Al-Garadi, A.~Badawy, A.~Mohamed, T.~Khattab, A.~K. Al-Ali,
	K.~A. Harras, and M.~Guizani, ``Design challenges of multi-{UAV} systems in
	cyber-physical applications: A comprehensive survey and future directions,''
	\emph{IEEE Communications Surveys \& Tutorials}, vol.~21, no.~4, pp.
	3340--3385, Fourthquarter 2019.
	
	\bibitem{ferreira2019reinforcement}
	P.~V.~R. Ferreira, R.~Paffenroth, A.~M. Wyglinski, T.~M. Hackett, S.~G. Bilen,
	R.~C. Reinhart, and D.~J. Mortensen, ``Reinforcement learning for satellite
	communications: from {LEO} to deep space operations,'' \emph{IEEE
		Communications Magazine}, vol.~57, no.~5, pp. 70--75, May 2019.
	
	\bibitem{werth2020silo}
	M.~Werth, J.~Lucas, T.~Kyono, I.~McQuaid, and J.~Fletcher, ``{SILO}: A machine
	learning dataset of synthetic ground-based observations of {LEO}
	satellites,'' in \emph{2020 IEEE Aerospace Conference}.\hskip 1em plus 0.5em
	minus 0.4em\relax Big Sky, MT, USA: IEEE, Mar. 2020, pp. 1--8.
	
	\bibitem{chen2020machine}
	H.~Chen, V.~Chandrasekar, R.~Cifelli, and P.~Xie, ``A machine learning system
	for precipitation estimation using satellite and ground radar network
	observations,'' \emph{IEEE Transactions on Geoscience and Remote Sensing},
	vol.~58, no.~2, pp. 982--994, Feb. 2020.
	
	\bibitem{kato2019optimizing}
	N.~Kato, Z.~M. Fadlullah, F.~Tang, B.~Mao, S.~Tani, A.~Okamura, and J.~Liu,
	``Optimizing space-air-ground integrated networks by artificial
	intelligence,'' \emph{IEEE Wireless Communications}, vol.~26, no.~4, pp.
	140--147, Aug. 2019.
	
	\bibitem{parkvall2017nr}
	S.~Parkvall, E.~Dahlman, A.~Furuskar, and M.~Frenne, ``{NR}: The new {5G} radio
	access technology,'' \emph{IEEE Communications Standards Magazine}, vol.~1,
	no.~4, pp. 24--30, Dec. 2017.
	
	\bibitem{zhou2018beam}
	P.~Zhou, X.~Fang, Y.~Fang, R.~He, Y.~Long, and G.~Huang, ``Beam management and
	self-healing for {mmWave} {UAV} mesh networks,'' \emph{IEEE Transactions on
		Vehicular Technology}, vol.~68, no.~2, pp. 1718--1732, Feb. 2018.
	
	\bibitem{zhao2018efficient}
	J.~Zhao and W.~Jia, ``Efficient channel tracking strategy for {mmWave UAV}
	communications,'' \emph{Electronics Letters}, vol.~54, no.~21, pp.
	1218--1220, Oct. 2018.
	
	\bibitem{mudonhi2018sdn}
	A.~Mudonhi, C.~Sacchi, and F.~Granelli, ``{SDN}-based multimedia content
	delivery in {5G MmWave} hybrid satellite-terrestrial networks,'' in
	\emph{2018 IEEE 29th Annual International Symposium on Personal, Indoor and
		Mobile Radio Communications (PIMRC)}.\hskip 1em plus 0.5em minus 0.4em\relax
	Bologna, Italy: IEEE, 2018, pp. 1--7.
	
	\bibitem{artiga2018shared}
	X.~Artiga, A.~P{\'e}rez-Neira, J.~Baranda, E.~Lagunas, S.~Chatzinotas,
	R.~Zetik, P.~Gorski, K.~Ntougias, D.~Perez, and G.~Ziaragkas, ``Shared access
	satellite-terrestrial reconfigurable backhaul network enabled by smart
	antennas at mmwave band,'' \emph{IEEE Network}, vol.~32, no.~5, pp. 46--53,
	Sep./Oct. 2018.
	
	\bibitem{tekbiyik2020reconfigurable}
	K.~Tekb{\i}y{\i}k, G.~K. Kurt, A.~R. Ekti, A.~G{\"o}r{\c{c}}in, and
	H.~Yanikomeroglu, ``Reconfigurable intelligent surface empowered terahertz
	communication for {LEO} satellite networks,'' \emph{arXiv preprint
		arXiv:2007.04281}, Jul. 2020.
	
	\bibitem{wang2019inter}
	Z.~Wang, R.~Malaney, and J.~Green, ``Inter-satellite quantum key distribution
	at terahertz frequencies,'' in \emph{ICC 2019-2019 IEEE International
		Conference on Communications (ICC)}.\hskip 1em plus 0.5em minus 0.4em\relax
	Shanghai, China: IEEE, 2019, pp. 1--7.
	
	\bibitem{mendrzik2018error}
	R.~Mendrzik, D.~Cabric, and G.~Bauch, ``Error bounds for {Terahertz} {MIMO}
	positioning of swarm {UAVs} for distributed sensing,'' in \emph{2018 IEEE
		International Conference on Communications Workshops (ICC Workshops)}.\hskip
	1em plus 0.5em minus 0.4em\relax Kansas City, MO, USA: IEEE, 2018, pp. 1--6.
	
	\bibitem{chu2019uav}
	Z.~Chu, W.~Hao, P.~Xiao, and J.~Shi, ``{UAV Assisted Spectrum Sharing
		Ultra-Reliable and Low-Latency Communications},'' in \emph{2019 IEEE Global
		Communications Conference (GLOBECOM)}.\hskip 1em plus 0.5em minus 0.4em\relax
	Waikoloa, HI, USA: IEEE, Dec. 2019, pp. 1--6.
	
	\bibitem{han2019uav}
	A.~Han, T.~Lv, and X.~Zhang, ``{UAV} beamwidth design for {Ultra-Reliable and
		Low-Latency Communications} with {NOMA},'' in \emph{2019 IEEE International
		Conference on Communications Workshops (ICC Workshops)}.\hskip 1em plus 0.5em
	minus 0.4em\relax Shanghai, China: IEEE, 2019, pp. 1--6.
	
	\bibitem{cho2019cross}
	W.~Cho and J.~P. Choi, ``Cross layer optimization of wireless control links in
	the software-defined {LEO} satellite network,'' \emph{IEEE Access}, vol.~7,
	pp. 113\,534--113\,547, Aug 2019.
	
	\bibitem{leyva2020leo}
	I.~Leyva-Mayorga, B.~Soret, M.~R{\"o}per, D.~W{\"u}bben, B.~Matthiesen,
	A.~Dekorsy, and P.~Popovski, ``{LEO} small-satellite constellations for {5G}
	and beyond-{5G} communications,'' \emph{Ieee Access}, vol.~8, pp.
	184\,955--184\,964, Oct. 2020.
	
	\bibitem{jameel2017massive}
	F.~Jameel, M.~A.~A. Haider, A.~A. Butt \emph{et~al.}, ``Massive {MIMO}: A
	survey of recent advances, research issues and future directions,'' in
	\emph{2017 International Symposium on Recent Advances in Electrical
		Engineering (RAEE)}.\hskip 1em plus 0.5em minus 0.4em\relax Islamabad,
	Pakistan: IEEE, 2017, pp. 1--6.
	
	\bibitem{molisch2017hybrid}
	A.~F. Molisch, V.~V. Ratnam, S.~Han, Z.~Li, S.~L.~H. Nguyen, L.~Li, and
	K.~Haneda, ``Hybrid beamforming for massive {MIMO}: A survey,'' \emph{IEEE
		Communications Magazine}, vol.~55, no.~9, pp. 134--141, Sep. 2017.
	
	\bibitem{chandhar2017massive}
	P.~Chandhar, D.~Danev, and E.~G. Larsson, ``Massive {MIMO} for communications
	with drone swarms,'' \emph{IEEE Transactions on Wireless Communications},
	vol.~17, no.~3, pp. 1604--1629, Mar. 2017.
	
	\bibitem{geraci2018supporting}
	G.~Geraci, A.~Garcia-Rodriguez, L.~G. Giordano, D.~L{\'o}pez-P{\'e}rez, and
	E.~Bj{\"o}rnson, ``Supporting {UAV} cellular communications through massive
	{MIMO},'' in \emph{2018 IEEE International Conference on Communications
		Workshops (ICC Workshops)}.\hskip 1em plus 0.5em minus 0.4em\relax Kansas
	City, MO, USA: IEEE, 2018, pp. 1--6.
	
	\bibitem{d2019cell}
	C.~D'Andrea, A.~Garcia-Rodriguez, G.~Geraci, L.~G. Giordano, and S.~Buzzi,
	``Cell-free massive {MIMO} for {UAV} communications,'' in \emph{2019 IEEE
		International Conference on Communications Workshops (ICC Workshops)}.\hskip
	1em plus 0.5em minus 0.4em\relax Shanghai, China: IEEE, 2019, pp. 1--6.
	
	\bibitem{you2020leo}
	L.~{You}, K.~{Li}, J.~{Wang}, X.~{Gao}, X.~{Xia}, and B.~{Otterstenx}, ``{LEO}
	satellite communications with {Massive MIMO},'' in \emph{ICC 2020 - 2020 IEEE
		International Conference on Communications (ICC)}, Dublin, Ireland, 2020, pp.
	1--6.
	
	\bibitem{Li2020downlink}
	K.-X. Li, L.~You, J.~Wang, X.~Gao, G.Tsinos, S.~Chatzinotas, and B.~Ottersten,
	``Downlink transmit design in massive {MIMO LEO} satellite communications,''
	\emph{arXiv preprint arXiv:2008.05343}, Aug. 2020.
	
	\bibitem{makki2020survey}
	B.~{Makki}, K.~{Chitti}, A.~{Behravan}, and M.~{Alouini}, ``A survey of {NOMA}:
	Current status and open research challenges,'' \emph{IEEE Open Journal of the
		Communications Society}, vol.~1, pp. 179--189, Jan. 2020.
	
	\bibitem{fang2020energy}
	F.~Fang, Y.~Xu, Q.-V. Pham, and Z.~Ding, ``Energy-efficient design of
	{IRS-NOMA} networks,'' \emph{IEEE Transactions on Vehicular Technology},
	vol.~69, no.~11, pp. 14\,088--14\,092, Nov. 2020.
	
	\bibitem{chu2020robust}
	J.~{Chu}, X.~{Chen}, C.~{Zhong}, and Z.~{Zhang}, ``Robust design for
	{NOMA}-based {Multi-Beam LEO} satellite {I}nternet of {T}hings,'' \emph{IEEE
		Internet of Things Journal}, 2020, in press.
	
	\bibitem{gao2020performance}
	Z.~{Gao}, A.~{Liu}, and X.~{Liang}, ``The performance analysis of downlink
	{NOMA} in {LEO} satellite communication system,'' \emph{IEEE Access}, vol.~8,
	pp. 93\,723--93\,732, May 2020.
	
	\bibitem{nasir2019uav}
	A.~A. {Nasir}, H.~D. {Tuan}, T.~Q. {Duong}, and H.~V. {Poor}, ``{UAV}-enabled
	communication using {NOMA},'' \emph{IEEE Transactions on Communications},
	vol.~67, no.~7, pp. 5126--5138, Jul. 2019.
	
	\bibitem{botsinis2018air}
	P.~{Botsinis}, D.~{Alanis}, C.~{Xu}, Z.~{Babar}, D.~{Chandra}, S.~X. {Ng}, and
	L.~{Hanzo}, ``Air-to-ground {NOMA} systems for the
	“{Internet-Above-the-Clouds}”,'' \emph{IEEE Access}, vol.~6, pp.
	47\,442--47\,460, Aug. 2018.
	
	\bibitem{hou2019multiple}
	T.~{Hou}, Y.~{Liu}, Z.~{Song}, X.~{Sun}, and Y.~{Chen}, ``Multiple antenna
	aided {NOMA} in {UAV} networks: A stochastic geometry approach,'' \emph{IEEE
		Transactions on Communications}, vol.~67, no.~2, pp. 1031--1044, Feb. 2019.
	
	\bibitem{pham2020sum}
	Q.-V. Pham, T.~Huynh-The, M.~Alazab, J.~Zhao, and W.-J. Hwang, ``Sum-rate
	maximization for {UAV}-assisted visible light communications using {NOMA}:
	Swarm intelligence meets machine learning,'' \emph{IEEE Internet of Things
		Journal}, vol.~7, no.~10, pp. 10\,375--10\,387, Oct. 2020.
	
	\bibitem{gong2020towards}
	S.~{Gong}, X.~{Lu}, D.~T. {Hoang}, D.~{Niyato}, L.~{Shu}, D.~I. {Kim}, and
	Y.~{Liang}, ``Towards smart wireless communications via intelligent
	reflecting surfaces: A contemporary survey,'' \emph{IEEE Communications
		Surveys Tutorials}, pp. 1--33, Jun. 2020.
	
	\bibitem{matthiesen2020intelligient}
	B.~Matthiesen, E.~Björnson, E.~De~Carvalho, and P.~Popovski, ``Intelligent
	reflecting surface operation under predictable receiver mobility: A
	continuous time propagation model,'' \emph{arXiv preprint arXiv:2006.06991},
	Aug. 2020.
	
	\bibitem{hua2020uav}
	M.~Hua, L.~Yang, Q.~Wu, C.~Pan, C.~Li, and A.~L. Swindlehurst, ``{UAV}-assisted
	intelligent reflecting surface symbiotic radio system,'' \emph{arXiv preprint
		arXiv:2007.14029}, Aug. 2020.
	
	\bibitem{ge2020joint}
	L.~Ge, P.~Dong, H.~Zhang, J.-B. Wang, and X.~You, ``Joint beamforming and
	trajectory optimization for intelligent reflecting surfaces-assisted {UAV}
	communications,'' \emph{IEEE Access}, vol.~8, pp. 78\,702--78\,712, Apr.
	2020.
	
	\bibitem{zhang2019reflections}
	Q.~Zhang, W.~Saad, and M.~Bennis, ``Reflections in the sky: Millimeter wave
	communication with {UAV}-carried intelligent reflectors,'' in \emph{2019 IEEE
		Global Communications Conference (GLOBECOM)}.\hskip 1em plus 0.5em minus
	0.4em\relax IEEE, Dec. 2019, pp. 1--6.
	
	\bibitem{bupe2015relief}
	P.~Bupe, R.~Haddad, and F.~Rios-Gutierrez, ``Relief and emergency communication
	network based on an autonomous decentralized {UAV} clustering network,'' in
	\emph{SoutheastCon 2015}.\hskip 1em plus 0.5em minus 0.4em\relax Fort
	Lauderdale, FL, USA: IEEE, 2015, pp. 1--8.
	
	\bibitem{gomez2016uav}
	C.~Gomez and H.~Purdie, ``{UAV}-based photogrammetry and geocomputing for
	hazards and disaster risk monitoring--a review,'' \emph{Geoenvironmental
		Disasters}, vol.~3, no.~1, p.~23, 2016.
	
	\bibitem{ejaz2020energy}
	W.~Ejaz, A.~Ahmed, A.~Mushtaq, and M.~Ibnkahla, ``Energy-efficient task
	scheduling and physiological assessment in disaster management using
	{UAV}-assisted networks,'' \emph{Computer Communications}, Apr. 2020.
	
	\bibitem{baranwal2019application}
	E.~Baranwal, P.~Seth, H.~Pande, S.~Raghavendra, and S.~Kushwaha, ``Application
	of unmanned aerial vehicle ({UAV}) for damage assessment of a cultural
	heritage monument,'' in \emph{International Conference on Unmanned Aerial
		System in Geomatics}.\hskip 1em plus 0.5em minus 0.4em\relax Roorkee, India:
	Springer, 2019, pp. 123--131.
	
	\bibitem{amit2016analysis}
	S.~N. K.~B. Amit, S.~Shiraishi, T.~Inoshita, and Y.~Aoki, ``Analysis of
	satellite images for disaster detection,'' in \emph{2016 IEEE International
		Geoscience and Remote Sensing Symposium (IGARSS)}.\hskip 1em plus 0.5em minus
	0.4em\relax Beijing, China: IEEE, 2016, pp. 5189--5192.
	
	\bibitem{doshi2018satellite}
	J.~Doshi, S.~Basu, and G.~Pang, ``From satellite imagery to disaster
	insights,'' \emph{arXiv preprint arXiv:1812.07033}, Dec. 2018.
	
	\bibitem{deepak2019overview}
	G.~Deepak, A.~Ladas, Y.~A. Sambo, H.~Pervaiz, C.~Politis, and M.~A. Imran, ``An
	overview of post-disaster emergency communication systems in the future
	networks,'' \emph{IEEE Wireless Communications}, vol.~26, no.~6, pp.
	132--139, Dec. 2019.
	
	\bibitem{hayat2017multi}
	S.~Hayat, E.~Yanmaz, T.~X. Brown, and C.~Bettstetter, ``Multi-objective {UAV}
	path planning for search and rescue,'' in \emph{2017 IEEE International
		Conference on Robotics and Automation (ICRA)}.\hskip 1em plus 0.5em minus
	0.4em\relax Singapore: IEEE, 2017, pp. 5569--5574.
	
	\bibitem{alotaibi2019lsar}
	E.~T. Alotaibi, S.~S. Alqefari, and A.~Koubaa, ``{LSAR}: Multi-{UAV}
	collaboration for search and rescue missions,'' \emph{IEEE Access}, vol.~7,
	pp. 55\,817--55\,832, Apr. 2019.
	
	\bibitem{nakadai2017development}
	K.~Nakadai, M.~Kumon, H.~G. Okuno, K.~Hoshiba, M.~Wakabayashi, K.~Washizaki,
	T.~Ishiki, D.~Gabriel, Y.~Bando, T.~Morito \emph{et~al.}, ``Development of
	microphone-array-embedded {UAV} for search and rescue task,'' in \emph{2017
		IEEE/RSJ International Conference on Intelligent Robots and Systems
		(IROS)}.\hskip 1em plus 0.5em minus 0.4em\relax Vancouver, BC, Canada: IEEE,
	2017, pp. 5985--5990.
	
	\bibitem{bracciale2016smartsos}
	L.~Bracciale, P.~Loreti, M.~Luglio, C.~Roseti, and F.~Zampognaro, ``{SMARTSOS}:
	Systems for maritime advanced rescue through satellite and optimized sensor
	networks,'' in \emph{2016 3rd International Conference on Information and
		Communication Technologies for Disaster Management (ICT-DM)}.\hskip 1em plus
	0.5em minus 0.4em\relax IEEE, 2016, pp. 1--5.
	
	\bibitem{7369958.2016}
	M.~{Rossi} and D.~{Brunelli}, ``Autonomous gas detection and mapping with
	unmanned aerial vehicles,'' \emph{IEEE Transactions on Instrumentation and
		Measurement}, vol.~65, no.~4, pp. 765--775, Apr. 2016.
	
	\bibitem{7759885.2016}
	Z.~{Zaheer}, A.~{Usmani}, E.~{Khan}, and M.~A. {Qadeer}, ``Aerial surveillance
	system using {UAV},'' in \emph{2016 Thirteenth International Conference on
		Wireless and Optical Communications Networks (WOCN)}, Hyderabad, India, 2016,
	pp. 1--7.
	
	\bibitem{8234482.2017}
	A.~{Saha}, A.~{Kumar}, and A.~K. {Sahu}, ``{FPV} drone with {GPS} used for
	surveillance in remote areas,'' in \emph{2017 Third International Conference
		on Research in Computational Intelligence and Communication Networks
		(ICRCICN)}, Kolkata, India, 2017, pp. 62--67.
	
	\bibitem{dai2018quality}
	R.~Dai, S.~Fotedar, M.~Radmanesh, and M.~Kumar, ``Quality-aware {UAV} coverage
	and path planning in geometrically complex environments,'' \emph{Ad Hoc
		Networks}, vol.~73, pp. 95--105, May 2018.
	
	\bibitem{8700598.2019}
	S.~R. {Pokhrel}, J.~{Jin}, and H.~L. {Vu}, ``Mobility-aware multipath
	communication for unmanned aerial surveillance systems,'' \emph{IEEE
		Transactions on Vehicular Technology}, vol.~68, no.~6, pp. 6088--6098, Jun.
	2019.
	
	\bibitem{garg2018uav}
	S.~{Garg}, A.~{Singh}, S.~{Batra}, N.~{Kumar}, and L.~T. {Yang},
	``{UAV}-empowered edge computing environment for cyber-threat detection in
	smart vehicles,'' \emph{IEEE Network}, vol.~32, no.~3, pp. 42--51, Jun. 2018.
	
	\bibitem{Bacco2018smart}
	M.~{Bacco}, A.~{Berton}, E.~{Ferro}, C.~{Gennaro}, A.~{Gotta}, S.~{Matteoli},
	F.~{Paonessa}, M.~{Ruggeri}, G.~{Virone}, and A.~{Zanella}, ``Smart farming:
	Opportunities, challenges and technology enablers,'' in \emph{2018 IoT
		Vertical and Topical Summit on Agriculture - Tuscany (IOT Tuscany)}, Tuscany,
	Italy, 2018, pp. 1--6.
	
	\bibitem{Islam2019bigdata}
	M.~N. {Islam Sarker}, M.~{Wu}, B.~{Chanthamith}, S.~{Yusufzada}, D.~{Li}, and
	J.~{Zhang}, ``Big data driven smart agriculture: Pathway for sustainable
	development,'' in \emph{2019 2nd International Conference on Artificial
		Intelligence and Big Data (ICAIBD)}, Chengdu, China, China, 2019, pp. 60--65.
	
	\bibitem{boursianis2020internet}
	A.~D. Boursianis, M.~S. Papadopoulou, P.~Diamantoulakis, A.~Liopa-Tsakalidi,
	P.~Barouchas, G.~Salahas, G.~Karagiannidis, S.~Wan, and S.~K. Goudos,
	``Internet of things ({IoT}) and agricultural unmanned aerial vehicles
	({UAVs}) in smart farming: A comprehensive review,'' \emph{Internet of
		Things}, p. 100187, Mar. 2020.
	
	\bibitem{maddikunta2020unmanned}
	P.~K.~R. Maddikunta, S.~Hakak, M.~Alazab, S.~Bhattacharya, T.~R. Gadekallu,
	W.~Z. Khan, and Q.-V. Pham, ``Unmanned aerial vehicles in smart agriculture:
	Applications, requirements, and challenges,'' \emph{IEEE Sensors Journal},
	2021.
	
	\bibitem{tam2020monitoring}
	N.~T. Tam, H.~T. Dat, P.~M. Tam, V.~T. Trinh, N.~T. Hung, Q.-T. Huynh, and
	J.~Jo, ``Monitoring agriculture areas with satellite images and deep
	learning,'' \emph{Applied Soft Computing}, p. 106565, Oct. 2020.
	
	\bibitem{murugan2016fusion}
	D.~Murugan, A.~Garg, T.~Ahmed, and D.~Singh, ``Fusion of drone and satellite
	data for precision agriculture monitoring,'' in \emph{2016 11th International
		Conference on Industrial and Information Systems (ICIIS)}.\hskip 1em plus
	0.5em minus 0.4em\relax Roorkee, India: IEEE, 2016, pp. 910--914.
	
	\bibitem{murugan2017development}
	D.~Murugan, A.~Garg, and D.~Singh, ``Development of an adaptive approach for
	precision agriculture monitoring with drone and satellite data,'' \emph{IEEE
		Journal of Selected Topics in Applied Earth Observations and Remote Sensing},
	vol.~10, no.~12, pp. 5322--5328, Dec. 2017.
	
	\bibitem{Chowdhury20206G}
	M.~Z. {Chowdhury}, M.~{Shahjalal}, S.~{Ahmed}, and Y.~M. {Jang}, ``{6G}
	wireless communication systems: Applications, requirements, technologies,
	challenges, and research directions,'' \emph{IEEE Open Journal of the
		Communications Society}, vol.~1, pp. 957--975, Jul. 2020.
	
	\bibitem{8460024.2018}
	C.~{Grasso} and G.~{Schembra}, ``Design of a {UAV}-based videosurveillance
	system with tactile internet constraints in a {5G} ecosystem,'' in \emph{2018
		4th IEEE Conference on Network Softwarization and Workshops (NetSoft)},
	Montreal, QC, Canada, 2018, pp. 449--455.
	
	\bibitem{9040423.2020}
	Y.~{Jin}, Z.~{Qian}, and W.~{Yang}, ``{UAV} cluster-based video surveillance
	system optimization in heterogeneous communication of smart cities,''
	\emph{IEEE Access}, vol.~8, pp. 55\,654--55\,664, Mar. 2020.
	
	\bibitem{ullah2019uav}
	S.~Ullah, K.-I. Kim, K.~H. Kim, M.~Imran, P.~Khan, E.~Tovar, and F.~Ali,
	``{UAV}-enabled healthcare architecture: Issues and challenges,''
	\emph{Future Generation Computer Systems}, vol.~97, pp. 425--432, Jan. 2019.
	
	\bibitem{9148581.2020}
	C.~{Tang}, C.~{Zhu}, X.~{Wei}, J.~J. P.~C. {Rodrigues}, M.~{Guizani}, and
	W.~{Jia}, ``{UAV} placement optimization for internet of medical things,'' in
	\emph{2020 International Wireless Communications and Mobile Computing
		(IWCMC)}, Limassol, Cyprus, Cyprus, 2020, pp. 752--757.
	
	\bibitem{9162738.2020}
	R.~{Gupta}, A.~{Shukla}, P.~{Mehta}, P.~{Bhattacharya}, S.~{Tanwar},
	S.~{Tyagi}, and N.~{Kumar}, ``{VAHAK}: A blockchain-based outdoor delivery
	scheme using {UAV} for healthcare 4.0 services,'' in \emph{IEEE INFOCOM 2020
		- IEEE Conference on Computer Communications Workshops (INFOCOM WKSHPS)},
	Aug., 2020, pp. 255--260.
	
	\bibitem{8355261.2018}
	M.~I. {Malik}, I.~{Mcateer}, P.~{Hannay}, and Z.~{Baig}, ``Preparing for secure
	wireless medical environment in 2050: A vision,'' \emph{IEEE Access}, vol.~6,
	pp. 25\,666--25\,674, May 2018.
	
	\bibitem{anema2020shaping}
	A.~Anema, N.~D. Preston, M.~Platz, and C.~Unnithan, ``Shaping the future of
	global health: A review of canadian space technology applications in
	healthcare,'' in \emph{Space Capacity Building in the XXI Century}.\hskip 1em
	plus 0.5em minus 0.4em\relax Springer, Apr. 2020, pp. 193--205.
	
	\bibitem{menouar2017uav}
	H.~Menouar, I.~Guvenc, K.~Akkaya, A.~S. Uluagac, A.~Kadri, and A.~Tuncer,
	``{UAV}-enabled intelligent transportation systems for the smart city:
	Applications and challenges,'' \emph{IEEE Communications Magazine}, vol.~55,
	no.~3, pp. 22--28, Mar. 2017.
	
	\bibitem{8326145.2018}
	C.~{Kyrkou}, S.~{Timotheou}, P.~{Kolios}, T.~{Theocharides}, and C.~G.
	{Panayiotou}, ``Optimized vision-directed deployment of {UAV}s for rapid
	traffic monitoring,'' in \emph{2018 IEEE International Conference on Consumer
		Electronics (ICCE)}, Las Vegas, NV, USA, 2018, pp. 1--6.
	
	\bibitem{9076661.2020}
	R.~A. {Nazib} and S.~{Moh}, ``Routing {P}rotocols for {U}nmanned {A}erial
	{V}ehicle-{A}ided {V}ehicular {A}d {H}oc {N}etworks: A survey,'' \emph{IEEE
		Access}, vol.~8, pp. 77\,535--77\,560, Apr. 2020.
	
	\bibitem{samir2019trajectory}
	M.~Samir, S.~Sharafeddine, C.~Assi, T.~M. Nguyen, and A.~Ghrayeb, ``Trajectory
	planning and resource allocation of multiple {UAVs} for data delivery in
	vehicular networks,'' \emph{IEEE Networking Letters}, vol.~1, no.~3, pp.
	107--110, Sep. 2019.
	
	\bibitem{wu2020optimal}
	H.~Wu, F.~Lyu, C.~Zhou, J.~Chen, L.~Wang, and X.~Shen, ``Optimal {UAV} caching
	and trajectory in aerial-assisted vehicular networks: A learning-based
	approach,'' \emph{IEEE Journal on Selected Areas in Communications}, vol.~38,
	no.~12, pp. 2783--2797, Dec. 2020.
	
	\bibitem{hadiwardoyo2020three}
	S.~A. Hadiwardoyo, C.~T. Calafate, J.-C. Cano, K.~Krinkin, D.~Klionskiy,
	E.~Hern{\'a}ndez-Orallo, and P.~Manzoni, ``Three dimensional {UAV}
	{P}ositioning for {D}ynamic {UAV}-to-{C}ar {C}ommunications,''
	\emph{Sensors}, vol.~20, no.~2, p. 356, Jan. 2020.
	
	\bibitem{9110438.2020}
	O.~{Bouachir}, M.~{Aloqaily}, I.~A. {Ridhawi}, O.~{Alfandi}, and H.~B.
	{Salameh}, ``{UAV}-{A}ssisted {V}ehicular {C}ommunication for {D}ensely
	{C}rowded {E}nvironments,'' in \emph{NOMS 2020 - 2020 IEEE/IFIP Network
		Operations and Management Symposium}, Budapest, Hungary, 2020, pp. 1--4.
	
	\bibitem{dovis2020recent}
	F.~Dovis, L.~Ruotsalainen, R.~Toledo-Moreo, Z.~Z.~M. Kassas, and V.~Gikas,
	``Recent advancement on the use of global navigation satellite system-based
	positioning for intelligent transport systems [guest editorial],'' \emph{IEEE
		Intelligent Transportation Systems Magazine}, vol.~12, no.~3, pp. 6--9, Jul.
	2020.
	
	\bibitem{7451189.2016}
	V.~{Sharma}, M.~{Bennis}, and R.~{Kumar}, ``{UAV}-assisted heterogeneous
	networks for capacity enhancement,'' \emph{IEEE Communications Letters},
	vol.~20, no.~6, pp. 1207--1210, Jun. 2016.
	
	\bibitem{8533634.2019}
	M.~{Mozaffari}, A.~{Taleb Zadeh Kasgari}, W.~{Saad}, M.~{Bennis}, and
	M.~{Debbah}, ``Beyond {5G} with {UAVs}: Foundations of a {3D} wireless
	cellular network,'' \emph{IEEE Transactions on Wireless Communications},
	vol.~18, no.~1, pp. 357--372, Jan. 2019.
	
	\bibitem{7289337.2016}
	M.~{De Sanctis}, E.~{Cianca}, G.~{Araniti}, I.~{Bisio}, and R.~{Prasad},
	``Satellite communications supporting internet of remote things,'' \emph{IEEE
		Internet of Things Journal}, vol.~3, no.~1, pp. 113--123, Feb. 2016.
	
	\bibitem{9034074.2020}
	S.~{Zhang}, C.~{Xiang}, and S.~{Xu}, ``6{G}: Connecting everything by 1000
	times price reduction,'' \emph{IEEE Open Journal of Vehicular Technology},
	vol.~1, pp. 107--115, Mar. 2020.
	
	\bibitem{letaief2019roadmap}
	K.~B. Letaief, W.~Chen, Y.~Shi, J.~Zhang, and Y.-J.~A. Zhang, ``The roadmap to
	{6G: AI} empowered wireless networks,'' \emph{IEEE Communications Magazine},
	vol.~57, no.~8, pp. 84--90, Aug. 2019.
	
	\bibitem{yang20196g}
	P.~Yang, Y.~Xiao, M.~Xiao, and S.~Li, ``{6G} wireless communications: Vision
	and potential techniques,'' \emph{IEEE Network}, vol.~33, no.~4, pp. 70--75,
	Jul. 2019.
	
\end{thebibliography}


\end{document}